\documentclass[aps,prd,showpacs,nofootinbib,floatfix,eqsecnum]{revtex4}
\usepackage{amsmath,amssymb,latexsym,graphicx,multirow,psfrag}

\newcommand{\Cs}{\mathcal{C}}
\newcommand{\Is}{\mathcal{I}}
\newcommand{\Js}{\mathcal{J}} 
\newcommand{\Ms}{\mathcal{M}}
\newcommand{\Ns}{\mathcal{N}}
\newcommand{\Rs}{\mathcal{R}}
\newcommand{\Ss}{\mathcal{S}}

\allowdisplaybreaks

\begin{document}
\title{Compact binary systems in scalar-tensor gravity. III. Scalar waves and energy flux}
\author{Ryan N.\ Lang$^{1,2,}$}
\email{lang@phys.ufl.edu}
\affiliation{$^1$Department of Physics, University of Florida, Gainesville, Florida 32611, USA \\
$^2$Department of Physics, University of Illinois at Urbana-Champaign, Urbana, Illinois 61801, USA}

\begin{abstract}
 We derive the scalar waveform generated by a binary of nonspinning compact objects (black holes or neutron stars) in a general class of scalar-tensor theories of gravity.  The waveform is accurate to 1.5 post-Newtonian order [$O((v/c)^3)$] beyond the leading-order tensor gravitational waves (the ``Newtonian quadrupole'').  To solve the scalar-tensor field equations, we adapt the direct integration of the relaxed Einstein equations formalism developed by Will, Wiseman, and Pati.  The internal gravity of the compact objects is treated with an approach developed by Eardley.  We find that the scalar waves are described by the same small set of parameters which describes the equations of motion and tensor waves.  For black hole--black hole binaries, the scalar waveform vanishes, as expected from previous results which show that these systems in scalar-tensor theory are indistinguishable from their general relativistic counterparts.  For black hole--neutron star binaries, the scalar waveform simplifies considerably from the generic case, essentially depending on only a single parameter up to first post-Newtonian order.  With both the tensor and scalar waveforms in hand, we calculate the total energy flux carried by the outgoing waves.  This quantity is computed to first post-Newtonian order relative to the ``quadrupole formula'' and agrees with previous, lower order calculations.
\end{abstract}
\maketitle

\section{Introduction}
\label{sec:intro}
We are entering the age of gravitational-wave (GW) astronomy.  The Advanced LIGO interferometric detectors in Louisiana and Washington \cite{h10a} will come online in 2015, followed later by Advanced Virgo in Italy \cite{virgo}, KAGRA in Japan \cite{kagra}, and possibly a third LIGO detector in India \cite{india}.  This network of detectors will measure GWs in the ``high-frequency'' band, $f = $ 1--$10^3$ Hz.  The International Pulsar Timing Array collaboration is currently using pulsar timing residuals to search for GWs in the ``very-low-frequency'' band, $f = 10^{-9}$--$10^{-7}$ Hz \cite{m14}.  In between these two bands, GW detection requires interferometry in space.  This science is the target of the European Space Agency ``L3'' mission, due to launch in 2034, most likely with an instrument resembling the well-known eLISA design \cite{a13}.

The primary sources for all of these experiments are compact binaries comprising white dwarfs, neutron stars, and black holes.  We expect to learn a great deal about the astrophysics of these systems, ranging from simple event rates to very precise information on their masses, spins, orientations, and sky locations.  We can also use the gravitational waves they emit to test Einstein's theory of general relativity (GR).

General relativity is nearing its one-hundredth birthday and continues to pass all tests with flying colors.  Deviations from GR have been strongly constrained by measurements of the Solar System and of binary pulsar systems \cite{w14}.  However, GR has not yet been tested in the strong-field, dynamical regime we expect to probe with the GW observatories.  In particular, as compact binaries emit gravitational radiation, they lose energy to that radiation, causing the binary orbit to decay.  Eventually, the two objects merge, producing a final burst of gravitational waves before settling into a quiescent state.  The combination of strong gravitational fields and short time scales involved in this inspiral and merger process represents a limit in which general relativity might break down.

One peculiarity of gravitational-wave detection is the need for very accurate template waveforms with which to compare the data.  While perhaps some very strong signals in the eLISA detector could be picked out by eye, most GW signals will be heavily buried in noise.  Matched filtering with a template bank is often the only way to recover the actual GW data.  While it is possible to conduct searches without templates (and sometimes necessary, in the case of unmodeled burst searches), templates are necessary to fully realize the detection capability of an instrument.  Furthermore, finding the best-fit template allows one to precisely determine the system parameters for purposes of astrophysics, as well as search for deviations from general relativity.

With this application in mind, we have undertaken an effort to produce waveform models for a particular alternative to general relativity, scalar-tensor theory.  This theory dates back over 50 years and yet remains extremely relevant today.  For instance, many so-called $f(R)$ theories, which modify the action of general relativity to allow arbitrary functions of the Ricci scalar, can be expressed in the form of a scalar-tensor theory \cite{dt10}.  These $f(R)$ theories may explain the acceleration of the Universe without resorting to dark energy.  Scalar-tensor theory is also a potential low-energy limit of string theory \cite{fm03}.

The term ``scalar-tensor theory'' is really a catch-all for a collection of theories, possibly including multiple scalar fields in addition to the tensor field (metric) of general relativity.  We limit ourselves to a single scalar.  The action we choose is

\begin{equation}
S = \frac{1}{16\pi}\int \left[\phi R-\frac{\omega(\phi)}{\phi}g^{\mu \nu}\partial_\mu \phi \partial_\nu \phi\right]\sqrt{-g}\ d^4x+S_m(\mathfrak{m},g_{\mu \nu}) \, ,
\label{eq:action}
\end{equation}
where $g_{\mu \nu}$ is the spacetime metric, $g$ is its determinant, $R$ is the Ricci scalar derived from this metric, $\phi$ is the scalar field, and $\omega$ is the scalar-tensor coupling.  Note that $\omega = \omega(\phi)$ is not a constant; that is, we are not restricting our attention to Brans-Dicke theory.  We do, however, restrict ourselves only to massless scalar fields (i.e., those without a potential).  We have also written $S_m$ to represent the matter action.  Note that it depends only on the matter fields $\mathfrak{m}$ and the metric; the scalar field $\phi$ does not couple directly to the matter.  This means that \eqref{eq:action} is expressed in the ``Jordan'' frame.  All of our work will be done in this frame.  An alternative representation is the ``Einstein'' frame, related to the Jordan frame by a conformal transformation \cite{de92}.  Regardless of in which frame the calculation is done, the final result---the gravitational waveform---will be the same.

The action generates the following field equations:
\begin{subequations}
\begin{align}
G_{\mu \nu} &= \frac{8\pi}{\phi}T_{\mu \nu}+\frac{\omega(\phi)}{\phi^2}\left(\phi_{,\mu}\phi_{,\nu}-\frac{1}{2}g_{\mu\nu}\phi_{,\lambda}\phi^{,\lambda}\right)+\frac{1}{\phi}(\phi_{;\mu\nu}-g_{\mu\nu}\Box_g \phi) \, ,
\label{eq:tensoreqn}\\
\Box_g \phi &= \frac{1}{3+2\omega(\phi)}\left(8\pi T - 16\pi\phi\frac{\partial T}{\partial \phi}-\frac{d\omega}{d\phi}\phi_{,\lambda}\phi^{,\lambda}\right) \, .
\label{eq:scalareqn}
\end{align}
\end{subequations}
Here $G_{\mu \nu}$ is the Einstein tensor, $T_{\mu \nu}$ is the stress energy of matter and nongravitational fields, and $T \equiv g^{\alpha \beta}T_{\alpha \beta}$ is its trace.  We use commas to denote ordinary derivatives.  Semicolons denote covariant derivatives (taken using $g_{\mu \nu}$ in the usual way), and $\Box_g \equiv g^{\alpha \beta}\partial_\alpha \partial_\beta$ is the d'Alembertian with indices raised by the metric.  These conventions will be used throughout the paper.  

The computation of the gravitational waveform in this set of scalar-tensor theories spans three papers, including this one.  We focus on the inspiral phase, when the compact objects can be regarded as separate bodies, and derive the waveform using the post-Newtonian (PN) approximation.  This approximation is an expansion in powers of $v/c \sim (Gm/rc^2)^{1/2}$, with each power representing half a post-Newtonian order, and is valid until just before the end of the inspiral phase.  In the first paper (\cite{mw13}, hereafter paper I), Mirshekari and Will computed the equations of motion for a compact binary to order $(v/c)^5$ (2.5PN) beyond the leading (``Newtonian'') term.  This result is a necessary precursor for computing the waveform, as well as extremely interesting in its own right.  In the second paper (\cite{l14}, hereafter paper II), we derived the tensor gravitational waveform to order $(v/c)^4$ (2PN) beyond the leading term.  In both of these papers, the final results were shown to reduce to the general relativistic versions in the appropriate limit.

In this paper, we compute the final piece of the puzzle, which has no GR analog: the scalar waveform.  A gravitational-wave detector will measure both tensor and scalar waves.  In paper II, we showed that the separation between masses $\xi^i$ in a detector will obey the equation

\begin{equation}
\ddot{\xi}^i = -R_{0i0j}\xi^j \, ,
\end{equation}
where

\begin{equation}
R_{0i0j} = -\frac{1}{2} \ddot{\tilde{h}}^{ij}_\text{TT} -\frac{1}{2}\ddot{\Psi}(\hat{N}^i\hat{N}^j-\delta^{ij}) \, .
\end{equation}
Here $\hat{N}^i$ is the direction from the source to the detector, $\delta^{ij}$ is the Kronecker delta, and $\tilde{h}^{ij}$ and $\Psi$ are redefined tensor and scalar fields which we use in this series.  They are written out explicitly in Sec.\ \ref{sec:DIRE} below.  The subscript ``TT'' designates the transverse-traceless projection of the tensor field.  The tensor field has the usual two polarizations expected in general relativity, $+$ and $\times$, though the exact time dependence of the waves will take the new form computed in paper II.  The scalar waves, on the other hand, will appear as a transverse ``breathing'' mode.  The detection of a third polarization state like this one would be a smoking gun that general relativity is somehow incorrect. 

The scalar waveform actually begins at $-0.5$PN order, keeping the same definition of ``0PN'' that we use for the tensor waveform in both GR and scalar-tensor theory.  This feature is due to the presence of a nonvanishing ``scalar dipole moment,'' which also creates 1.5PN radiation-reaction effects in the equations of motion and various effects in the tensor waves.  (Technically, there is a monopole piece in the far field at $-1$PN order; however, it is time independent and not wavelike. The leading-order monopole contribution to the waves thus enters at 0PN order.)  Because of this ``shifting'' of post-Newtonian orders, it is much more difficult to calculate the scalar waveform at the same order as the tensor waveform.  Therefore, we compute the scalar waveform only to 1.5PN order.

The challenge in all three papers is to solve the scalar-tensor field equations \eqref{eq:tensoreqn}--\eqref{eq:scalareqn}.  To do so, we use a method called ``direct integration of the relaxed Einstein equations'' (DIRE), based on the original framework of Epstein and Wagoner \cite{ew75} and then extended by Will, Wiseman, and Pati \cite{w92, ww96, pw00, pw02}.  It is easily adapted to scalar-tensor theories.  In the adapted DIRE method, the scalar-tensor field equations are first rewritten in a ``relaxed'' form: flat-spacetime wave equations for the ``gravitational field'' $\tilde{h}^{\mu \nu}$ and the modified scalar field $\Psi$.  The wave equations are simplified by the choice of a particular coordinate system, represented by a gauge condition on $\tilde{h}^{\mu \nu}$.  Together, the wave equations and gauge condition contain all the content of the full field equations.  The wave equations can be solved formally by using a retarded Green's function.  When converting to a more useful form, we evaluate these formal solutions differently depending on whether the source and field points are close to the compact objects (in the ``near zone'') or far away (in the ``radiation zone'').  The four different methods of evaluation are described in detail in \cite{ww96,pw00}.

Since scalar-tensor theories do not obey the strong equivalence principle, the motion and gravitational-wave emission of a binary depend on the internal composition of its constituent bodies.  To handle this effect, we have adopted the approach of Eardley \cite{e75}.  We treat the matter stress-energy tensor as a sum of delta functions located at the position of each compact object.  However, instead of assigning each body a constant mass, we let the mass be a function of the scalar field, $M_A = M_A(\phi)$.  This gives the matter action an indirect dependence on $\phi$, even though we still work in the Jordan frame.  This is the origin of the term $\partial T/\partial \phi$ in \eqref{eq:scalareqn}.  In the final waveform, we express the varying mass using the ``sensitivity,''

\begin{equation}
s_A \equiv \left(\frac{d \ln M_A(\phi)}{d \ln \phi}\right)_0 \, ,
\end{equation}
as well as derivatives of this quantity.  (The subscript 0 means that the derivative should be evaluated using the asymptotic value of the scalar field, $\phi_0$.)  In the weak-field limit, the sensitivity is proportional to the Newtonian self-gravitational energy per unit mass of the body.  For neutron stars, the sensitivity depends on the mass and equation of state of the star, with typical values 0.1--0.3 \cite{wz89,z92}.  For black holes, $s = 0.5$, and all derivatives vanish.  Since we assume that the sensitivities are constant in time, our work does not capture any ``dynamical scalarization'' effects \cite{bppl13,pbpl13,stob14}; however, we expect such effects to be relevant only at the end of inspiral, when the post-Newtonian approximation breaks down. 

The scalar waves we find depend on the same relatively small set of parameters that characterizes the equations of motion and tensor waves.  These parameters are defined in terms of the coupling function (or more precisely, its Taylor expansion coefficients) and the sensitivities and sensitivity derivatives of the compact objects.  For black hole--black hole binaries, the scalar waveform vanishes completely.  This result is not unexpected.  Papers I and II showed that black hole binaries in scalar-tensor theory cannot be distinguished from their general relativistic counterparts by studying their motion or tensor wave emission.  For mixed (black hole--neutron star) binaries, the scalar waves simplify considerably.  After a mass rescaling, the waves up to 1PN order only depend on two parameters: the same parameter $Q$ which characterizes the equations of motion and the tensor waves for mixed binaries, plus an overall amplitude factor.  At 1.5PN order, the expression becomes more complicated.

Having computed both the tensor and scalar radiation from the system, we also calculate the total energy flux carried off by the waves.  This energy loss backreacts on the system, causing the radius of the orbit to shrink and the frequency of the orbit and GWs to increase (``chirp'').  The expression we derive can be used to generate useful template waveforms for GW detectors.  Because of the strange nature of post-Newtonian counting in scalar-tensor theory, specifically the existence of a negative-order dipole contribution to the scalar waves, computing the energy flux at $N$th post-Newtonian order actually requires knowing at least some pieces of the scalar waveform at $(N+1/2)$th order.  For this reason, we stop our calculation of the flux at first post-Newtonian (1PN) order, where ``0PN'' is defined as the lowest order tensor flux (i.e., the ``quadrupole formula'').  We plan to continue the calculation to higher order in future work.

The outline of the paper is as follows: Section \ref{sec:review} reviews necessary results from papers I and II, including the relaxed field equations, the Eardley approach to the matter source, and the definition of various ``potentials'' with which we express intermediate results.  Section \ref{sec:NZintegrals} describes the calculation of the near-zone contribution to the scalar waveform.  We first describe the general technique, which requires the computation of ``scalar multipole moments.''  We then review the construction of the source $\tau_s$ in the near zone and give explicit expressions for it.  Next, we calculate the scalar multipole moments.  The techniques for doing so are identical to those used in calculating the ``Epstein-Wagoner moments'' of paper II.  We then simplify the scalar moments to two-body systems and convert them to relative coordinates.  We also discuss how the equations of motion from paper I are used to expand time derivatives of the moments.

Section \ref{sec:RZintegrals} presents the radiation-zone contribution to the scalar waves.  We first discuss the formalism for finding this contribution and review results from paper II which are needed to construct $\tau_s$ in the radiation zone.  We then compute the waveform produced by this source.  These terms begin at 1PN order, half an order lower than in the tensor case.  They include scalar ``tail'' terms which depend on the entire history of the binary.  We call such terms ``hereditary'' terms.  Unlike the tensor case, there are no nontail hereditary terms.

In Sec.\ \ref{sec:scalarwaves}, we present the full 1.5PN scalar waveform for a nonspinning compact binary in massless scalar-tensor theory.  Finally, in Sec.\ \ref{sec:flux}, we compute the rate at which energy is carried away from the system by both tensor and scalar waves.

In this paper, we use units in which $c = 1$.  We do not set $G = 1$; as we shall see, the effective Newtonian gravitational constant depends on the asymptotic value of the scalar field.  Greek indices run over four spacetime values (0, 1, 2, 3), while Latin indices run over three spatial values (1, 2, 3).  We use the Einstein summation convention, in which repeated indices are summed over.  We use a multi-index notation for products of vector components: $x^{ijk} \equiv x^ix^jx^k$.  A capital letter superscript denotes a product of that dimensionality: $x^L \equiv x^{k_1}x^{k_2}\cdots x^{k_l}$.  Angular brackets around indices denote symmetric, trace-free (STF) products (see Appendix B of paper II for details).  Finally, we use standard notation for symmetrized indices, e.g. $x^{(i}y^{j)} \equiv (x^iy^j+x^jy^i)/2$. 

\section{Preliminaries}
\label{sec:review}
In this section, we review some key concepts from papers I and II with which the reader should be familiar before continuing on to the next sections.  For brevity, we omit many details.  They can be found in papers I and II.

\subsection{Reduced field equations}
\label{sec:DIRE}
The adapted DIRE method solves the scalar-tensor field equations \eqref{eq:tensoreqn}--\eqref{eq:scalareqn} by rewriting them in an equivalent, but more easily manageable, form.  We first assume that far away from the compact binary, the metric reduces to the Minkowski metric, $\eta_{\mu \nu}$, and the scalar field tends to a constant $\phi_0$.  We introduce a rescaled scalar field,

\begin{equation}
\varphi \equiv \frac{\phi}{\phi_0} \equiv 1 + \Psi\, .
\end{equation}
Calculating the form of $\Psi$ at a distant gravitational-wave detector is the primary goal of this paper.  We also define a new tensor field of interest, the ``gravitational field'' $\tilde{h}^{\mu \nu}$,

\begin{equation}
\tilde{h}^{\mu \nu} \equiv \eta^{\mu \nu} - \tilde{\mathfrak{g}}^{\mu \nu} \, .
\label{eq:htilde}
\end{equation}
The ``gothic'' metric is given by

\begin{equation}
\tilde{\mathfrak{g}}^{\mu \nu} \equiv \sqrt{-\tilde{g}}\tilde{g}^{\mu \nu} \, ,
\end{equation}
where

\begin{equation}
\tilde{g}_{\mu \nu} \equiv \varphi g_{\mu \nu} 
\label{eq:gtilde}
\end{equation}
is a conformally transformed version of the metric and $\tilde{g}$ is its determinant.  In general relativity, the DIRE method defines a gravitational field in much the same way.  However, since there is no scalar field, the regular metric $g_{\mu \nu}$ and its determinant $g$ take the place of the conformally transformed versions.  For convenience, we define the components of the gravitational-field tensor to be 

\begin{subequations}
\begin{align}
\tilde{h}^{00} &\equiv N \, ,\\
\tilde{h}^{0i} &\equiv K^i \, ,\\
\tilde{h}^{ij} &\equiv B^{ij} \, ,\\
\tilde{h}^{ii} &\equiv B \, .
\end{align}
\end{subequations}

We are free to pick coordinates in which the field equations take on a simpler form.  We choose the Lorenz gauge condition

\begin{equation}
{\tilde{h}^{\mu \nu}}_{\hphantom{\mu\nu}, \nu} = 0 \, .
\label{eq:gauge}
\end{equation}
Then the field equation \eqref{eq:tensoreqn} reduces to

\begin{equation}
\Box_\eta \tilde{h}^{\mu \nu} = -16\pi \tau^{\mu \nu} \, ,
\label{eq:tensorwave}
\end{equation}
where $\Box_\eta \equiv \eta^{\alpha \beta}\partial_\alpha \partial_\beta$ is the flat-spacetime wave operator and the source is

\begin{equation}
\tau^{\mu\nu} \equiv (-g)\frac{\varphi}{\phi_0}T^{\mu \nu} + \frac{1}{16\pi}(\Lambda^{\mu\nu} + \Lambda_s^{\mu \nu}) \, .
\end{equation}
Here $T^{\mu \nu}$ is the stress energy of matter and nongravitational fields; in our case, it is generated by the compact source.  We will discuss it further in Sec.\ \ref{sec:eardley}.  The two terms $\Lambda^{\mu\nu}$ and $\Lambda_s^{\mu\nu}$ depend on the gravitational (tensor) and scalar fields.  Exact expressions for them can be found in paper II.  Our choice of variables ensures that $\Lambda^{\mu\nu}$ maintains the same basic form as in general relativity; compare Eqs.\ (3.4) in paper I with Eqs.\ (4.4) in \cite{pw00}.  The $\Lambda_s^{\mu\nu}$ term is new to scalar-tensor theory.  The gauge condition \eqref{eq:gauge} implies a conservation law for the source,

\begin{equation}
{\tau^{\mu\nu}}_{,\nu} = 0 \, .
\label{eq:conservation}
\end{equation}
The scalar field equation \eqref{eq:scalareqn} can also be written as a flat-spacetime wave equation,

\begin{equation}
\Box_\eta \Psi = -8\pi \tau_s \, ,
\label{eq:scalarwave}
\end{equation}
with source

\begin{equation}
\tau_s \equiv -\frac{1}{3+2\omega}\sqrt{-g}\frac{\varphi}{\phi_0}\left(T-2\varphi\frac{\partial T}{\partial \varphi}\right)+\frac{1}{16\pi}\frac{d}{d\varphi}\left[\ln\left(\frac{3+2\omega}{\varphi^2}\right)\right]\varphi_{,\alpha}\varphi_{,\beta}\tilde{\mathfrak{g}}^{\alpha\beta}-\frac{1}{8\pi}\tilde{h}^{\alpha\beta}\varphi_{,\alpha\beta} \, .
\label{eq:taus}
\end{equation}
Recall that $T$ is the trace of $T^{\mu \nu}$.

The wave equations \eqref{eq:tensorwave} and \eqref{eq:scalarwave} can be solved formally in all spacetime by using a retarded Green's function,

\begin{subequations}
\begin{align}
\tilde{h}^{\mu \nu}(t,\mathbf{x}) &= 4\int \frac{\tau^{\mu \nu}(t',\mathbf{x}')\delta(t'-t+|\mathbf{x}-\mathbf{x}'|)}{|\mathbf{x}-\mathbf{x}'|}d^4x' \, ,\label{eq:hintegral}\\
\Psi(t,\mathbf{x}) &= 2\int \frac{\tau_s(t',\mathbf{x}')\delta(t'-t+|\mathbf{x}-\mathbf{x}'|)}{|\mathbf{x}-\mathbf{x}'|}d^4x' \, .
\label{eq:phiintegral}
\end{align}
\end{subequations}
The integrations take place over the past flat-spacetime null cone $\Cs$ emanating from the field point $(t,\mathbf{x})$.  To obtain explicit solutions, we divide the spacetime into two regions.  We define the characteristic size of the source as $\Ss$ and assume the bodies move at velocities $v \ll 1$.  Then the ``near zone'' is defined as the area with $|\mathbf{x}| = R < \Rs$, where $\Rs \sim \Ss/v$ is the characteristic wavelength of gravitational radiation from the system.  (We use capital $R$ to denote the distance from the binary's center of mass to a field point in order to avoid confusion later with $r$, the orbital separation of the binary.)  Everything outside the near zone ($R > \Rs$) is the ``radiation zone.''

There are a total of four different ways to evaluate \eqref{eq:hintegral}--\eqref{eq:phiintegral}, depending on which of the two zones contains the field and source points.  For instance, in paper I, the integrals were evaluated for field points in the near zone.  The integration was split into two pieces, one piece for source points in the near zone and one piece for source points in the radiation zone.  (The latter turned out not to contribute at the post-Newtonian order considered.)  In paper II, \eqref{eq:hintegral} was evaluated for field points in the radiation zone, i.e., far away from the source where gravitational waves are measured.  The integral was again split into two pieces, one for source points in the near zone and one for source points in the radiation zone.  In this paper, we evaluate \eqref{eq:phiintegral} for field points in the radiation zone.  Integration over the near-zone source points is described in Sec.\ \ref{sec:NZintegrals}.  Integration over the radiation-zone source points is described in Sec.\ \ref{sec:RZintegrals}.

Because the radius $\Rs$ of the boundary between zones is completely arbitrary, no terms in the final expressions for the fields should depend on it.  It was shown in \cite{ww96,pw00} that, for general relativity, the terms dependent on $\Rs$ which arise from the near-zone integral cancel exactly with the terms dependent on $\Rs$ which are generated by the radiation-zone integral.  In our work, we simply assume that this property holds and throw away all terms which depend on $\Rs$.

\subsection{Matter source and potentials}
\label{sec:eardley}
Following paper I, the compact source can be described in terms of ``$\sigma$ densities'' \cite{bd89},

\begin{subequations}
\begin{align}
\sigma &\equiv T^{00} + T^{ii} \, ,\\
\sigma^i &\equiv T^{0i} \, , \\
\sigma^{ij} &\equiv T^{ij} \, , \\
\sigma_s &\equiv -T + 2\varphi \frac{\partial T}{\partial \varphi} \, .
\end{align}
\end{subequations}
From these, we can define a number of Poisson-like potentials.  For example, given a generic Poisson integral for a function $f(t,\mathbf{x})$,

\begin{equation}
P(f) \equiv \frac{1}{4\pi}\int_\Ms \frac{f(t,\mathbf{x}')}{|\mathbf{x}-\mathbf{x}'|} d^3x' \, ,
\label{eq:Ppot}
\end{equation}
the simplest potentials are

\begin{subequations}
\begin{equation}
U_\sigma \equiv P(4\pi\sigma) = \int_\Ms \frac{\sigma(t,\mathbf{x}')}{|\mathbf{x}-\mathbf{x}'|}d^3x' \, ,
\end{equation}
\begin{equation}
U_{s\sigma} \equiv P(4\pi\sigma_s) = \int_\Ms \frac{\sigma_s(t,\mathbf{x}')}{|\mathbf{x}-\mathbf{x}'|}d^3x' \, .
\end{equation}
\end{subequations}
The integrals are taken over a constant-time hypersurface $\Ms$ at time $t$ out to radius $\Rs$, the boundary of the near zone.  The $\sigma$ subscript clarifies that these potentials use the $\sigma$ densities.  Expressions for all other $\sigma$-density potentials can be found in paper I, Eqs.\ (3.12)--(3.13).  Note that the generic Poisson integral has the property

\begin{equation}
\nabla^2 P(f) = -f \, .
\label{eq:delsquaredP}
\end{equation}
This will be very useful throughout the calculation.  Using the $\sigma$ densities and the associated potentials, we can solve the field equations for a generic system.  To get answers specific to a system of compact objects, we must study the compact stress energy $T^{\mu \nu}$ more closely.

Since a compact object is gravitationally bound, its total mass depends on its internal gravitational energy.  This, in turn, depends on the effective local value of the gravitational coupling.  In scalar-tensor theory, the coupling is controlled by the value of the scalar field $\phi$ in the vicinity of the body (scaling like $1/\phi$).  To deal with this complication, we use the approach of Eardley \cite{e75}.  In his method, we consider the compact objects to be point masses, with a mass $M(\phi)$ that is a function of the scalar field.  The stress-energy tensor is then given by

\begin{equation}
\begin{split}
T^{\mu \nu}(x^\alpha) &= (-g)^{-1/2}\sum_A \int d\tau \, M_A(\phi)u_A^\mu u_A^\nu \delta^4(x^\alpha-x_A^\alpha(\tau)) \\
&= (-g)^{-1/2}\sum_A M_A(\phi) u_A^\mu u_A^\nu (u_A^0)^{-1}\delta^3(\mathbf{x}-\mathbf{x}_A) \, .
\end{split}
\end{equation}
Here $u_A^\mu$ is the four-velocity of body $A$ and $\tau$ is the proper time measured along its world line.  (This is the only instance in which we use the symbol $\tau$ for this purpose.)  This construction assumes that the dynamical time scale of the body is short compared to an orbital time scale.  

We expand $M_A(\phi)$ about the asymptotic value of the scalar field, $\phi_0$,

\begin{equation}
\begin{split}
M_A(\phi) &= M_{A0} + \left(\frac{dM_A}{d\phi}\right)_0 \delta \phi + \frac{1}{2}\left(\frac{d^2M_A}{d\phi^2}\right)_0 \delta \phi^2 + \frac{1}{6}\left(\frac{d^3M_A}{d\phi^3}\right)_0 \delta \phi^3 +\cdots\\
&= m_A\left[1+s_A\Psi+\frac{1}{2}(s_A^2+s_A'-s_A)\Psi^2+\frac{1}{6}(s_A''+3s_A' s_A-3s_A'+s_A^3-3s_A^2+2s_A)\Psi^3+O(\Psi^4)\right]\\
&\equiv m_A[1+\Ss(s_A;\Psi)] \, ,
\end{split}
\end{equation}
where $m_A \equiv M_{A0}$.  We define the sensitivity and its derivatives as

\begin{subequations}
\begin{align}
s_A &\equiv \left(\frac{d\ln M_A(\phi)}{d\ln \phi}\right)_0 \, ,\\
s_A' &\equiv \left(\frac{d^2\ln M_A(\phi)}{d(\ln \phi)^2}\right)_0 \, ,\\
s_A'' &\equiv \left(\frac{d^3\ln M_A(\phi)}{d(\ln \phi)^3}\right)_0 \, ,
\end{align}
\end{subequations}
and so on.  Note that $s_A'$ has the opposite sign of the equivalent quantity in \cite{w93,abwz12}.  For later convenience, we also define the quantities 

\begin{subequations}
\begin{align}
a_{sA} &\equiv s_A^2+s_A'-\frac{1}{2}s_A \, ,\\
a_{sA}' &\equiv s_A''+2s_As_A'-\frac{1}{2}s_A' \, , \\
b_{sA} &\equiv a_{sA}'-a_{sA}+s_Aa_{sA} \, .
\end{align}
\end{subequations}
If we define a new density

\begin{equation}
\rho^* \equiv \sum_A m_A\delta^3(\mathbf{x}-\mathbf{x}_A) \, ,
\label{eq:rhostar}
\end{equation}
the stress energy becomes

\begin{equation}
T^{\mu \nu} = \rho^*(-g)^{-1/2}u^0v^\mu v^\nu[1+\Ss(s;\Psi)] \, ,
\label{eq:Twithsensitivities}
\end{equation}
where $v^\mu = (1,\mathbf{v})$ is the ordinary velocity.  The various velocities and the sensitivity $s$ technically should have body labels, but they will each pick one up when multiplied by the delta function in $\rho^*$.  As shown in paper I, \eqref{eq:Twithsensitivities} can be used to rewrite the $\sigma$ densities in terms of the $\rho^*$ density as a post-Newtonian expansion. 

We can also define new potentials based on the $\rho^*$ density.  For instance,

\begin{subequations}
\begin{align}
U &\equiv \int_\Ms\frac{\rho^*(t,\mathbf{x}')}{|\mathbf{x}-\mathbf{x}'|}d^3x' \, ,\\
U_s &\equiv \int_\Ms\frac{(1-2s(\mathbf{x}'))\rho^*(t,\mathbf{x}')}{|\mathbf{x}-\mathbf{x}'|}d^3x' \, .
\end{align}
\end{subequations}
Note that \eqref{eq:delsquaredP} implies that $\nabla^2 U = -4\pi \rho^*$ and $\nabla^2 U_s = -4\pi \rho^*(1-2s)$.  More generally,

\begin{subequations}
\begin{align}
\Sigma(f) &\equiv \int_\Ms\frac{\rho^*(t,\mathbf{x}')f(t,\mathbf{x}')}{|\mathbf{x}-\mathbf{x}'|}d^3x' = P(4\pi\rho^*f) \, , \label{eq:sigmapot}\\
\Sigma^i(f) &\equiv \int_\Ms\frac{\rho^*(t,\mathbf{x}')v^{\prime i}f(t,\mathbf{x}')}{|\mathbf{x}-\mathbf{x}'|}d^3x' = P(4\pi\rho^*v^if) \, , \label{eq:sigmaipot}\\
\Sigma^{ij}(f) &\equiv \int_\Ms\frac{\rho^*(t,\mathbf{x}')v^{\prime ij}f(t,\mathbf{x}')}{|\mathbf{x}-\mathbf{x}'|}d^3x' = P(4\pi\rho^*v^{ij}f) \, , \label{eq:sigmaijpot}\\
\Sigma_s(f) &\equiv \int_\Ms\frac{(1-2s(\mathbf{x}'))\rho^*(t,\mathbf{x}')f(t,\mathbf{x}')}{|\mathbf{x}-\mathbf{x}'|}d^3x' = P(4\pi(1-2s)\rho^*f) \, , \label{eq:sigmaspot}\\
X(f) &\equiv \int_\Ms\rho^*(t,\mathbf{x}')f(t,\mathbf{x}')|\mathbf{x}-\mathbf{x}'| d^3x' \, , \label{eq:Xpot} \\
X_s(f) &\equiv \int_\Ms (1-2s(\mathbf{x}'))\rho^*(t,\mathbf{x}')f(t,\mathbf{x}')|\mathbf{x}-\mathbf{x}'| d^3x' \, ,
\label{eq:Xspot}
\end{align}
\end{subequations}
so that $U = \Sigma(1)$ and $U_s = \Sigma_s(1)$.  The other potentials we use in this paper are

\begin{align}
V^i &\equiv \Sigma^i(1) \, , & V_s^i &\equiv \Sigma_s(v^i) \, , & \Phi_1^{ij} &\equiv \Sigma^{ij}(1) \, , & \Phi_1 &\equiv \Sigma^{ii}(1) \, , \nonumber \\
\Phi^s_1 &\equiv \Sigma_s(v^2) \, , & \Phi_2 &\equiv \Sigma(U) \, , & \Phi^s_2 &\equiv \Sigma_s(U) \, ,  & \Phi^s_{2s} &\equiv \Sigma_s(U_s) \, , \nonumber \\ 
X &\equiv X(1) \, , & X_s &\equiv X_s(1) \, , & P_2^{ij} &\equiv P(U^{,i}U^{,j}) \, , & P_{2s}^{ij} &\equiv P(U_s^{,i}U_s^{,j}) \, . \nonumber \\
\end{align}
Equations (5.13)--(5.22) in paper I show how to convert between many $\sigma$-density and $\rho^*$-density potentials (e.g., $U_\sigma$ and $U$).

\section{Near-zone contribution to the scalar waveform}
\label{sec:NZintegrals}

In this section, we calculate the near-zone contribution to the scalar waveform.  This is, by far, the more difficult and time consuming of the two contributions.  We begin by discussing the general formalism for evaluating the near-zone integral, namely the calculation of scalar multipole moments.  We also discuss the counting of post-Newtonian orders, which is a bit more subtle than in the tensor wave case.  Next, we discuss the construction of the source $\tau_s$.  Third, we present the calculation of the scalar multipole moments.  We provide some details but refer the reader to paper II for more.  Finally, we conclude this section by rewriting the scalar multipole moments explicitly for two-body systems in relative coordinates.  We hold off on presenting the final scalar waveform generated by the near-zone integral until Sec.\ \ref{sec:scalarwaves}.

\subsection{General structure of near-zone calculation}

The goal of this entire section is to evaluate \eqref{eq:phiintegral} for near-zone source points and radiation-zone field points.  We can simplify the problem even further by considering only a subset of the radiation zone, the ``far-away zone.''  In the far-away zone, $R \gg \Rs$, so we only need to keep the leading $1/R$ part of the field.  A distant gravitational-wave detector measuring the scalar waves from the system lies in the far-away zone.  With this simplification, the near-zone contribution to the waveform is given by
\begin{equation}
\begin{split}
\Psi_\Ns (t,\mathbf{x}) &= \frac{2}{R}\sum_{m=0}^\infty \frac{1}{m!}\frac{\partial^m}{\partial t^m}\int_\Ms\tau_s(\tau,\mathbf{x}')(\mathbf{\hat{N}}\cdot\mathbf{x'})^m d^3x' \\
&= \frac{2}{R}\sum_{m=0}^\infty \hat{N}^{k_1}\cdots \hat{N}^{k_m}\frac{1}{m!}\frac{d^m}{dt^m}\Is_s^{k_1\cdots k_m}(\tau) \, .
\label{eq:scalarfaraway}
\end{split}
\end{equation}
Here $\Ns$ is the three-dimensional hypersurface representing the intersection of the past null cone $\Cs$ and the near-zone world tube.  The actual integration takes place over $\Ms$, the intersection of the near-zone world tube with a hypersurface of constant {\em retarded} time $\tau = t-R$.  The direction from the source to the detector is $\mathbf{\hat{N}} \equiv \mathbf{x}/R$.  Finally, we define the scalar multipole moments as 
\begin{equation}
\Is_s^{k_1\cdots k_m}(\tau) \equiv \int_\Ms\tau_s(\tau,\mathbf{x})x^{k_1}\cdots x^{k_m}d^3x \, .
\end{equation}
In paper II, these were also known as $M_s^M$; here, we will only use the $\Is_s^M$ notation.  The role of these moments in finding the near-zone contribution to the scalar waveform is analogous to the role of the Epstein-Wagoner moments [paper II, Eqs.\ (2.22)] in finding the near-zone contribution to the tensor waveform.  The Epstein-Wagoner construction is a bit more complicated due to a convenient rearrangement using the conservation law \eqref{eq:conservation}.  That approach requires the calculation of  ``surface moments'' for the two- and three-index case.  No such complications exist in the computation of the scalar moments.

In the tensor case, the lowest order moment is the two-index, or quadrupole, moment.  Its contribution to the tensor waves is given by

\begin{equation}
\tilde{h}_\Ns^{ij}(t,\mathbf{x}) = \frac{2}{R}\frac{d^2}{dt^2}I_\text{EW}^{ij}(\tau) \, ,
\label{eq:tensorquad}
\end{equation}
where

\begin{equation}
I_\text{EW}^{ij} = \int_\Ms \tau^{00} x^{ij} d^3x + I^{ij}_{\text{EW (surf)}} \, .
\end{equation}
(The second term is the surface moment, defined in paper II.)  In the scalar case, the lowest order moment is the zero-index, or monopole, moment.  Note that the sources that enter these two moments are of the same post-Newtonian order, $\tau^{00} \sim O(\rho) \sim \tau_s$.  But because of the two time derivatives in \eqref{eq:tensorquad}, the lowest order tensor field contains a factor of $v^2$ that the lowest order scalar field does not.  This means that the lowest order tensor field is one post-Newtonian order higher than the lowest order scalar field.  However, in paper II, we defined the lowest order tensor field to be ``0PN'' order.  While it would be a simple matter to redefine the post-Newtonian scale to start with the lowest order of all the waves, we instead keep the previous definition.  In this way, our usage of ``0PN'' matches the usual usage in general relativity (as well as all other studies of scalar-tensor theory). 

As a consequence of this choice, the lowest order piece of the scalar monopole moment generates a $-1$PN scalar field in the far-away zone.  (As it turns out, this piece is time independent and therefore uninteresting; however, higher order pieces of the monopole moment generate 0PN and higher order scalar waves.)   The dipole moment generates $-0.5$PN and higher order scalar waves.  The quadrupole moment generates 0PN and higher order waves, and the pattern continues to higher-index moments.  Therefore, in order to generate a final waveform at $N$th post-Newtonian order, the $m$-index moment, if it contributes to the $N$PN term at all, must be calculated to $(N+1-m/2)$th post-Newtonian order beyond its own leading-order term.

Because of this complication, we compute the scalar waveform only to 1.5PN order.  Computing it to 2PN order, as we did for the tensor waveform in paper II, would require first constructing the monopole moment to 3PN order beyond its leading-order term.  The source $\tau_s$ has several times as many terms at 3PN order than at 2.5PN order.  In addition, many of these terms become quite complicated to integrate.  For instance, many include ``triangle potentials'' and ``quadrangle potentials'' \cite{pw02}, which are difficult to integrate even when multiplied by a compact source (usually the easiest case, as we shall see below).  Higher order pieces of the scalar waveform will be considered in future work.

\subsection{Source $\tau_s$}

To calculate the scalar multipole moments, we first need an expression for the source $\tau_s$ to 2.5PN relative order, or $O(\rho\epsilon^{5/2})$.  Here we have introduced a post-Newtonian counting parameter $\epsilon \sim m/r \sim v^2$, where $m$ is the total mass of the binary, $r$ is the orbital separation, and $v$ is the magnitude of the relative velocity.  To simplify some of the discussion, we divide $\tau_s$ into three pieces, one compact and two noncompact,

\begin{equation}
\tau_s \equiv \tau_{s,\text{C}}+\tau_{s,{\text F1}}+\tau_{s,\text{F2}} \, ,
\end{equation}
corresponding to the first, second, and third terms in \eqref{eq:taus}.  We label the noncompact terms ``F'' for ``field.''  Writing out each term, we find

\begin{subequations}
\begin{equation}
\tau_{s,\text{C}} = G\sigma_s\left[\zeta-(2\lambda_1+\zeta)\Psi+\frac{1}{2}\zeta N+\frac{1}{\zeta}(-\lambda_2+4\lambda_1^2+2\zeta\lambda_1+\zeta^2)\Psi^2-\frac{1}{2}(2\lambda_1+\zeta)\Psi N-\frac{1}{2}\zeta B-\frac{1}{8}\zeta N^2 +\cdots\right] \, ,
\label{eq:tausC}
\end{equation}

\begin{equation}
\tau_{s,{\text F1}} = \frac{1}{8\pi}\left[\frac{1}{\zeta}(\lambda_1-\zeta)(\nabla \Psi)^2-\frac{1}{\zeta}(\lambda_1-\zeta)\dot{\Psi}^2+\frac{1}{\zeta^2}(\lambda_2-2\lambda_1^2+\zeta^2)\Psi(\nabla \Psi)^2 +\cdots\right] \, ,
\label{eq:tausF1}
\end{equation}

\begin{equation}
\tau_{s,\text{F2}} = -\frac{1}{8\pi}(N\ddot{\Psi}+2K^i\dot{\Psi}^{,i}+B^{ij}\Psi^{,ij}) \, .
\label{eq:tausF2}
\end{equation}
\end{subequations}
We have defined the quantity

\begin{equation}
G \equiv \frac{1}{\phi_0}\frac{4+2\omega_0}{3+2\omega_0} \, ,
\end{equation}
where $\omega_0 \equiv \omega(\phi_0)$.  The definition ensures that for a perfect fluid with no internal gravitational binding energy (i.e., zero sensitivities), the metric component $g_{00} = -1+2GU_\sigma$ as in general relativity.  We do not set $G$ equal to 1, since it depends on the asymptotic value of the scalar field $\phi$, which could potentially vary in time over the history of the Universe.  The other parameters in \eqref{eq:tausC}--\eqref{eq:tausF2} are

\begin{subequations}
\begin{align}
\zeta &\equiv \frac{1}{4+2\omega_0} \, , \\
\lambda_1 &\equiv \frac{(d\omega/d\varphi)_0\zeta}{3+2\omega_0} \, , \\
\lambda_2 &\equiv \frac{(d^2\omega/d\varphi^2)_0\zeta^2}{3+2\omega_0} \, .
\end{align}
\end{subequations}
The expression \eqref{eq:tausF2} for $\tau_{s,\text{F2}}$ is exact.  Equations \eqref{eq:tausC}--\eqref{eq:tausF1} have been kept to the PN order necessary for the near-zone integral.  These general expressions will also be useful when we compute the radiation-zone integral in Sec.\ \ref{sec:RZintegrals}.

We wish to convert \eqref{eq:tausC}--\eqref{eq:tausF2} to a more explicit form valid in the near zone.  This requires finding the fields $N$, $K^i$, $B^{ij}$, and $\Psi$ in the near zone.  This calculation was the focus of much of paper I.  Finding the near-zone fields typically requires integrations over both the near and radiation zones.  However, the radiation-zone integrals do not contribute at the order to which we work in this series of papers.  The near-zone integrals are given by
\begin{subequations}
\begin{align}
\tilde{h}_\Ns ^{\mu \nu}(t,\mathbf{x}) &= 4\sum_{m=0}^\infty \frac{(-1)^m}{m!}\frac{\partial^m}{\partial t^m}\int_\Ms \tau^{\mu \nu}(t,\mathbf{x}')|\mathbf{x}-\mathbf{x}'|^{m-1}d^3x' \label{eq:hNN} \, , \\
\Psi_\Ns (t,\mathbf{x}) &= 2\sum_{m=0}^\infty \frac{(-1)^m}{m!}\frac{\partial^m}{\partial t^m}\int_\Ms \tau_s(t,\mathbf{x}')|\mathbf{x}-\mathbf{x}'|^{m-1}d^3x' \, .
\label{eq:phiNN}
\end{align}
\end{subequations}
Here $\Ms$ is the intersection of the near-zone world tube with the hypersurface $t = \text{const}$ (i.e., not the retarded time).  Note that in the near zone, the slow-motion approximation $v \ll 1$ means that each time derivative corresponds to an increase of one-half post-Newtonian order.

In the near zone, it turns out that $N\sim O(\epsilon)$, $K^i \sim O(\epsilon^{3/2})$, $B^{ij} \sim B \sim O(\epsilon^2)$, and $\Psi \sim O(\epsilon)$.  Equations \eqref{eq:hNN} and \eqref{eq:phiNN} are solved in an iterative manner.  At lowest order, $N$ and $\Psi$ are determined purely in terms of the compact piece of the sources $\tau^{00}$ and $\tau_s$.  At the next order, these fields are plugged into the field pieces of the sources.  The process continues until we have the fields (and thus the sources) to the desired post-Newtonian order.  Paper I presents expressions for $N$, $K^i$, $B^{ij}$, and $\Psi$ to high post-Newtonian order.  All are expressed in terms of $\sigma$-density potentials.

We can find $\tau_s$ to the order we need in this paper by plugging these expressions into \eqref{eq:tausC}--\eqref{eq:tausF2}.  We must also convert $\sigma$ to $\rho^*$ and the $\sigma$-density potentials to the $\rho^*$-density potentials using the relations in paper I.  The final results are

\begin{subequations}
\begin{align}
\tau_{s,\text{C}} &= G\zeta\rho^* \left\{\vphantom{\frac{1}{2}}1-2s \right.  \nonumber \\
& \qquad +[-G(4\lambda_1-\zeta)(1-2s) - 4G\zeta a_s]U_s -G(1-\zeta)(1-2s)U -\frac{1}{2}(1-2s)v^2  \nonumber \\
& \qquad +\left[G^2\left(20\lambda_1^2-9\zeta \lambda_1+\frac{1}{2}\zeta^2-4\lambda_2\right)(1-2s)+4G^2\zeta(5\lambda-2\zeta)a_s-4G^2\zeta^2b_s\right]U_s^2  \nonumber \\
& \qquad +[G^2(1-\zeta)(4\lambda_1-\zeta)(1-2s)+4G^2\zeta(1-\zeta)a_s]\Phi^s_2   \nonumber \\
& \qquad +[G^2(8\lambda_1^2+2\zeta\lambda_1-2\zeta^2+\zeta)(1-2s) + 4G^2\zeta(2\lambda_1+\zeta)a_s]\Phi^s_{2s} +\left[-\frac{1}{2}G(4\lambda_1-\zeta)(1-2s)-2G\zeta a_s\right]\ddot{X}_s    \nonumber \\
& \qquad +\frac{1}{2}G^2(1-\zeta)^2(1-2s)U^2 -\frac{3}{2}G(1-\zeta)(1-2s)\Phi_1-\frac{1}{2}G(1-\zeta)(1-2s)\ddot{X} \nonumber \\
& \qquad +[G^2(1-\zeta)(4\lambda_1-\zeta)(1-2s)+4G^2\zeta(1-\zeta)a_s]UU_s +\left[\frac{1}{2}G(4\lambda_1-\zeta)(1-2s)+2G\zeta a_s\right]\Phi^s_1  \nonumber \\
& \qquad +[4G^2\zeta(4\lambda_1-\zeta)(1-2s)+16G^2\zeta^2 a_s]\Sigma(a_sU_s) +\left[\frac{1}{2}G(4\lambda_1-\zeta)(1-2s)+2G\zeta a_s\right]v^2U_s  \nonumber \\
& \qquad \left.-\frac{3}{2}G(1-\zeta)(1-2s)v^2U + G^2(1-\zeta)^2(1-2s)\Phi_2+4G(1-\zeta)(1-2s)v^iV^i-\frac{1}{8}(1-2s)v^4 \right\}  \nonumber \\
& \quad +G\rho^*\left\{[(4\lambda_1-\zeta)(1-2s)+4\zeta a_s]\dot{\Is}_s(t) + \left[-\frac{1}{3}(4\lambda_1-\zeta)(1-2s)-\frac{4}{3}\zeta a_s\right]x^j\dddot{\Is}_s^j(t) \right.  \nonumber \\
& \qquad \left. + \left[\frac{1}{6}(4\lambda_1-\zeta)(1-2s)+\frac{2}{3}\zeta a_s\right]\dddot{\Is}_s^{kk}(t)+\frac{2}{3}\zeta(1-2s)\dddot{\Is}^{kk}(t) \right\} \, ,
\label{eq:tausCpotential}
\end{align}

\begin{equation}
\begin{split}
\tau_{s,{\text F1}} &= G^2\zeta(\lambda_1-\zeta)\left[\frac{1}{2\pi}(\nabla U_s)^2-\frac{1}{2\pi}\nabla U_s\cdot \nabla \Phi^s_1-\frac{1}{\pi}G(1-\zeta)\nabla U_s\cdot \nabla \Phi^s_2 -\frac{1}{\pi}G(2\lambda_1+\zeta)\nabla U_s\cdot \nabla \Phi^s_{2s} \right.\\
&\left.\qquad-\frac{4}{\pi}G\zeta\nabla U_s\cdot \nabla\Sigma(a_sU_s)+\frac{1}{2\pi}\nabla U_s\cdot \nabla \ddot{X}_s-\frac{1}{2\pi}\dot{U}_s^2\right] \\
&\quad +\frac{1}{\pi}G^3\zeta(-4\lambda_1^2+4\zeta\lambda_1-\zeta^2+\lambda_2)U_s(\nabla U_s)^2 \\
&\quad + \frac{1}{3\pi}G(\lambda_1-\zeta)U_s^{,i}\dddot{\Is}_s^i(t) \, ,
\end{split}
\label{eq:tausF1potential}
\end{equation}

\begin{equation}
\begin{split}
\tau_{s,\text{F2}} &= G^2\zeta(1-\zeta)\left[-\frac{1}{\pi}U\ddot{U}_s-\frac{2}{\pi}V^i\dot{U}_s^{,i}-\frac{1}{\pi}\Phi_1^{ij}U_s^{,ij}-\frac{1}{\pi}G(1-\zeta)P_2^{ij}U_s^{,ij}+\frac{1}{2\pi}G(1-\zeta)\Phi_2 \nabla^2 U_s \right.\\
&\left.\qquad -\frac{1}{4\pi}G(1-\zeta)U^2\nabla^2 U_s-\frac{1}{\pi}G\zeta P_{2s}^{ij}U_s^{,ij}+\frac{1}{2\pi}G\zeta \Phi^s_{2s}\nabla^2 U_s-\frac{1}{4\pi}G\zeta U_s^2 \nabla^2 U_s\right] \\
&\quad+\frac{1}{2\pi}G\zeta U_s^{,ij}\dddot{\Is}^{ij}(t) \, .
\end{split}
\label{eq:tausF2potential}
\end{equation}
\end{subequations}
The highest order (2.5PN) pieces of these expressions contain ``regular'' and scalar multipole moments.  Since we are constructing $\tau_s$ for the purpose of calculating the scalar multipole moments, the process is naturally iterative.  The regular multipole moments are given by

\begin{equation}
\Is^{k_1\cdots k_m}(t) \equiv \int_\Ms\tau^{00}(t,\mathbf{x})x^{k_1}\cdots x^{k_m}d^3x \, .
\end{equation}
In particular, we need the quadrupole moment $\Is^{ij}$.  Because it only appears in the highest order terms, we only require its lowest order form,

\begin{equation}
\Is^{ij} = G(1-\zeta)\sum_A m_Ax_A^{ij} \, .
\end{equation}
Similarly, we only need the lowest order forms of the scalar moments $\Is_s^i$ and $\Is_s^{ij}$.  However, the scalar monopole moment $\Is_s$ is time independent to lowest order, so we need it to relative 1PN order.  All of the necessary expressions are written out below.  Note that $\tau_{s,\text{C}}$ begins at $O(\rho)$.  Relative to this, $\tau_{s,{\text F1}}$ begins at 1PN order, or $O(\rho\epsilon)$, and $\tau_{s,\text{F2}}$ begins at 2PN order.

\subsection{Zero-index moment $\Is_s$}
\label{sec:zeroindex}
We begin with the zero-index moment,

\begin{equation}
\Is_s = \int_\Ms \tau_s\ d^3x \, .
\end{equation}
To evaluate it to the necessary order, we require $\tau_s$ all the way to $O(\rho\epsilon^{5/2})$.  The lowest order piece of the moment will generate a $-1$PN scalar field, while the highest order piece we calculate will generate 1.5PN scalar waves.

Since $\rho^*$ [Eq.\ \eqref{eq:rhostar}] contains delta functions, the compact moment can be written down by inspection,

\begin{align}
\Is_{s,\text{C}} &= G\zeta\sum_A m_A\left\{ 1-2s_A-\frac{1}{2}(1-2s_A)v_A^2 \right. \nonumber \\
&\qquad -\sum_{B\neq A}\frac{Gm_B}{r_{AB}}[(1-\zeta)(1-2s_A)+(4\lambda_1-\zeta)(1-2s_A)(1-2s_B)+4\zeta a_{sA}(1-2s_B)]  \nonumber \\
&\qquad -\frac{1}{8}(1-2s_A)v_A^4 \nonumber \\
&\qquad +\sum_{B\neq A}\frac{Gm_B}{r_{AB}}\left\{\left[-\frac{3}{2}(1-\zeta)(1-2s_A)+\frac{1}{2}(4\lambda_1-\zeta)(1-2s_A)(1-2s_B)+2\zeta a_{sA}(1-2s_B)\right]v_A^2 \right.  \nonumber \\
&\qquad \quad-2(1-\zeta)(1-2s_A)v_B^2+4(1-\zeta)(1-2s_A)\mathbf{v}_A\cdot\mathbf{v}_B  \nonumber \\
&\qquad \quad +\left[\frac{1}{2}(1-\zeta)(1-2s_A)+\frac{1}{2}(4\lambda_1-\zeta)(1-2s_A)(1-2s_B)+2\zeta a_{sA}(1-2s_B)\right]\mathbf{a}_B\cdot \mathbf{x}_{AB}  \nonumber \\
&\qquad \quad +\left[\frac{1}{2}(1-\zeta)(1-2s_A)+\frac{1}{2}(4\lambda_1-\zeta)(1-2s_A)(1-2s_B)+2\zeta a_{sA}(1-2s_B)\right](\mathbf{v}_B\cdot \mathbf{\hat{n}}_{AB})^2  \nonumber \\
&\qquad \quad +\sum_{C\neq A}\frac{Gm_C}{r_{AC}}\left[\frac{1}{2}(1-\zeta)^2(1-2s_A)+(1-\zeta)(4\lambda_1-\zeta)(1-2s_A)(1-2s_C)+4\zeta(1-\zeta)a_{sA}(1-2s_C)\right. \nonumber \\
&\qquad \qquad +\left(20\lambda_1^2-9\zeta \lambda_1+\frac{1}{2}\zeta^2-4\lambda_2\right)(1-2s_A)(1-2s_B)(1-2s_C)+4\zeta(5\lambda_1-2\zeta)a_{sA}(1-2s_B)(1-2s_C)  \nonumber \\
&\left.\qquad \qquad -4\zeta^2 b_{sA}(1-2s_B)(1-2s_C)\vphantom{\frac{1}{2}}\right]  \nonumber \\
&\qquad \quad +\sum_{C\neq B}\frac{Gm_C}{r_{BC}}\left[(1-\zeta)^2(1-2s_A)+(1-\zeta)(4\lambda_1-\zeta)(1-2s_A)(1-2s_B)+4\zeta(1-\zeta)a_{sA}(1-2s_B) \right.  \nonumber \\
&\qquad \qquad +(8\lambda_1^2+2\zeta\lambda_1-2\zeta^2+\zeta)(1-2s_A)(1-2s_B)(1-2s_C)+4\zeta(2\lambda_1+\zeta)a_{sA}(1-2s_B)(1-2s_C)  \nonumber \\
  &\left.\left.\qquad \qquad +4\zeta(4\lambda_1-\zeta)(1-2s_A)a_{sB}(1-2s_C)+16\zeta^2a_{sA}a_{sB}(1-2s_C)\right]\vphantom{\frac{1}{2}}\right\}  \nonumber \\
&\qquad +\sum_B Gm_B\left\{\vphantom{\frac{1}{3}}4(1-\zeta)(1-2s_A)\mathbf{v}_B\cdot\mathbf{a}_B \right. \nonumber \\
&\qquad \quad +\left[\frac{4}{3}(1-\zeta)(1-2s_A)+\frac{1}{3}(4\lambda_1-\zeta)(1-2s_A)(1-2s_B)+\frac{4}{3}\zeta a_{sA}(1-2s_B)\right]\mathbf{x}_B\cdot \dot{\mathbf{a}}_B \nonumber \\
&\qquad \quad +\left[-\frac{1}{3}(4\lambda_1-\zeta)(1-2s_A)(1-2s_B)-\frac{4}{3}\zeta a_{sA}(1-2s_B)\right]\mathbf{x}_A\cdot \dot{\mathbf{a}}_B  \nonumber \\
&\qquad \quad +\sum_{C\neq B}\frac{Gm_C}{r_{BC}^2}\dot{r}_{BC}\left[(1-\zeta)(4\lambda_1-\zeta)(1-2s_A)(1-2s_B)+4\zeta(1-\zeta)a_{sA}(1-2s_B)\right.  \nonumber \\
&\qquad \qquad  +(4\lambda_1-\zeta)(2\lambda_1+\zeta)(1-2s_A)(1-2s_B)(1-2s_C)+4\zeta(2\lambda_1+\zeta)a_{sA}(1-2s_B)(1-2s_C) \nonumber \\
&\left.\left.\left.\qquad \qquad  +4\zeta(4\lambda_1-\zeta)(1-2s_A)a_{sB}(1-2s_C)+16\zeta^2a_{sA}a_{sB}(1-2s_C)\right]\vphantom{\frac{1}{3}}\right\} \right\} \, .
\end{align}
Here body $A$ (for example) has position $\mathbf{x}_A$, velocity $\mathbf{v}_A$, and acceleration $\mathbf{a}_A$.  The distance between bodies $A$ and $B$ is $r_{AB}$, and the unit vector $\mathbf{\hat{n}}_{AB} = \mathbf{x}_{AB}/r_{AB}$ points from body $B$ to body $A$.  This expression contains terms at 0PN, 1PN, 2PN, and 2.5PN orders relative to itself.  (As discussed above, with our counting scheme, it generates a scalar field at $-1$PN, 0PN, 1PN, and 1.5PN orders.)

To calculate the contribution from the noncompact parts of $\tau_s$, we first notice that four terms in $\tau_{s,\text{F2}}$ involve $\nabla^2 U_s = -4\pi \rho^*(1-2s)$, making these terms effectively compact terms.  They can then be evaluated as trivially as the contribution from $\tau_{s,\text{C}}$.  The rest of the contributions from $\tau_{s,{\text F1}}$ and $\tau_{s,\text{F2}}$ require evaluating integrals involving the potentials.  To do so, we use the same techniques described in Sec.\ IVA of paper II.  The source $\tau_{s,{\text F1}}$ contains the lowest order (1PN) field integral,

\begin{equation}
\int_\Ms (\nabla U_s)^2 d^3x \, ,
\end{equation}
where we have left off the constants for the time being.  This is easily evaluated by integrating by parts.  The surface term is evaluated by expanding the potential $U_s$ and its derivative in inverse powers of the surface radius $\Rs$:

\begin{align}
U_s &= \sum_A m_A(1-2s_A)\left(\frac{1}{\Rs}+\frac{\hat{n}^ax_A^a}{\Rs^2}+\frac{1}{2}\frac{(3\hat{n}^{ab}-\delta^{ab})x_A^{ab}}{\Rs^3}+\cdots\right) \sim \frac{1}{\Rs}+\frac{\hat{n}^a}{\Rs^2}+\frac{1+\hat{n}^{ab}}{\Rs^3}+\cdots \, , 
\label{eq:Uexpansion}\\
U_s^{,k} &\sim \frac{\hat{n}^k}{\Rs^2}+\frac{1+\hat{n}^{ck}}{\Rs^3}+\frac{\hat{n}^c+\hat{n}^{cdk}}{\Rs^4}+\cdots \, .
\label{eq:Uexpansion2}
\end{align}
In these expressions only, $\hat{n}^a$ represents a unit normal to the surface.  Plugging \eqref{eq:Uexpansion}--\eqref{eq:Uexpansion2} into the surface integral reveals no pieces independent of $\Rs$.  Recall that we ignore all pieces which depend on $\Rs$ because they will cancel with radiation-zone contributions in the final waveform.  The volume integral can be reduced to a compact integral using $\nabla^2 U_s = -4\pi \rho^*(1-2s)$.  Thus the final value of the original integral is 

\begin{equation}
\int_\Ms (\nabla U_s)^2 d^3x = 4\pi \sum_A \sum_{B\neq A} \frac{m_A(1-2s_A)m_B(1-2s_B)}{r_{AB}} \, .
\label{eq:1PNintegral}
\end{equation}

Most of the 2PN integrals arising from $\tau_{s,{\text F1}}$ can be calculated in the same way, by integrating by parts and turning the remaining volume integral into a compact integral.  The surface terms always vanish. The exception is the integral $\int_\Ms \dot{U}_s^2 d^3x$.  To tackle it, we write $\dot{U}_s = -V_s^{i,i}$ and integrate by parts.  The remaining volume integral requires a bit more work than in the previously discussed cases.  Written out, it is

\begin{equation}
\sum_A\sum_{B\neq A}m_A(1-2s_A)m_B(1-2s_B)v_A^jv_B^i\int_\Ms\left[-3\frac{(x-x_A)^{ij}}{|\mathbf{x}-\mathbf{x}_A|^5}+\frac{\delta^{ij}}{|\mathbf{x}-\mathbf{x}_A|^3}+\frac{4\pi}{3}\delta^{ij}\delta^3(\mathbf{x}-\mathbf{x}_A)\right]\frac{1}{|\mathbf{x}-\mathbf{x}_B|}d^3x \, .
\label{eq:example}
\end{equation}
The delta-function term is required to ensure the right answer when $\dot{U}_s^{,k}$ is integrated in a sphere around the point mass position $\mathbf{x}_A$. [See paper II, Eqs.\ (4.20).] 

Several of the 2PN integrals arising from $\tau_{s,\text{F2}}$, as well as many of the integrals we will encounter in higher order moments, can also be written in a form like \eqref{eq:example}.  They always involve two potentials.  The strategy to solve them is simple: First, integrate by parts, as we did above, so that only one potential is undifferentiated.  Next, evaluate any integrals which can easily be converted to a compact integral.  The remaining pieces should be of a form similar to \eqref{eq:example}.  

To evaluate integrals like \eqref{eq:example}, we change integration variables from $\mathbf{x}$ to $\mathbf{y} = \mathbf{x}-\mathbf{x}_A$.  We check to see if the surface terms so generated contribute anything independent of $\Rs$.  (Sometimes terms which are potentially independent of $\Rs$ vanish upon integrating over the surface.)  We also check the case $A = B$ to see if it contributes.  To solve the case $A \neq B$, we make use of the following relation,

\begin{equation}
\frac{1}{|\mathbf{y}+\mathbf{x}_{AB}|}= \sum_{l,m}\frac{4\pi}{2l+1}\frac{(-r_<)^l}{r_>^{l+1}}Y_{l m}^*(\mathbf{\hat{n}}_{AB})Y_{l m}(\mathbf{\hat{y}}) \, ,
\label{eq:harmonicexpansion}
\end{equation}
where $Y_{l m}$ are the spherical harmonics, and $r_{<(>)}$ denotes the lesser (greater) of $r_{AB}$ and $y$.  We then substitute this expansion into the volume integral and then express all products of $\hat{y}^i$ in terms of STF products $\hat{y}^{\langle L'\rangle}$ (see Appendix B of paper II).  Here $\langle L'\rangle$ denotes an $l'$-dimensional STF combination.  We can then perform the angular integration using

\begin{equation}
\sum_m \int Y^*_{l m}(\mathbf{\hat{n}}_{AB})Y_{l m}(\mathbf{\hat{y}})\hat{y}^{\langle L'\rangle}d^2\Omega_y = \hat{n}_{AB}^{\langle L\rangle}\delta^{l l'} \, .
\label{eq:angularintegral}
\end{equation}
Finally, the radial integral is evaluated using

\begin{equation}
\int_0^\Rs \frac{r_<^l}{r_>^{l+1}}y^q dy = \frac{2l+1}{(l+q+1)(l-q)}r_{AB}^q \, ,
\label{eq:radialintegral}
\end{equation}
where we have dropped terms dependent on $\Rs$.  See paper II for a more detailed example of this technique.  Note that we can reuse many intermediate results from paper II.  We have to be slightly careful, however: In paper II, we dropped terms which could not survive a transverse-traceless projection at the end of the calculation.  In the scalar case, we have to make sure we reuse only the purest results, before any terms were dropped.

The integrals arising from $\tau_{s,\text{F2}}$ involving $P_2^{ij}$ and $P_{2s}^{ij}$ are a bit more complicated.  Luckily, we have also tackled similar terms in paper II.  We can write out the first integral as

\begin{equation}
\begin{split}
\int_\Ms P_2^{ij}U_s^{,ij} d^3x &= \int_\Ms d^3x \left[\frac{1}{4\pi}\int_\Ms\frac{1}{|\mathbf{x}-\mathbf{x}'|}U^{\prime ,i}U^{\prime ,j}d^3x'\right] \\
&\qquad \times \sum_A m_A(1-2s_A)\left[3\frac{(x-x_A)^{ij}}{|\mathbf{x}-\mathbf{x}_A|^5}-\frac{\delta^{ij}}{|\mathbf{x}-\mathbf{x}_A|^3}-\frac{4\pi}{3}\delta^{ij}\delta^3(\mathbf{x}-\mathbf{x}_A)\right] \, ,
\end{split}
\end{equation}
where $U'$ is the usual potential written as a function of $\mathbf{x}'$.  We can integrate over $\mathbf{x}$ (unprimed) first, using the change-of-variable technique discussed above, except with $\mathbf{x}_B\rightarrow \mathbf{x}'$.  Dropping the primes on the remaining integration variable, the result is

\begin{equation}
\begin{split}
\int_\Ms P_2^{ij}U_s^{,ij} d^3x  &= \int_\Ms d^3x\ U^{,i}U^{,j}\sum_A m_A(1-2s_A)\left[\frac{1}{2}\frac{(x_A-x)^{ij}}{|\mathbf{x}_A-\mathbf{x}|^3}-\frac{1}{2}\frac{\delta^{ij}}{|\mathbf{x}_A-\mathbf{x}|}\right] \\
&= \int_\Ms d^3x\ U^{,i} U^{,j} \sum_A m_A(1-2s_A)\left(-\frac{1}{2}X^{A,ij}\right) \, ,
\end{split}
\end{equation}
where $X^A \equiv |\mathbf{x}-\mathbf{x}_A|$.  This integral can be evaluated by integrating by parts several times and converting to compact integrals (including using the relation $\nabla^2 X^A = 2U^A$).  The second integral, involving $P_{2s}^{ij}$, is evaluated in exactly the same way.

Finally, the 2.5PN integrals, which involve only one potential times a function of time (one of the other moments), can be evaluated simply by integrating by parts.  In these cases the surface integrals are the only contributions, and they do not vanish.  We find the $\Rs$-independent pieces

\begin{align}
\int_\Ms U_s^{,i}d^3x &= \frac{4\pi}{3}\sum_A m_A(1-2s_A)x_A^i \, , \\
\int_\Ms U_s^{,ij}d^3x &= -\frac{4\pi}{3}\sum_A m_A(1-2s_A)\delta^{ij} \, .
\end{align}
The final field pieces of the monopole moment are given by

\begin{align}
\Is_{s,{\text F1}} &= G^2\zeta(\lambda_1-\zeta)\sum_A\sum_{B\neq A}\frac{m_A(1-2s_A)m_B(1-2s_B)}{r_{AB}}\left[2-2(\mathbf{v}_A\cdot\mathbf{\hat{n}}_{AB})^2+2\mathbf{a}_A\cdot\mathbf{x}_{AB} \right.\nonumber \\
&\left.\qquad +(\mathbf{v}_A\cdot\mathbf{\hat{n}}_{AB})(\mathbf{v}_B\cdot\mathbf{\hat{n}}_{AB})-\mathbf{v}_A\cdot\mathbf{v}_B\right]\nonumber \\
&\quad +G^3\zeta\sum_A\sum_{B\neq A}\sum_{C\neq A}\frac{m_Am_Bm_C}{r_{AB}r_{AC}}[-4(1-\zeta)(\lambda_1-\zeta)(1-2s_A)(1-2s_B) \nonumber \\
&\qquad +(12\zeta\lambda_1+2\zeta^2-16\lambda_1^2+2\lambda_2)(1-2s_A)(1-2s_B)(1-2s_C)-16\zeta(\lambda_1-\zeta)a_{sA}(1-2s_B)(1-2s_C)] \nonumber \\
&\quad +\frac{4}{9}G^2\zeta(\lambda_1-\zeta)\sum_{A,B}m_A(1-2s_A)m_B(1-2s_B)(\mathbf{x}_B\cdot\dot{\mathbf{a}}_A) \, ,
\end{align}

\begin{equation}
\begin{split}
\Is_{s,\text{F2}} &= G^2\zeta(1-\zeta)\sum_A \sum_{B\neq A}m_A(1-2s_A)\frac{m_B}{r_{AB}}\left[-2(\mathbf{v}_A\cdot\mathbf{\hat{n}}_{AB})^2+2v_A^2+2\mathbf{a}_A\cdot \mathbf{x}_{AB}+4(\mathbf{v}_A\cdot\mathbf{\hat{n}}_{AB})(\mathbf{v}_B\cdot \mathbf{\hat{n}}_{AB}) \vphantom{-4\mathbf{v}_A\cdot\mathbf{v}_B-2(\mathbf{v}_B\cdot \mathbf{\hat{n}}_{AB})^2+2v_B^2+2\sum_{C\neq B}Gm_C[1-\zeta+\zeta(1-2s_B)(1-2s_C)]\frac{\hat{n}_{BC}^i}{r_{BC}^2}x_{AB}^i}\right. \\
&\left.\qquad -4\mathbf{v}_A\cdot\mathbf{v}_B-2(\mathbf{v}_B\cdot \mathbf{\hat{n}}_{AB})^2+2v_B^2+2\sum_{C\neq B}Gm_C[1-\zeta+\zeta(1-2s_B)(1-2s_C)]\frac{\hat{n}_{BC}^i}{r_{BC}^2}x_{AB}^i\right] \\
&\quad +G^2\zeta(1-\zeta)\sum_{A,B}m_A(1-2s_A)m_B\left[-4\mathbf{v}_B\cdot\mathbf{a}_B-\frac{4}{3}\mathbf{x}_B\cdot \dot{\mathbf{a}}_B\right] \, .
\label{eq:IsF2}
\end{split}
\end{equation}
We see that $\Is_{s,{\text F1}}$ contains terms at 1PN, 2PN, and 2.5PN order (relative to the compact moment), while $\Is_{s,\text{F2}}$ contains terms at 2PN and 2.5PN order.

It is interesting to note that the integral of $\tau_{s,\text{F2}}$ can also be calculated by integrating the original expression from \eqref{eq:taus} directly by parts, i.e., without substituting in for the fields.  Using the conservation law \eqref{eq:conservation}, we find

\begin{equation}
-\frac{1}{8\pi}\int_\Ms \tilde{h}^{\alpha \beta}\varphi_{,\alpha \beta}\ d^3x = -\frac{1}{8\pi}\left[\frac{d^2}{dt^2}\int_\Ms \tilde{h}^{00}\Psi\ d^3x + 2\oint_{\partial \Ms}\tilde{h}^{0i}\dot{\Psi}\ d^2S^i+\oint_{\partial \Ms}\tilde{h}^{ij}\Psi^{,i}\ d^2S^j+\oint_{\partial \Ms}\dot{\tilde{h}}^{0i}\Psi\ d^2S^i\right] \, .
\end{equation}
The notation $\partial \Ms$ means that the surface integrals are evaluated on a sphere of radius $\Rs$ bounding the hypersurface $\Ms$.  The first and third surface integrals contribute nothing.  The second surface integral reproduces the 2.5PN terms in \eqref{eq:IsF2}.  The volume integral can be evaluated using the previous change-of-variable technique or by applying $\nabla^2 X = 2U$ and integrating by parts twice.  The final result is

\begin{equation}
2G^2\zeta(1-\zeta)\sum_A \sum_{B\neq A} m_A(1-2s_A)m_B\ddot{r}_{AB} \, .
\label{eq:easierIsF1}
\end{equation}
This can be shown to be equal to the 2PN terms in \eqref{eq:IsF2}.  This alternative method is therefore an important check of part of our result.  However, it also has other utility.  The form in \eqref{eq:easierIsF1} makes it easier to write down the two-body moment, as we will do in an upcoming subsection.  Furthermore, the alternative method will help save considerable effort when we extend this calculation to higher order.  It is much simpler to calculate the integral of $\tilde{h}^{00}\Psi$ and then take two time derivatives than to calculate directly the integral of $\tilde{h}^{\alpha\beta}\varphi_{,\alpha\beta}$.

\subsection{Higher-order scalar multipole moments}

The rest of the moments are calculated in much the same way as $\Is_s$.  For the one-index (dipole) moment, we only need $\tau_s$ to $O(\rho\epsilon^2)$.  The compact moment can be easily written down by adding a factor of $x_A^i$ to the monopole version,

\begin{align}
\Is_{s,\text{C}}^i &= G\zeta\sum_A m_Ax_A^i\left\{ 1-2s_A-\frac{1}{2}(1-2s_A)v_A^2 \right. \nonumber \\
&\qquad -\sum_{B\neq A}\frac{Gm_B}{r_{AB}}[(1-\zeta)(1-2s_A)+(4\lambda_1-\zeta)(1-2s_A)(1-2s_B)+4\zeta a_{sA}(1-2s_B)] \nonumber \\
&\qquad -\frac{1}{8}(1-2s_A)v_A^4\nonumber \\
&\qquad +\sum_{B\neq A}\frac{Gm_B}{r_{AB}}\left\{\left[-\frac{3}{2}(1-\zeta)(1-2s_A)+\frac{1}{2}(4\lambda_1-\zeta)(1-2s_A)(1-2s_B)+2\zeta a_{sA}(1-2s_B)\right]v_A^2 \right. \nonumber \\
&\qquad \quad-2(1-\zeta)(1-2s_A)v_B^2+4(1-\zeta)(1-2s_A)\mathbf{v}_A\cdot\mathbf{v}_B \nonumber \\
&\qquad \quad +\left[\frac{1}{2}(1-\zeta)(1-2s_A)+\frac{1}{2}(4\lambda_1-\zeta)(1-2s_A)(1-2s_B)+2\zeta a_{sA}(1-2s_B)\right]\mathbf{a}_B\cdot \mathbf{x}_{AB} \nonumber \\
&\qquad \quad +\left[\frac{1}{2}(1-\zeta)(1-2s_A)+\frac{1}{2}(4\lambda_1-\zeta)(1-2s_A)(1-2s_B)+2\zeta a_{sA}(1-2s_B)\right](\mathbf{v}_B\cdot \mathbf{\hat{n}}_{AB})^2 \nonumber \\
&\qquad \quad +\sum_{C\neq A}\frac{Gm_C}{r_{AC}}\left[\frac{1}{2}(1-\zeta)^2(1-2s_A)+(1-\zeta)(4\lambda_1-\zeta)(1-2s_A)(1-2s_C)+4\zeta(1-\zeta)a_{sA}(1-2s_C)\right.\nonumber \\
&\qquad \qquad +\left(20\lambda_1^2-9\zeta \lambda_1+\frac{1}{2}\zeta^2-4\lambda_2\right)(1-2s_A)(1-2s_B)(1-2s_C)+4\zeta(5\lambda_1-2\zeta)a_{sA}(1-2s_B)(1-2s_C) \nonumber \\
&\left.\qquad \qquad -4\zeta^2 b_{sA}(1-2s_B)(1-2s_C)\vphantom{\frac{1}{2}}\right] \nonumber \\
&\qquad \quad +\sum_{C\neq B}\frac{Gm_C}{r_{BC}}\left[(1-\zeta)^2(1-2s_A)+(1-\zeta)(4\lambda_1-\zeta)(1-2s_A)(1-2s_B)+4\zeta(1-\zeta)a_{sA}(1-2s_B) \right. \nonumber \\
&\qquad \qquad +(8\lambda_1^2+2\zeta\lambda_1-2\zeta^2+\zeta)(1-2s_A)(1-2s_B)(1-2s_C)+4\zeta(2\lambda_1+\zeta)a_{sA}(1-2s_B)(1-2s_C) \nonumber \\
&\left.\left.\left.\qquad \qquad +4\zeta(4\lambda_1-\zeta)(1-2s_A)a_{sB}(1-2s_C)+16\zeta^2a_{sA}a_{sB}(1-2s_C)\right]\vphantom{\frac{1}{2}}\right\}  \right\} \, .
\label{eq:dipolecompact}
\end{align}

The field integrals are evaluated using the same techniques as for the monopole moment.  One specific integral is worth discussing briefly.  From $\tau_{s,\text{F2}}$, we get

\begin{equation}
\int_\Ms P_2^{kl}U_s^{,kl} x^i d^3x = \int_\Ms d^3x\ U^{,k}U^{,l} \sum_A m_A(1-2s_A)\left[-\frac{1}{4}\Psi^{A,kli}+X^{A,(k}\delta^{l)i}-\frac{1}{2}X^{A,kl}x_A^i\right] \, ,
\end{equation}
where on the right-hand side we have already performed an integration using the change-of-variables technique.  A similar integral exists with $P_{2s}^{kl}$ in place of $P_2^{kl}$.  Here $\Psi^A \equiv |\mathbf{x}-\mathbf{x}_A|^3/3$.  This integral is a bit more difficult than the similar one computed for the zero-index moment.  Each piece can be integrated by parts multiple times.  The third term produces only vanishing surface integrals and integrals which can be turned into compact integrals.  The first and second produce other volume integrals; however, when the two terms are added, these unevaluated volume integrals cancel.  (A similar cancellation occurred within the integrals involving $P_2^{ij}$ in paper II.)  

The final F1 contribution to the dipole moment is

\begin{equation}
\begin{split}
\Is_{s,{\text F1}}^i &= G^2\zeta(\lambda_1-\zeta)\sum_A\sum_{B\neq A}\frac{m_A(1-2s_A)m_B(1-2s_B)}{r_{AB}}\left[2x_A^i-\frac{1}{2}v_A^2x_A^i+\frac{1}{2}v_B^2x_A^i-\frac{1}{2}(\mathbf{v}_A\cdot\mathbf{\hat{n}}_{AB})^2x_A^i \right. \\
&\qquad -\frac{3}{2}(\mathbf{v}_B\cdot\mathbf{\hat{n}}_{AB})^2x_A^i+(\mathbf{v}_A\cdot\mathbf{x}_{AB})v_A^i+\frac{1}{2}(\mathbf{a}_A\cdot\mathbf{x}_{AB})x_A^i-\frac{3}{2}(\mathbf{a}_B\cdot\mathbf{x}_{AB})x_A^i+\frac{1}{2}r_{AB}^2a_A^i  \\
&\left.\qquad +(\mathbf{v}_A\cdot\mathbf{\hat{n}}_{AB})(\mathbf{v}_B\cdot\mathbf{\hat{n}}_{AB})x_A^i-(\mathbf{v}_A\cdot\mathbf{v}_B)x_A^i-(\mathbf{v}_B\cdot\mathbf{x}_{AB})v_A^i\vphantom{\frac{1}{2}}\right] \\
&\quad +G^3\zeta \sum_A\sum_{B\neq A}\frac{m_Am_B}{r_{AB}}x_A^i\left[\sum_{C\neq A}\frac{m_C}{r_{AC}}[-2(1-\zeta)(\lambda_1-\zeta)(1-2s_A)(1-2s_B) \right. \\
&\qquad \quad+(-12\lambda_1^2+10\zeta\lambda_1+2\lambda_2)(1-2s_A)(1-2s_B)(1-2s_C)-8\zeta(\lambda_1-\zeta)a_{sA}(1-2s_B)(1-2s_C)]\\
&\qquad +\sum_{C\neq B}\frac{m_C}{r_{BC}}[-2(1-\zeta)(\lambda_1-\zeta)(1-2s_A)(1-2s_B)-2(\lambda_1-\zeta)(2\lambda_1+\zeta)(1-2s_A)(1-2s_B)(1-2s_C) \\
&\left.\qquad \quad -8\zeta(\lambda_1-\zeta)(1-2s_A)a_{sB}(1-2s_C)]\vphantom{\sum_A\sum_{B\neq A}\frac{m_Am_B}{r_{AB}}x_A^i\left[\sum_{C\neq A}\frac{m_C}{r_{AC}}[-2(1-\zeta)(\lambda_1-\zeta)(1-2s_A)(1-2s_B)\right.}\right] \, .
\end{split}
\end{equation}
The F2 contribution is

\begin{align}
\Is_{s,\text{F2}}^i &= G^2\zeta(1-\zeta)\sum_A\sum_{B\neq A}m_A(1-2s_A)\frac{m_B}{r_{AB}}\left[\vphantom{\sum_{C\neq B}\frac{Gm_C}{r_{BC}}[1-\zeta+\zeta(1-2s_B)(1-2s_C)]\left[\frac{\hat{n}_{BC}^k}{r_{BC}}x_{AB}^k(x_A^i+x_B^i)+3\frac{\hat{n}_{BC}^i}{r_{BC}}r_{AB}^2\right]}-(\mathbf{v}_A\cdot\mathbf{\hat{n}}_{AB})^2x_A^i-(\mathbf{v}_A\cdot\mathbf{\hat{n}}_{AB})^2x_B^i+2(\mathbf{v}_A\cdot\mathbf{x}_{AB})v_A^i+v_A^2x_A^i+v_A^2x_B^i \right. \nonumber \\
&\qquad +(\mathbf{a}_A\cdot\mathbf{x}_{AB})x_A^i+(\mathbf{a}_A\cdot\mathbf{x}_{AB})x_B^i+r_{AB}^2a_A^i+2(\mathbf{v}_A\cdot\mathbf{\hat{n}}_{AB})(\mathbf{v}_B\cdot\mathbf{\hat{n}}_{AB})x_A^i+2(\mathbf{v}_A\cdot\mathbf{\hat{n}}_{AB})(\mathbf{v}_B\cdot\mathbf{\hat{n}}_{AB})x_B^i \nonumber \\
&\qquad -2(\mathbf{v}_A\cdot\mathbf{x}_{AB})v_B^i-2(\mathbf{v}_B\cdot\mathbf{x}_{AB})v_A^i-2(\mathbf{v}_A\cdot\mathbf{v}_B)x_A^i -2(\mathbf{v}_A\cdot\mathbf{v}_B)x_B^i-(\mathbf{v}_B\cdot\mathbf{\hat{n}}_{AB})^2x_A^i-(\mathbf{v}_B\cdot\mathbf{\hat{n}}_{AB})^2x_B^i\nonumber \\
&\qquad +2(\mathbf{v}_B\cdot\mathbf{x}_{AB})v_B^i+v_B^2x_A^i+v_B^2x_B^i \nonumber \\
&\left.\qquad +\sum_{C\neq B}\frac{Gm_C}{r_{BC}}[1-\zeta+\zeta(1-2s_B)(1-2s_C)]\left[\frac{\hat{n}_{BC}^k}{r_{BC}}x_{AB}^k(x_A^i+x_B^i)+3\frac{\hat{n}_{BC}^i}{r_{BC}}r_{AB}^2\right]\right] \, .
\end{align}
This has been checked by starting from the $\tilde{h}^{\alpha\beta}\varphi_{,\alpha\beta}$ term in \eqref{eq:taus} and applying manipulations similar to those that led to Eq.\  \eqref{eq:easierIsF1}.  

The two-index (quadrupole) moment requires $\tau_s$ to 1.5PN order, or $O(\rho\epsilon^{3/2})$.  This means we do not need to consider $\tau_{s,\text{F2}}$ in this or any further moments, since it begins at $O(\rho\epsilon^2)$.  It turns out that $\tau_s$ has no contributions at 1.5PN order, so we essentially use it at 1PN order.  The complete two-index moment is given by

\begin{equation}
\begin{split}
\Is_s^{ij} &= G\zeta\sum_A m_A x_A^{ij}\left\{\vphantom{-\sum_{B\neq A}\frac{Gm_B}{r_{AB}}[(1-\zeta)(1-2s_A)+(4\lambda_1-\zeta)(1-2s_A)(1-2s_B)+4\zeta a_{sA}(1-2s_B)]}1-2s_A-\frac{1}{2}(1-2s_A)v_A^2 \right. \\
&\left.\qquad -\sum_{B\neq A}\frac{Gm_B}{r_{AB}}[(1-\zeta)(1-2s_A)+(4\lambda_1-\zeta)(1-2s_A)(1-2s_B)+4\zeta a_{sA}(1-2s_B)]\right\} \\
&\quad +G^2\zeta(\lambda_1-\zeta)\sum_A\sum_{B\neq A} \frac{m_A(1-2s_A)m_B(1-2s_B)}{r_{AB}}\left[2x_A^{ij}-r_{AB}^2\delta^{ij}\right] \, .
\end{split}
\end{equation}
The last line represents the F1 contribution.  Note the presence of a $\delta^{ij}$ term. We ignored such terms in paper II because they would not survive a TT projection; in the scalar case, we must keep all terms.

The three-index moment also uses $\tau_s$ to $O(\rho\epsilon)$.  It is given by

\begin{equation}
\begin{split}
\Is_s^{ijk} &= G\zeta\sum_A m_A x_A^{ijk}\left\{\vphantom{-\sum_{B\neq A}\frac{Gm_B}{r_{AB}}[(1-\zeta)(1-2s_A)+(4\lambda_1-\zeta)(1-2s_A)(1-2s_B)+4\zeta a_{sA}(1-2s_B)]}1-2s_A-\frac{1}{2}(1-2s_A)v_A^2 \right. \\
&\left.\qquad -\sum_{B\neq A}\frac{Gm_B}{r_{AB}}[(1-\zeta)(1-2s_A)+(4\lambda_1-\zeta)(1-2s_A)(1-2s_B)+4\zeta a_{sA}(1-2s_B)]\right\} \\
&\quad +G^2\zeta(\lambda_1-\zeta)\sum_A\sum_{B\neq A}\frac{m_A(1-2s_A)m_B(1-2s_B)}{r_{AB}}\left[2x_A^{ijk}-3r_{AB}^2\delta^{(ij}x_A^{k)}\right] \, .
\end{split}
\end{equation}
The four-index moment needs $\tau_s$ to $O(\rho\epsilon^{1/2})$, which means that it just uses its lowest order compact piece,

\begin{equation}
\Is_s^{ijkl} = G\zeta\sum_A m_A(1-2s_A)x_A^{ijkl} \, .
\end{equation}
Similarly, the five-index moment is

\begin{equation}
\Is_s^{ijklm} = G\zeta\sum_A m_A(1-2s_A)x_A^{ijklm} \, .
\end{equation}

\subsection{Two-body scalar moments}
\label{sec:twobody}
So far we have written down expressions for the scalar multipole moments which are fully general and can be applied to a system with any number of compact objects.  Now we convert the moments to the two-body case relevant for compact binaries.  We define the masses $m_A$ of the bodies as $m_1$ and $m_2$.  They have positions $\mathbf{x}_1$ and $\mathbf{x}_2$ and velocities $\mathbf{v}_1$ and $\mathbf{v}_2$.  We also define the usual quantities $m \equiv m_1+m_2$, $\mu \equiv m_1m_2/m$, $\eta \equiv \mu/m$, and $\delta m \equiv m_1-m_2$.  We note that $(\delta m/m)^2 = 1-4\eta$.

It will be useful to have the moments expressed in terms of relative variables, $r = r_{12}$, $\mathbf{x} = \mathbf{x}_{1}-\mathbf{x}_2$, $\mathbf{\hat{n}} = \mathbf{\hat{n}}_{12} = \mathbf{x}/r$, and $\mathbf{v} = \mathbf{v}_1-\mathbf{v}_2$.  To make this transformation, we must first find explicit expressions for $\mathbf{x}_1$ and $\mathbf{x}_2$ in terms of relative variables to 2PN order.  In paper II, we stopped at 1.5PN order, but here we require the 2PN pieces to plug into the lowest order term of \eqref{eq:dipolecompact}.  We first define the center of mass,

\begin{equation}
X_\text{CM}^i = \frac{1}{m}\int_\Ms \tau^{00} x^i d^3x \, .
\end{equation}
This integral can be evaluated by the techniques of Sec.\ \ref{sec:zeroindex}.  Now if $X_\text{CM}^i$ satisfied the normal conservation statement $m X_\text{CM}^i(t) = mX_\text{CM}^i(0)+P_\text{CM}^i t$ with a conserved $P_\text{CM}^i$, we could choose a frame in which $X_\text{CM}^i = 0$ and use that equation to solve for $\mathbf{x}_1$ and $\mathbf{x}_2$.  However, the presence of dipole radiation reaction at 1.5PN order in the equations of motion means that $\ddot{X}_\text{CM}^i$ is nonzero at this order, giving an apparent 1.5PN ``kick'' of the system's center of mass.  But it turns out that the 1.5PN contribution to $\ddot{X}_\text{CM}^i$ is itself a double total time derivative which, when integrated twice, can be absorbed into a 1.5PN correction of the definition of $X_\text{CM}^i$.  Since we need the transformations from $\mathbf{x}_1$ and $\mathbf{x}_2$ to $\mathbf{x}$ (and thus $X_\text{CM}^i$) to 2PN order, we must include this 1.5PN correction term.

When solving for the bodies' positions, we need to simultaneously solve for their velocities $\mathbf{v}_1$ and $\mathbf{v}_2$ in terms of relative variables.  We need these only up to 1.5PN order.  The final results for the positions are

\begin{subequations}
\begin{align}
x_1^i &= \frac{m_2}{m}x^i + \delta_x^i \, , \label{eq:x1}\\
x_2^i &= -\frac{m_1}{m}x^i + \delta_x^i \, ,
\label{eq:x2}
\end{align}
\end{subequations}
with

\begin{equation}
\begin{split}
\delta_x^i &= \frac{1}{2}\eta\frac{\delta m}{m}\left(v^2-\frac{G\alpha m}{r}\right)x^i+\frac{2}{3}\eta\zeta\Ss_-\left(\Ss_++\frac{\delta m}{m}\Ss_-\right)G\alpha mv^i \\
&\quad+\eta\frac{\delta m}{m}\left[\frac{3}{8}(1-4\eta)v^4+\left(\frac{19}{8}+\frac{3}{2}\eta+\frac{3}{2}\bar{\gamma}\right)v^2\frac{G\alpha m}{r}+\left(-\frac{1}{8}+\frac{3}{4}\eta\right)\dot{r}^2\frac{G\alpha m}{r}+\left(\frac{7}{4}-\frac{1}{2}\eta+\bar{\gamma}\right)\left(\frac{G\alpha m}{r}\right)^2\right]x^i \\
&\quad +2\eta^2\bar{\beta}_-\left(\frac{G\alpha m}{r}\right)^2x^i-\eta\frac{\delta m}{m}\left(\frac{7}{4}+\bar{\gamma}\right)G\alpha m\dot{r}v^i+O(\epsilon^{5/2}) \, .
\label{eq:deltax}
\end{split}
\end{equation}
We have defined

\begin{equation}
\alpha \equiv 1-\zeta + \zeta(1-2s_1)(1-2s_2) \, .
\label{eq:alpha}
\end{equation}
The quantity $G\alpha$ is the effective gravitational constant in the 0PN equations of motion; see paper I.  We also introduce some redefined sensitivity parameters,
\begin{subequations}
\begin{align}
\Ss_+ &\equiv \alpha^{-1/2}(1-s_1-s_2) \, ,\\
\Ss_- &\equiv \alpha^{-1/2}(s_2-s_1) \, .
\label{eq:S+S-}
\end{align}
\end{subequations}
These represent the sum and difference of the bodies' ``scalar charges.''  Finally, we use the following parameters from papers I and II:

\begin{subequations}
\begin{align}
\bar{\gamma} &\equiv -2\alpha^{-1}\zeta(1-2s_1)(1-2s_2) \, , \\
\bar{\beta}_1 &\equiv \alpha^{-2}\zeta(1-2s_2)^2(\lambda_1(1-2s_1)+2\zeta s_1') \, , \\
\bar{\beta}_2 &\equiv \alpha^{-2}\zeta(1-2s_1)^2(\lambda_1(1-2s_2)+2\zeta s_2') \, , \\
\bar{\beta}_+ &\equiv \frac{1}{2}(\bar{\beta}_1+\bar{\beta}_2) \, , \\
\bar{\beta}_- & \equiv \frac{1}{2}(\bar{\beta}_1-\bar{\beta}_2) \, .
\end{align}
\label{eq:STparameters1}
\end{subequations}
The first term in \eqref{eq:deltax} is 1PN order, the second 1.5PN order, and the rest 2PN order.  The velocities are given by
\begin{subequations}
\begin{align}
v_1^i &= \frac{m_2}{m}v^i + \delta_v^i \, ,\\
v_2^i &= -\frac{m_1}{m}v^i + \delta_v^i \, ,
\end{align}
\end{subequations}
with

\begin{equation}
\delta_v^i = \frac{1}{2}\eta\frac{\delta m}{m}\left[\left(v^2-\frac{G\alpha m}{r}\right)v^i-\frac{G\alpha m}{r^2}\dot{r}x^i\right]-\frac{2}{3}\eta\zeta\Ss_-\left(\Ss_++\frac{\delta m}{m}\Ss_-\right)\left(\frac{G\alpha m}{r}\right)^2\hat{n}^i+O(\epsilon^2) \, .
\label{eq:vdelta}
\end{equation}
Here, the first term is 1PN order and the second term is 1.5PN order. 

To simplify the moments, we introduce additional scalar-tensor parameters from papers I and II,

\begin{subequations}
\begin{align}
\bar{\delta}_1 &\equiv \alpha^{-2}\zeta(1-\zeta)(1-2s_1)^2 \, ,\\
\bar{\delta}_2 &\equiv \alpha^{-2}\zeta(1-\zeta)(1-2s_2)^2 \, ,\\
\bar{\chi}_1 &\equiv \alpha^{-3}\zeta(1-2s_2)^3[(\lambda_2-4\lambda_1^2+\zeta\lambda_1)(1-2s_1)-6\zeta\lambda_1s_1'+2\zeta^2s_1''] \, , \\
\bar{\chi}_2 &\equiv \alpha^{-3}\zeta(1-2s_1)^3[(\lambda_2-4\lambda_1^2+\zeta\lambda_1)(1-2s_2)-6\zeta\lambda_1s_2'+2\zeta^2s_2''] \, .
\end{align}
\label{eq:STparameters2}
\end{subequations}
Just as with $\bar{\beta}$, we use the notation

\begin{subequations}
\begin{align}
\xi_+ &\equiv \frac{1}{2}(\xi_1+\xi_2) \, ,\\
\xi_- &\equiv \frac{1}{2}(\xi_1-\xi_2) \, ,
\end{align}
\end{subequations}
where $\xi$ is $\bar{\delta}$ or $\bar{\chi}$.  (This notation should not be confused with $\Ss_+$ and $\Ss_-$.)  We also need expressions for the accelerations to lowest post-Newtonian order,
\begin{subequations}
\begin{align}
a_1^i &= -\frac{G\alpha m_2}{r^2}\hat{n}^i \, ,\\
a_2^i &= \frac{G\alpha m_1}{r^2}\hat{n}^i \, .
\end{align}
\end{subequations}
Using all these expressions and definitions, we simplify the scalar moments (combining the C, F1, and F2 terms if originally separated) to find the following two-body moments,

\begin{subequations}
\begin{align}
\Is_s &= G\zeta m \alpha^{1/2}\left(\Ss_++\frac{\delta m}{m}\Ss_-\right) \nonumber \\
&\quad +G\zeta\mu\alpha^{1/2}\left\{-\frac{1}{2}\left(\Ss_+-\frac{\delta m}{m}\Ss_-\right)v^2+\left[-2\Ss_++\frac{8}{\bar{\gamma}}(\Ss_+\bar{\beta}_++\Ss_-\bar{\beta}_-)\right]\frac{G\alpha m}{r} \vphantom{\left(\frac{G\alpha m}{r}\right)^2}\right. \nonumber \\
&\qquad +\left[-\frac{1}{8}(1-3\eta)\Ss_++\frac{1}{8}\frac{\delta m}{m}(1-9\eta)\Ss_-\right]v^4 \nonumber \\
&\qquad +\left[\frac{1}{2}(1-2\eta)\left(\Ss_+-\frac{\delta m}{m}\Ss_-\right)-\frac{2}{\bar{\gamma}}(1-2\eta)(\Ss_+\bar{\beta}_++\Ss_-\bar{\beta}_-)+\frac{2}{\bar{\gamma}}\frac{\delta m}{m}(\Ss_-\bar{\beta}_++\Ss_+\bar{\beta}_-)\right]v^2\frac{G\alpha m}{r}\nonumber \\
&\qquad +\left[\left(-\frac{7}{2}-\eta-2\bar{\gamma}\right)\Ss_++\frac{1}{2}\frac{\delta m}{m}(1+2\eta)\Ss_--\frac{2}{\bar{\gamma}}(1-2\eta)(\Ss_+\bar{\beta}_++\Ss_-\bar{\beta}_-)-\frac{2}{\bar{\gamma}}\frac{\delta m}{m}(\Ss_-\bar{\beta}_++\Ss_+\bar{\beta}_-)\right]\dot{r}^2\frac{G\alpha m}{r}\nonumber \\
&\qquad +\left[\left(-2-2\bar{\gamma}+3\bar{\beta}_+-3\frac{\delta m}{m}\bar{\beta}_-\right)\Ss_++\frac{\delta m}{m}(1+\bar{\beta}_+)\Ss_--\bar{\beta}_-\Ss_-\right. \nonumber \\
&\qquad \quad -\frac{10}{\bar{\gamma}}(\Ss_+\bar{\beta}_++\Ss_-\bar{\beta}_-)-\frac{2}{\bar{\gamma}}\frac{\delta m}{m}(\Ss_-\bar{\beta}_++\Ss_+\bar{\beta}_-)\nonumber \\
&\left.\left.\qquad \quad +\frac{16}{\bar{\gamma}^2}\left(\Ss_++\frac{\delta m}{m}\Ss_-\right)(\bar{\beta}_+^2-\bar{\beta}_-^2)+\frac{4}{\bar{\gamma}}(\Ss_+\bar{\chi}_++\Ss_-\bar{\chi}_-)-\frac{4}{\bar{\gamma}}\frac{\delta m}{m}(\Ss_-\bar{\chi}_++\Ss_+\bar{\chi}_-)\right]\left(\frac{G\alpha m}{r}\right)^2\right\} \nonumber \\
&\quad +G(\lambda_1-\zeta)\mu\bar{\gamma}\left[-\frac{1}{2}\frac{G\alpha m}{r}v^2+\frac{1}{2}\frac{G\alpha m}{r}\dot{r}^2+\frac{1}{2}\left(\frac{G\alpha m}{r}\right)^2\right] \nonumber \\
&\quad +G\zeta\mu\alpha^{1/2}\left(\frac{G\alpha m}{r}\right)^2\dot{r}\left\{\frac{8}{3}\zeta\Ss_+^3+\frac{14}{3}\frac{\delta m}{m}\zeta\Ss_+^2\Ss_-+\frac{4}{3}(1+\eta)\zeta\Ss_+\Ss_-^2-\frac{2}{3}\frac{\delta m}{m}(1-2\eta)\zeta\Ss_-^3 \right. \nonumber \\
&\qquad +\frac{1}{\bar{\gamma}^2}\left[\frac{64}{3}\zeta\Ss_+^3+\frac{112}{3}\frac{\delta m}{m}\zeta \Ss_+^2\Ss_-+\frac{32}{3}\zeta\Ss_+\Ss_-^2-\frac{16}{3}\frac{\delta m}{m}\zeta\Ss_-^3+16\zeta^2\Ss_+^5+32\frac{\delta m}{m}\zeta^2\Ss_+^4\Ss_- \right. \nonumber \\
&\left.\qquad \quad -32\frac{\delta m}{m}\zeta^2\Ss_+^2\Ss_-^3-16\zeta^2\Ss_+\Ss_-^4\right]\bar{\beta}_+\nonumber \\
&\qquad +\frac{1}{\bar{\gamma}^2}\left[\frac{64}{3}\frac{\delta m}{m}\zeta\Ss_+^3+\frac{112}{3}\zeta \Ss_+^2\Ss_-+\frac{32}{3}\frac{\delta m}{m}\zeta\Ss_+\Ss_-^2-\frac{16}{3}\zeta\Ss_-^3+16\zeta^2\Ss_+^4\Ss_-+32\frac{\delta m}{m}\zeta^2\Ss_+^3\Ss_-^2 \right. \nonumber \\
&\left.\qquad \quad -32\frac{\delta m}{m}\zeta^2\Ss_+\Ss_-^4-16\zeta^2\Ss_-^5\right]\bar{\beta}_-\nonumber \\
&\qquad -\frac{64}{\bar{\gamma}^3}\zeta\Ss_+\left(\Ss_+^2+2\frac{\delta m}{m}\Ss_+\Ss_-+\Ss_-^2\right)\bar{\beta}_+^2-\frac{64}{\bar{\gamma}^3}\zeta\Ss_-\left(\frac{\delta m}{m}\Ss_+^2+2\Ss_+\Ss_-+\frac{\delta m}{m}\Ss_-^2\right)\bar{\beta}_-^2\nonumber \\
&\qquad -\frac{64}{\bar{\gamma}^3}\left(\frac{\delta m}{m}\zeta \Ss_+^3+3\zeta\Ss_+^2\Ss_-+3\frac{\delta m}{m}\zeta\Ss_+\Ss_-^2+\zeta\Ss_-^3\right)\bar{\beta}_+\bar{\beta}_- \nonumber \\
&\left.\qquad +(\lambda_1-\zeta)\frac{1}{\alpha^{1/2}}\left[\frac{16}{3}\Ss_+^2+4\frac{\delta m}{m}\Ss_+\Ss_--\frac{4}{3}\left(1-\frac{8}{3}\eta\right)\Ss_-^2-\frac{16}{\bar{\gamma}}\left(\Ss_++\frac{\delta m}{m}\Ss_-\right)(\Ss_+\bar{\beta}_++\Ss_-\bar{\beta}_-)\right]\right\} \, ,
\label{eq:IsTB}
\end{align}

\begin{align}
\Is_s^i &= G\zeta\mu\alpha^{1/2}\left\{2\Ss_-x^i+\left(\frac{\delta m}{m}\Ss_+-\eta\Ss_-\right)v^2x^i \right.  \nonumber \\
&\qquad +\left[\frac{1}{2}\frac{\delta m}{m}\Ss_++\left(-\frac{3}{2}+2\eta\right)\Ss_--\frac{4}{\bar{\gamma}}\frac{\delta m}{m}(\Ss_+\bar{\beta}_++\Ss_-\bar{\beta}_-)+\frac{4}{\bar{\gamma}}(\Ss_-\bar{\beta}_++\Ss_+\bar{\beta}_-)\right]\frac{G\alpha m}{r}x^i \nonumber \\
&\qquad +\frac{2}{3}G\zeta\alpha m \Ss_-\left(\Ss_++\frac{\delta m}{m}\Ss_-\right)^2v^i  \nonumber \\
&\qquad +\left[\frac{1}{2}\frac{\delta m}{m}(1-5\eta)\Ss_++\frac{1}{4}(1-7\eta+11\eta^2)\Ss_-\right]v^4x^i  \nonumber \\
&\qquad +\left[\frac{\delta m}{m}\left(\frac{15}{8}+\frac{7}{4}\eta+\frac{3}{2}\bar{\gamma}\right)\Ss_++\left(\frac{7}{8}-\frac{41}{4}\eta-3\eta^2+\frac{1}{2}(1-12\eta)\bar{\gamma}\right)\Ss_-\right.  \nonumber \\
&\left.\qquad \quad +\frac{2}{\bar{\gamma}}\frac{\delta m}{m}(1+\eta)(\Ss_+\bar{\beta}_++\Ss_-\bar{\beta}_-)-\frac{2}{\bar{\gamma}}(1-3\eta)(\Ss_-\bar{\beta}_++\Ss_+\bar{\beta}_-)\right]v^2\frac{G\alpha m}{r}x^i \nonumber \\
&\qquad +\left[\frac{\delta m}{m}\left(\frac{15}{8}+\frac{7}{4}\eta+\bar{\gamma}\right)\Ss_++\left(-\frac{1}{8}+\frac{5}{4}\eta-\eta^2\right)\Ss_--\frac{2}{\bar{\gamma}}\frac{\delta m}{m}\eta(\Ss_+\bar{\beta}_++\Ss_-\bar{\beta}_-) \right. \nonumber \\
&\left.\qquad \quad -\frac{2}{\bar{\gamma}}\eta(\Ss_-\bar{\beta}_++\Ss_+\bar{\beta}_-)\right]\dot{r}^2\frac{G\alpha m}{r}x^i  \nonumber \\
&\qquad +\left[\frac{\delta m}{m}\left(\frac{5}{4}+\frac{1}{2}\eta+\bar{\gamma}-\bar{\beta}_++\frac{\delta m}{m}\bar{\beta}_-\right)\Ss_++\left(-\frac{7}{4}-\frac{11}{2}\eta+2\eta^2-(1+4\eta)\bar{\gamma}+(1+2\eta)\bar{\beta}_+ \right. \right. \nonumber \\
&\left.\qquad \quad -(1-2\eta)\frac{\delta m}{m}\bar{\beta}_-\right)\Ss_- +\frac{4}{\bar{\gamma}}\frac{\delta m}{m}(1-\eta)(\Ss_+\bar{\beta}_++\Ss_-\bar{\beta}_-)-\frac{4}{\bar{\gamma}}(1+\eta)(\Ss_-\bar{\beta}_++\Ss_+\bar{\beta}_-)  \nonumber \\
&\left.\qquad \quad +\frac{32}{\bar{\gamma}^2}\eta\Ss_-(\bar{\beta}_+^2-\bar{\beta}_-^2)-\frac{4}{\bar{\gamma}}\frac{\delta m}{m}(\Ss_+\bar{\chi}_++\Ss_-\bar{\chi}_-)+\frac{4}{\bar{\gamma}}(1-2\eta)(\Ss_-\bar{\chi}_++\Ss_+\bar{\chi}_-)\right]\left(\frac{G\alpha m}{r}\right)^2 x^i \nonumber \\
&\left.\qquad +\left[\frac{\delta m}{m}\left(-\frac{7}{4}-\bar{\gamma}\right)\Ss_++\left(\frac{9}{4}+7\eta+(1+4\eta)\bar{\gamma}\right)\Ss_-\right]G\alpha m\dot{r}v^i\right\} \nonumber \\
&\quad +G(\lambda_1-\zeta)\mu \bar{\gamma}\frac{\delta m}{m}\left[\frac{1}{4}\frac{G\alpha m}{r}v^2x^i-\frac{1}{4}\frac{G\alpha m}{r}\dot{r}^2x^i-\frac{1}{2}\left(\frac{G\alpha m}{r}\right)^2x^i+\frac{1}{2}G\alpha m\dot{r}v^i\right] \, ,
\label{eq:IsiTB}
\end{align}

\begin{equation}
\begin{split}
\Is_s^{ij} &= G\zeta\mu\alpha^{1/2}\left\{\left(\Ss_+-\frac{\delta m}{m}\Ss_-\right)x^{ij} \right.\\
&\qquad +\left[-\frac{1}{2}(1-3\eta)\Ss_++\frac{1}{2}\frac{\delta m}{m}(1+3\eta)\Ss_-\right]v^2x^{ij} \\
&\left.\qquad +\left[-(1-2\eta)\Ss_++\frac{\delta m}{m}(1-2\eta)\Ss_-+\frac{4}{\bar{\gamma}}(1-2\eta)(\Ss_+\bar{\beta}_++\Ss_-\bar{\beta}_-)-\frac{4}{\bar{\gamma}}\frac{\delta m}{m}(\Ss_-\bar{\beta}_++\Ss_+\bar{\beta}_-)\right]\frac{G\alpha m}{r}x^{ij}\right\} \\
&\quad +G^2(\lambda_1-\zeta)\mu\alpha mr\bar{\gamma}\delta^{ij} \\
&\quad +\frac{8}{3}G^2\zeta^2\alpha^{3/2}\mu\eta m\Ss_-^2\left(\Ss_++\frac{\delta m}{m}\Ss_-\right)x^{(i}v^{j)} \, ,
\end{split}
\label{eq:IsijTB}
\end{equation}

\begin{equation}
\begin{split}
\Is_s^{ijk} &= G\zeta\mu\alpha^{1/2}\left\{\left[-\frac{\delta m}{m}\Ss_++(1-2\eta)\Ss_-\right]x^{ijk} +\left[\frac{1}{2}\frac{\delta m}{m}(1+\eta)\Ss_++\left(-\frac{1}{2}+\frac{1}{2}\eta+5\eta^2\right)\Ss_-\right]v^2x^{ijk}  \right. \\
&\qquad +\left[\frac{\delta m}{m}\left(1-\frac{5}{2}\eta\right)\Ss_++\left(-1+\frac{9}{2}\eta-6\eta^2\right)\Ss_--\frac{4}{\bar{\gamma}}\frac{\delta m}{m}(1-\eta)(\Ss_+\bar{\beta}_++\Ss_-\bar{\beta}_-)\right. \\
&\left.\left.\qquad \quad+\frac{4}{\bar{\gamma}}(1-3\eta)(\Ss_-\bar{\beta}_++\Ss_+\bar{\beta}_-)\right]\frac{G\alpha m}{r}x^{ijk}\right\} \\
&\quad -\frac{3}{2}G^2(\lambda_1-\zeta)\mu\alpha\bar{\gamma}mr\frac{\delta m}{m}\delta^{(ij}x^{k)} \, ,
\label{eq:IsijkTB}
\end{split}
\end{equation}

\begin{equation}
\Is_s^{ijkl} = G\zeta\mu\alpha^{1/2}\left[(1-3\eta)\Ss_+-\frac{\delta m}{m}(1-\eta)\Ss_-\right]x^{ijkl} \, ,
\label{eq:IsijklTB}
\end{equation}

\begin{equation}
\Is_s^{ijklm} = G\zeta\mu\alpha^{1/2}\left[-\frac{\delta m}{m}(1-2\eta)\Ss_++(1-4\eta+2\eta^2)\Ss_-\right]x^{ijklm} \, .
\label{eq:IsijklmTB}
\end{equation}

\end{subequations}

The final step is to take time derivatives.  The equation for $\Psi_\Ns$, \eqref{eq:scalarfaraway}, shows that we must take $m$ derivatives of the $m$-index moment.  Along the way, we need to substitute the relative equation of motion for each acceleration $a^i$.  We take this result from paper I, 

\begin{equation}
a^i = -\frac{G\alpha m}{r^2}\hat{n}^i+\frac{G\alpha m}{r^2}(A_\text{PN}\hat{n}^i+B_\text{PN}\dot{r}v^i)+\frac{8}{5}\eta\frac{(G\alpha m)^2}{r^3}(A_\text{1.5PN}\dot{r}\hat{n}^i-B_\text{1.5PN}v^i)+\frac{G\alpha m}{r^2}(A_\text{2PN}\hat{n}^i+B_\text{2PN}\dot{r}v^i) \, .
\end{equation}
We keep it to 2PN order; however, the 2PN terms are not needed for finding the scalar waves.  We will need them for computation of the energy flux in Sec.\ \ref{sec:flux}.  The coefficients $A_\text{PN}$, $B_\text{PN}$, $A_\text{1.5PN}$, $B_\text{1.5PN}$, $A_\text{2PN}$, and $B_\text{2PN}$ are given in paper I, Eqs.\ (1.5).  Of these, $A_\text{PN}$, $A_\text{2PN}$, and $B_\text{2PN}$ depend on time, while the others are constant.   From $x^i = r\hat{n}^i$, we can also find

\begin{equation}
\ddot{r}=\frac{v^2}{r}-\frac{\dot{r}^2}{r}-\frac{G\alpha m}{r^2}+\frac{G\alpha m}{r^2}(A_\text{PN}+\dot{r}^2B_\text{PN}) \, ,
\end{equation}
where we have kept only the terms needed in this paper.  We hold off on presenting the final results from the near-zone integral until Sec.\ \ref{sec:scalarwaves}.

\section{Radiation-zone contribution to the scalar waveform}
\label{sec:RZintegrals}

So far, our calculation of the scalar waveform has only considered the contribution from the near zone, $\Psi_\Ns$.  We must also calculate the contribution from the radiation zone, $\Psi_{\Cs-\Ns}$.  (Recall that $\Cs$ is the past null cone emanating from a field point and that $\Ns$ is its intersection with the near-zone world tube.  Then $\Cs-\Ns$ denotes the remainder of the past null cone, i.e., its intersection with the radiation zone.)  The first step is to calculate the source $\tau_s$ in the radiation zone.  By definition, the radiation zone does not include the compact objects, so $\tau_s$ will consist only of field terms.  Therefore, to find $\tau_s$, we must first compute the fields $\tilde{h}^{\mu \nu}$ and $\Psi$ in the radiation zone.  The near-zone contributions to these are found using

\begin{subequations}
\begin{align}
\tilde{h}_\Ns ^{\mu\nu}(t,\mathbf{x}) &= 4\sum_{m=0}^\infty\frac{(-1)^m}{m!}\left(\frac{1}{R}M^{\mu\nu k_1\cdots k_m}(\tau)\right)_{,k_1\cdots k_m} \label{eq:hRN}\, ,\\
\Psi_\Ns(t,\mathbf{x}) &= 2\sum_{m=0}^\infty\frac{(-1)^m}{m!}\left(\frac{1}{R}\Is_s^{k_1\cdots k_m}(\tau)\right)_{,k_1\cdots k_m} \label{eq:phiRN}\, ,
\end{align}
\end{subequations}
where  

\begin{equation}
M^{\mu\nu k_1\cdots k_m}(\tau) \equiv \int_\Ms\tau^{\mu\nu}(\tau,\mathbf{x})x^{k_1}\cdots x^{k_m}d^3x \, ,
\end{equation}
so that $M^{00M}(\tau) = \Is^M(\tau)$.  Here $\Ms$ is again the intersection of the near-zone world tube with a hypersurface of constant {\em retarded} time $\tau = t-R$.  Equation \eqref{eq:hRN} reduces to the Epstein-Wagoner construction for the tensor waves when $\mu = i$, $\nu = j$, and $R \gg \Rs$ (so that only $1/R$ terms are kept).  Similarly, Eq.\ \eqref{eq:phiRN} reduces to \eqref{eq:scalarfaraway} when $R \gg \Rs$.  To construct $\tau_s$ in the radiation zone, we require these more general expressions, which include all components of $\tilde{h}^{\mu\nu}$ as well as terms which fall off faster than $1/R$.

The evaluation of \eqref{eq:hRN}--\eqref{eq:phiRN} was detailed in paper II.  The final result for the radiation-zone fields is 

\begin{subequations}
\begin{align}
N  &= \tilde{h} ^{00} = 4G(1-\zeta)\frac{m+E}{R}+2\left(\frac{\Is^{ij}}{R}\right)_{,ij}+\cdots \label{eq:NRN} \, ,\\
K^i &= \tilde{h} ^{0i} = -2\left(\frac{\dot{\Is}^{ij}-\epsilon^{ija}\Js^a}{R}\right)_{,j}+ \cdots \label{eq:KRN}\, ,\\
B^{ij} &= \tilde{h} ^{ij} = 2\frac{\ddot{\Is}^{ij}}{R} + \cdots \label{eq:BRN}\, ,\\
\Psi &= 2G\zeta\frac{m_s+m_{s1}}{R}-2\left(\frac{\Is_s^i}{R}\right)_{,i}+\left(\frac{\Is_s^{ij}}{R}\right)_{,ij}+\cdots \, ,
\label{eq:PsiRN}
\end{align}
\end{subequations}
where the angular-momentum integral $\Js^{iQ}$ is defined in paper I, Eq.\ (3.7b).  We have split $\Is = m+E$ into its 0PN piece, the time-independent $m$, and its 1PN piece $E$.  Similarly, $\Is_s$ is split into its 0PN piece, the time-independent $m_s$, and its 1PN piece $m_{s1}$.  This split is made explicit because all of the other moments are only needed to lowest post-Newtonian order.  The scalar moments can be found in Sec.\ \ref{sec:NZintegrals}, albeit to much higher order than we need at this juncture.  The regular moments are written out explicitly in paper II.  

In paper II, we described the post-Newtonian orders of the terms in \eqref{eq:NRN}--\eqref{eq:PsiRN} only relative to one another.  Here we take a different approach and use the same explicit PN counting we use for the final waves.  In this system, the lowest order terms, those involving $m$ and $m_s$, are $-1$PN order.  The next highest order terms are those in $\Psi$ involving the dipole.  They are $-0.5$PN order; the pieces which scale like $1/R$ are the lowest order scalar waves generated by the near-zone integral.  Finally, all of the other terms are 0PN order.  Note that in paper II, we carried out this expansion to one-half higher order (i.e., 0.5PN).  Those higher-order terms were needed to find the 2PN tensor waveform.  Since we are only calculating the scalar waveform to 1.5PN order, we do not need them here.  

The radiation-zone fields also include contributions from the radiation zone.  It turns out that these pieces begin at 0.5PN order, one-half order higher than we express \eqref{eq:NRN}--\eqref{eq:PsiRN}.  Some of those pieces are the very scalar waves we are looking to calculate.  However, we do not need them to construct the source.

In any case, we can now use \eqref{eq:NRN}--\eqref{eq:PsiRN} to find an explicit expression for $\tau_s$.  We plug them into the generic field sources \eqref{eq:tausF1}--\eqref{eq:tausF2}.  When doing so, we must be careful because the post-Newtonian counting is not the same as in the near zone.  For instance, $K^i$ is 0.5PN higher order than $N$ in the near zone but 1PN higher order in the radiation zone.  Furthermore, unlike in the near zone, time derivatives and spatial derivatives both increase a term's post-Newtonian order in the same way (adding one-half PN order).

Once we have the source, we can find the scalar waves using
\begin{equation}
\begin{split}
\Psi_{\Cs-\Ns }(t,\mathbf{x}) &= 2\int_{\tau-2\Rs}^\tau d\tau'\int_0^{2\pi}d\phi' \int_{1-\xi}^1 \frac{\tau_s(\tau'+R',\mathbf{x}')}{t-\tau'-\mathbf{\hat{N}}'\cdot \mathbf{x}}[R'(\tau',\Omega')]^2d\cos\theta'\\
&\quad +2\int_{-\infty}^{\tau-2\Rs}d\tau'\oint\frac{\tau_s(\tau'+R',\mathbf{x}')}{t-\tau'-\mathbf{\hat{N}}'\cdot\mathbf{x}}[R'(\tau',\Omega')]^2d^2\Omega' \, .
\label{eq:phiRR}
\end{split}
\end{equation}
To express this in a more useful way, we note that the terms of the source all end up having the generic form

\begin{equation}
\tau_s(l, n) = \frac{1}{4\pi}\frac{f(\tau)}{R^n}\hat{N}^{\langle L\rangle} \, .
\label{eq:stdsource}
\end{equation}
If we restrict ourselves to sources of this form, \eqref{eq:phiRR} can be rewritten as

\begin{equation}
\Psi_{\Cs-\Ns} = \frac{2}{R}\hat{N}^{\langle L\rangle}\left[\int_0^\Rs f(\tau-2s)A(s,R)\ ds + \int_\Rs^\infty f(\tau-2s)B(s,R)\ ds\right] \, ,
\label{eq:ABintegrals}
\end{equation}
where

\begin{subequations}
\begin{align}
A(s,R) &\equiv \int_\Rs^{R+s} \frac{P_l(\xi)}{p^{n-1}}dp \, , \label{eq:A} \\
B(s,R) &\equiv \int_s^{R+s} \frac{P_l(\xi)}{p^{n-1}}dp \, .
\label{eq:B}
\end{align}
\end{subequations}
Here $P_l(\xi)$ are the Legendre polynomials with argument

\begin{equation}
\xi \equiv \frac{R+2s}{R}-\frac{2s(R+s)}{Rp} \, .
\end{equation}
Thus the computation of the scalar waves reduces to the evaluation of various $(l,n)$ integrals, all of which have previously been calculated for use in paper II.  Note that \eqref{eq:ABintegrals} is valid for field points in the entire radiation zone.  In order to find the fields in the far-away zone, we ignore all terms which fall off faster than $1/R$.

For simplicity, we will do the calculation order by order.  At lowest order, $\tau_s$ consists solely of the $(\nabla \Psi)^2$ term in \eqref{eq:tausF1}.  When evaluating the spatial derivatives, we remember that the moments are functions of retarded time, $\tau = t-R$, so that, for instance, $\partial_c\Is^{ab} = -\dot{\Is}^{ab}\hat{N}^c$.  Then we find

\begin{equation}
\tau_s = \frac{1}{2\pi}G^2\zeta(\lambda_1-\zeta)\frac{m_s^2}{R^4} +\cdots \, .
\end{equation}
This term requires evaluating the $l = 0, n = 4$ integrals.  The result is the 0.5PN piece of $\Psi$ which is written out in paper II, Eq.\ (6.23).  It has no pieces proportional to $1/R$ and thus does not represent scalar waves in the far-away zone.

At the next highest order, the $(\nabla \Psi)^2$ term again contributes, this time a cross term between the monopole piece of $\Psi$ and the dipole piece of $\Psi$. In addition, we get a contribution from the term proportional to $N\ddot{\Psi}$ in \eqref{eq:tausF2}; this uses the monopole piece of $N$ and the dipole piece of $\Psi$.  The result is

\begin{equation}
\tau_s = \cdots +\frac{1}{4\pi}\left[-4G(\lambda_1-\zeta)\frac{m_s}{R^2}\hat{N}^i\left(-2\frac{\Is_s^i}{R^3}-2\frac{\dot{\Is}_s^i}{R^2}-\frac{\ddot{\Is}_s^i}{R}\right)-4G(1-\zeta)\frac{m}{R^2}\hat{N}^i\left(\frac{\ddot{\Is}_s^i}{R}+\dddot{\Is}_s^i\right)\right] +\cdots \, .
\end{equation}
We have to evaluate \eqref{eq:ABintegrals} for $l = 1$ and $n =$ 2--5.  Performing these integrals and adding up the results, we find for this piece of the scalar field

\begin{equation}
\Psi_{\Cs-\Ns} = \frac{4G(1-\zeta)m}{R}\hat{N}^i\left[\ddot{\Is}_s^i+\int_0^\infty ds\ \dddot{\Is}_s^i(\tau-s) \ln\frac{s}{2R+s}\right] \, .
\label{eq:tail1}
\end{equation}
As in paper II, we have made a slight change of variable from \eqref{eq:ABintegrals}: $s \rightarrow s/2$.  Note that only the F2 piece of the source (the $N\ddot{\Psi}$ term) contributes to the final answer.

The expression \eqref{eq:tail1} is a 1PN contribution to the scalar waveform.  Therefore, we only need to go one-half order higher to find the 1.5PN waveform.  In this case, the source will again include contributions from the $(\nabla \Psi)^2$ term, this time featuring monopole-monopole (i.e., $m_s$--$m_{s1}$), monopole-quadrupole, and dipole-dipole couplings.  It also has a contribution from the $N\ddot{\Psi}$ term, this time a coupling between the monopole piece of $N$ and the quadrupole piece of $\Psi$.  In addition, we find that new terms contribute, specifically those proportional to $\dot{\Psi}^2$ (a dipole-dipole coupling, from $\tau_{s,{\text F1}}$) and $B^{ij}\Psi^{,ij}$ (from $\tau_{s,\text{F2}}$).  Adding these up, and changing the products of unit vectors to STF tensors, we find

\begin{equation}
\begin{split}
\tau_s &= \cdots + \frac{1}{4\pi}\left\{2G(\lambda_1-\zeta)\frac{m_s}{R^2}\left[\left(9\frac{\Is_s^{ij}}{R^4}+9\frac{\dot{\Is}_s^{ij}}{R^3}+4\frac{\ddot{\Is}_s^{ij}}{R^2}+\frac{\dddot{\Is}_s^{ij}}{R}\right)\hat{N}^{\langle ij\rangle} +2G\zeta \frac{m_{s1}}{R^2}+2G\zeta \frac{\dot{m}_{s1}}{R}+\frac{1}{3}\frac{\ddot{\Is}_s^{kk}}{R^2}+\frac{1}{3}\frac{\dddot{\Is}_s^{kk}}{R}\right] \right. \\
&\qquad +\frac{\lambda_1-\zeta}{\zeta}\left[\left(6\frac{\Is_s^i\Is_s^j}{R^6}+6\frac{\Is_s^i\dot{\Is}_s^j}{R^5}+6\frac{\dot{\Is}_s^i\Is_s^j}{R^5}+4\frac{\Is_s^i\ddot{\Is}_s^j}{R^4}+4\frac{\dot{\Is}_s^i\dot{\Is}_s^j}{R^4}+4\frac{\ddot{\Is}_s^i\Is_s^j}{R^4}+2\frac{\dot{\Is}_s^i\ddot{\Is}_s^j}{R^3}+2\frac{\ddot{\Is}_s^i\dot{\Is}_s^j}{R^3}\right)\hat{N}^{\langle ij\rangle} \right.\\
&\left.\qquad \quad +4\frac{\Is_s^k\Is_s^k}{R^6}+8\frac{\Is_s^k\dot{\Is}_s^k}{R^5}+\frac{8}{3}\frac{\Is_s^k\ddot{\Is}_s^k}{R^4}+\frac{10}{3}\frac{\dot{\Is}_s^k\dot{\Is}_s^k}{R^4}+\frac{4}{3}\frac{\dot{\Is}_s^k\ddot{\Is}_s^k}{R^3}\right] \\
&\qquad -2G(1-\zeta)\frac{m}{R^2}\left[\left(3\frac{\ddot{\Is}_s^{ij}}{R^2}+3\frac{\dddot{\Is}_s^{ij}}{R}+\overset{(4)}{\Is}\vphantom{\Is}_s^{ij}\right)\hat{N}^{\langle ij\rangle}+2G\zeta\ddot{m}_{s1}+\frac{1}{3}\overset{(4)}{\Is}\vphantom{\Is}_s^{kk}\right] \\
&\left.\qquad -6G\zeta\frac{m_s}{R^4}\ddot{\Is}^{ij}\hat{N}^{\langle ij\rangle}
\vphantom{2G(\lambda_1-\zeta)\frac{m_s}{R^2}\left[\left(9\frac{\Is_s^{ij}}{R^4}+9\frac{\dot{\Is}_s^{ij}}{R^3}+4\frac{\ddot{\Is}_s^{ij}}{R^2}+\frac{\dddot{\Is}_s^{ij}}{R}\right)\hat{N}^{\langle ij\rangle} +2G\zeta \frac{m_{s1}}{R^2}+2G\zeta \frac{\dot{m}_{s1}}{R}+\frac{1}{3}\frac{\ddot{\Is}_s^{kk}}{R^2}+\frac{1}{3}\frac{\dddot{\Is}_s^{kk}}{R}\right]} \right\} +\cdots \, .
\end{split}
\end{equation}
We must evaluate \eqref{eq:ABintegrals} for $l = 0, n =$ 2--6; and $l = 2, n =$ 2--6.  The final result is

\begin{equation}
\begin{split}
\Psi_{\Cs-\Ns} &= \frac{2}{R}\hat{N}^{ij}\left\{G(1-\zeta)m\left[\frac{3}{2}\dddot{\Is}_s^{ij}+\int_0^\infty ds\ \overset{(4)}{\Is}\vphantom{\Is}_s^{ij}(\tau-s)\ln \frac{s}{2R+s}\right]-\frac{1}{2}G\zeta m_s \dddot{\Is}^{ij}\right\} \\
&\quad +\frac{2}{R}\left\{G(\lambda_1-\zeta)m_s\left[-2G\zeta \dot{m}_{s1}-\frac{1}{3}\dddot{\Is}_s^{kk}\right]+\frac{2}{9}\frac{\lambda_1-\zeta}{\zeta}\Is_s^k\dddot{\Is}_s^k \right. \\
&\left.\qquad +G(1-\zeta)m\left[-\frac{1}{2}\dddot{\Is}_s^{kk}+2G\zeta\int_0^\infty ds\ \ddot{m}_{s1}(\tau-s)\ln \frac{s}{2R+s}\right]+\frac{1}{6}G\zeta m_s\dddot{\Is}^{kk} \right\} \, .
\end{split}
\label{eq:tail2}
\end{equation}
In this case, both the F1 and F2 pieces of $\tau_s$ contribute.  However, we shall see in the next section that the F1 piece (proportional to $\lambda_1-\zeta$) cancels with a piece of the near-zone contribution to the waveform.

\section{Final scalar waveform}
\label{sec:scalarwaves}

To find the final scalar waveform, we add the contributions from the near and radiation zones.  The near-zone contribution is found by inserting the differentiated two-body scalar multipole moments into \eqref{eq:scalarfaraway}.  We ignore the $-1$PN contribution from the monopole moment, since it does not depend on time (and thus does not have a wavelike character).  For the radiation-zone pieces \eqref{eq:tail1} and \eqref{eq:tail2}, we write the moments $\Is^M$, $\Is_s^M$, and their derivatives explicitly in terms of relative two-body variables.  However, unlike in paper II, we do not bring all of the terms inside the integrals.

The final scalar waveform can be written as a post-Newtonian expansion,

\begin{equation}
\Psi = \frac{2G\zeta\mu\alpha^{1/2}}{R}[\Psi_{-0.5}+\Psi_0+\Psi_{0.5}+\Psi_{1,\Ns}+\Psi_{1,\Cs-\Ns}+\Psi_{1.5,\Ns}+\Psi_{1.5,\Cs-\Ns}] \, ,
\label{eq:finalscalarwaves}
\end{equation}
where the subscripts denote the PN order of each term.  For clarity, we have separated out the 1PN and 1.5PN near-zone contributions (marked $\Ns$) from the radiation-zone contributions at the same order (marked $\Cs-\Ns$).  We find

\begin{subequations}
\begin{equation}
\Psi_{-0.5} = 2\Ss_-(\mathbf{\hat{N}}\cdot \mathbf{v}) \, ,
\label{eq:Psiminus05}
\end{equation}

\begin{equation}
\Psi_0 = \left(\Ss_+-\frac{\delta m}{m}\Ss_-\right)\left[-\frac{G\alpha m}{r}(\mathbf{\hat{N}}\cdot \mathbf{\hat{n}})^2+(\mathbf{\hat{N}}\cdot \mathbf{v})^2-\frac{1}{2}v^2\right]+\frac{G\alpha m}{r}\left[-2\Ss_++\frac{8}{\bar{\gamma}}(\Ss_+\bar{\beta}_++\Ss_-\bar{\beta}_-)\right] \, ,
\label{eq:Psi0}
\end{equation}

\begin{equation}
\begin{split}
\Psi_{0.5} &= \left[-\frac{\delta m}{m}\Ss_++(1-2\eta)\Ss_-\right]\left[\frac{3}{2}\frac{G\alpha m}{r}\dot{r}(\mathbf{\hat{N}}\cdot\mathbf{\hat{n}})^3-\frac{7}{2}\frac{G\alpha m}{r}(\mathbf{\hat{N}}\cdot\mathbf{v})(\mathbf{\hat{N}}\cdot\mathbf{\hat{n}})^2+(\mathbf{\hat{N}}\cdot\mathbf{v})^3\right] \\
&\quad + \frac{G\alpha m}{r}\dot{r}(\mathbf{\hat{N}}\cdot\mathbf{\hat{n}})\left[-\frac{5}{2}\frac{\delta m}{m}\Ss_++\frac{3}{2}\Ss_-+\frac{4}{\bar{\gamma}}\frac{\delta m}{m}(\Ss_+\bar{\beta}_++\Ss_-\bar{\beta}_-)-\frac{4}{\bar{\gamma}}(\Ss_-\bar{\beta}_++\Ss_+\bar{\beta}_-)\right] \\
&\quad + (\mathbf{N}\cdot \mathbf{v})\left\{\left(\frac{\delta m}{m}\Ss_+-\eta \Ss_-\right)v^2\right. \\
&\left.\qquad +\frac{G\alpha m}{r}\left[\frac{1}{2}\frac{\delta m}{m}\Ss_++\left(-\frac{3}{2}+2\eta\right)\Ss_--\frac{4}{\bar{\gamma}}\frac{\delta m}{m}(\Ss_+\bar{\beta}_++\Ss_-\bar{\beta}_-)+\frac{4}{\bar{\gamma}}(\Ss_-\bar{\beta}_++\Ss_+\bar{\beta}_-)\right]\right\} \, ,
\end{split}
\label{eq:Psi05}
\end{equation}

\begin{align}
\Psi_{1,\Ns} &= \left[(1-3\eta)\Ss_+-\frac{\delta m}{m}(1-\eta)\Ss_-\right]\left[\frac{G\alpha m}{r}(\mathbf{\hat{N}}\cdot\mathbf{\hat{n}})^4\left(-\frac{5}{2}\dot{r}^2+\frac{1}{2}v^2+\frac{7}{6}\frac{G\alpha m}{r}\right)+7\frac{G\alpha m}{r}\dot{r}(\mathbf{\hat{N}}\cdot\mathbf{v})(\mathbf{\hat{N}}\cdot\mathbf{\hat{n}})^3 \right. \nonumber \\
&\left.\qquad -8\frac{G\alpha m}{r}(\mathbf{\hat{N}}\cdot\mathbf{v})^2(\mathbf{\hat{N}}\cdot\mathbf{\hat{n}})^2+(\mathbf{\hat{N}}\cdot\mathbf{v})^4\right]
  \nonumber \\
&\quad + (\mathbf{\hat{N}}\cdot\mathbf{\hat{n}})^2\left\{\frac{G\alpha m}{r}\dot{r}^2\left[(-3+9\eta)\Ss_++3\frac{\delta m}{m}\Ss_-+\frac{6}{\bar{\gamma}}(1-2\eta)(\Ss_+\bar{\beta}_++\Ss_-\bar{\beta}_-)-\frac{6}{\bar{\gamma}}\frac{\delta m}{m}(\Ss_-\bar{\beta}_++\Ss_+\bar{\beta}_-)\right]\right.  \nonumber \\
&\qquad +\frac{G\alpha m}{r}v^2\left[\left(\frac{1}{2}-7\eta-\bar{\gamma}\right)\Ss_++\frac{\delta m}{m}\left(-\frac{1}{2}+\eta+\bar{\gamma}\right)\Ss_--\frac{2}{\bar{\gamma}}(1-2\eta)(\Ss_+\bar{\beta}_++\Ss_-\bar{\beta}_-) \right. \nonumber \\
&\left.\qquad \quad +\frac{2}{\bar{\gamma}}\frac{\delta m}{m}(\Ss_-\bar{\beta}_++\Ss_+\bar{\beta}_-)\right]  \nonumber \\
&\qquad + \left(\frac{G\alpha m}{r}\right)^2\left[\left(4+\frac{5}{2}\eta+2\bar{\gamma}+2\bar{\beta}_+-2\frac{\delta m}{m}\bar{\beta}_-\right)\Ss_++\frac{\delta m}{m}\left(-4+\frac{1}{2}\eta-2\bar{\gamma}-2\bar{\beta}_++2\frac{\delta m}{m}\bar{\beta}_-\right)\Ss_- \right. \nonumber \\
&\left.\left.\qquad \quad -\frac{2}{\bar{\gamma}}(1-2\eta)(\Ss_+\bar{\beta}_++\Ss_-\bar{\beta}_-)+\frac{2}{\bar{\gamma}}\frac{\delta m}{m}(\Ss_-\bar{\beta}_++\Ss_+\bar{\beta}_-)\right]\right\} \nonumber \\
&\quad +(\mathbf{\hat{N}}\cdot\mathbf{v})(\mathbf{\hat{N}}\cdot\mathbf{\hat{n}})\frac{G\alpha m}{r}\dot{r}\left[(8-12\eta+2\bar{\gamma})\Ss_++\frac{\delta m}{m}(-8-2\bar{\gamma})\Ss_--\frac{8}{\bar{\gamma}}(1-2\eta)(\Ss_+\bar{\beta}_++\Ss_-\bar{\beta}_-) \right. \nonumber \\
&\left.\qquad +\frac{8}{\bar{\gamma}}\frac{\delta m}{m}(\Ss_-\bar{\beta}_++\Ss_+\bar{\beta}_-)\right] \nonumber \\
&\quad +(\mathbf{\hat{N}}\cdot\mathbf{v})^2\left\{v^2\left[-\frac{1}{2}(1-3\eta)\Ss_++\frac{1}{2}\frac{\delta m}{m}(1+3\eta)\Ss_-\right]\right.  \nonumber \\
&\left.\qquad +\frac{G\alpha m}{r}\left[-(1-2\eta)\left(\Ss_+-\frac{\delta m}{m}\Ss_-\right)+\frac{4}{\bar{\gamma}}(1-2\eta)(\Ss_+\bar{\beta}_++\Ss_-\bar{\beta}_-)-\frac{4}{\bar{\gamma}}\frac{\delta m}{m}(\Ss_-\bar{\beta}_++\Ss_+\bar{\beta}_-)\right]\right\}  \nonumber \\
&\quad +(\mathbf{\hat{N}}\cdot\mathbf{\hat{n}})\left[-\frac{2}{3}\zeta \Ss_-\left(\Ss_++\frac{\delta m}{m}\Ss_-\right)^2\left(\frac{G\alpha m}{r}\right)^2\right] \nonumber \\
&\quad +v^4\left[-\frac{1}{8}(1-3\eta)\Ss_++\frac{1}{8}\frac{\delta m}{m}(1-9\eta)\Ss_-\right]  \nonumber \\
&\quad +v^2\frac{G\alpha m}{r}\left[\frac{1}{2}(1-2\eta)\left(\Ss_+-\frac{\delta m}{m}\Ss_-\right)-\frac{2}{\bar{\gamma}}(1-2\eta)(\Ss_+\bar{\beta}_++\Ss_-\bar{\beta}_-)+\frac{2}{\bar{\gamma}}\frac{\delta m}{m}(\Ss_-\bar{\beta}_++\Ss_+\bar{\beta}_-)\right]  \nonumber \\
&\quad +\dot{r}^2\frac{G\alpha m}{r}\left[\left(-\frac{7}{2}-\eta-2\bar{\gamma}\right)\Ss_++\frac{1}{2}\frac{\delta m}{m}(1+2\eta)\Ss_--\frac{2}{\bar{\gamma}}(1-2\eta)(\Ss_+\bar{\beta}_++\Ss_-\bar{\beta}_-)-\frac{2}{\bar{\gamma}}\frac{\delta m}{m}(\Ss_-\bar{\beta}_++\Ss_+\bar{\beta}_-)\right] \nonumber \\
&\quad +\left(\frac{G\alpha m}{r}\right)^2\left[\left(-2-2\bar{\gamma}+3\bar{\beta}_+-3\frac{\delta m}{m}\bar{\beta}_-\right)\Ss_++\frac{\delta m}{m}(1+\bar{\beta}_+)\Ss_--\bar{\beta}_-\Ss_--\frac{10}{\bar{\gamma}}(\Ss_+\bar{\beta}_++\Ss_-\bar{\beta}_-)\right. \nonumber \\
&\left.\qquad  -\frac{2}{\bar{\gamma}}\frac{\delta m}{m}(\Ss_-\bar{\beta}_++\Ss_+\bar{\beta}_-)+\frac{16}{\bar{\gamma}^2}\left(\Ss_++\frac{\delta m}{m}\Ss_-\right)(\bar{\beta}_+^2-\bar{\beta}_-^2)+\frac{4}{\bar{\gamma}}(\Ss_+\bar{\chi}_++\Ss_-\bar{\chi}_-)-\frac{4}{\bar{\gamma}}\frac{\delta m}{m}(\Ss_-\bar{\chi}_++\Ss_+\bar{\chi}_-)\right] \, ,
\label{eq:Psi1N}
\end{align}

\begin{equation}
\Psi_{1,\Cs-\Ns} = 4\Ss_-\left(1+\frac{1}{2}\bar{\gamma}\right)\left\{-\left(\frac{G\alpha m}{r}\right)^2(\mathbf{\hat{N}}\cdot\mathbf{\hat{n}})+\int_0^\infty ds\left[\frac{(G\alpha m)^2}{r^3}(3\dot{r}(\mathbf{\hat{N}}\cdot\mathbf{\hat{n}})-(\mathbf{\hat{N}}\cdot\mathbf{v}))\right]_{\tau-s} \ln \frac{s}{2R+s}\right\} \, ,
\label{eq:Psi1R}
\end{equation}

\begin{align}
\Psi_{1.5,\Ns} &= \left[-\frac{\delta m}{m}(1-2\eta)\Ss_++(1-4\eta+2\eta^2)\Ss_-\right]\left[(\mathbf{\hat{N}}\cdot\mathbf{\hat{n}})^5\frac{G\alpha m}{r}\dot{r}\left(\frac{35}{8}\dot{r}^2-\frac{15}{8}v^2-\frac{15}{4}\frac{G\alpha m}{r}\right)\right. \nonumber \\
&\qquad +(\mathbf{\hat{N}}\cdot\mathbf{v})(\mathbf{\hat{N}}\cdot\mathbf{\hat{n}})^4\frac{G\alpha m}{r}\left(-\frac{115}{8}\dot{r}^2+\frac{23}{8}v^2+\frac{43}{6}\frac{G\alpha m}{r}\right) \nonumber \\
&\left.\qquad +20(\mathbf{\hat{N}}\cdot\mathbf{v})^2(\mathbf{\hat{N}}\cdot\mathbf{\hat{n}})^3\frac{G\alpha m}{r}\dot{r}-15(\mathbf{\hat{N}}\cdot\mathbf{v})^3(\mathbf{\hat{N}}\cdot\mathbf{\hat{n}})^2\frac{G\alpha m}{r}+(\mathbf{\hat{N}}\cdot\mathbf{v})^5\right] \nonumber \\
&\quad +(\mathbf{\hat{N}}\cdot\mathbf{\hat{n}})^3\frac{G\alpha m}{r}\dot{r}\left\{v^2\left[\frac{\delta m}{m}\left(\frac{9}{4}-\frac{15}{2}\eta-\frac{3}{2}\bar{\gamma}\right)\Ss_++\left(-\frac{9}{4}+12\eta+\frac{3}{2}\eta^2+\frac{3}{2}\bar{\gamma}-3\eta\bar{\gamma}\right)\Ss_-\right.\right.\nonumber \\
&\left.\qquad \quad -\frac{6}{\bar{\gamma}}\frac{\delta m}{m}(1-\eta)(\Ss_+\bar{\beta}_++\Ss_-\bar{\beta}_-)+\frac{6}{\bar{\gamma}}(1-3\eta)(\Ss_-\bar{\beta}_++\Ss_+\bar{\beta}_-)\right] \nonumber \\
&\qquad +\dot{r}^2\left[\frac{\delta m}{m}\left(-5+\frac{15}{2}\eta\right)\Ss_++\left(5-\frac{35}{2}\eta-\frac{5}{2}\eta^2\right)\Ss_- \right. \nonumber \\
&\left.\qquad \quad +\frac{10}{\bar{\gamma}}\frac{\delta m}{m}(1-\eta)(\Ss_+\bar{\beta}_++\Ss_-\bar{\beta}_-)-\frac{10}{\bar{\gamma}}(1-3\eta)(\Ss_-\bar{\beta}_++\Ss_+\bar{\beta}_-)\right]\nonumber \\
&\qquad +\frac{G\alpha m}{r}\left[\frac{\delta m}{m}\left(\frac{65}{6}-\frac{5}{2}\eta+4\bar{\gamma}+4\bar{\beta}_+-4\frac{\delta m}{m}\bar{\beta}_-\right)\Ss_+\right.\nonumber \\
&\qquad \quad +\left(-\frac{65}{6}+\frac{145}{6}\eta-4\bar{\gamma}+8\eta\bar{\gamma}-\frac{16}{3}\eta^2-4(1-2\eta)\bar{\beta}_++4(1-2\eta)\frac{\delta m}{m}\bar{\beta}_-\right)\Ss_- \nonumber \\
&\left.\left.\qquad \quad -\frac{20}{3\bar{\gamma}}\frac{\delta m}{m}(1-\eta)(\Ss_+\bar{\beta}_++\Ss_-\bar{\beta}_-)+\frac{20}{3\bar{\gamma}}(1-3\eta)(\Ss_-\bar{\beta}_++\Ss_+\bar{\beta}_-)\right]\right\} \nonumber \\
&\quad +(\mathbf{\hat{N}}\cdot\mathbf{v})(\mathbf{\hat{N}}\cdot\mathbf{\hat{n}})^2\frac{G\alpha m}{r}\left\{v^2\left[\frac{\delta m}{m}\left(-\frac{13}{4}+12\eta+\frac{5}{2}\bar{\gamma}\right)\Ss_++\left(\frac{13}{4}-\frac{37}{2}\eta-\frac{1}{2}\eta^2-\frac{5}{2}\bar{\gamma}+5\eta\bar{\gamma}\right)\Ss_- \right.\right.\nonumber \\
&\left.\qquad \quad +\frac{6}{\bar{\gamma}}\frac{\delta m}{m}(1-\eta)(\Ss_+\bar{\beta}_++\Ss_-\bar{\beta}_-)-\frac{6}{\bar{\gamma}}(1-3\eta)(\Ss_-\bar{\beta}_++\Ss_+\bar{\beta}_-)\right] \nonumber \\
&\qquad +\dot{r}^2\left[\frac{\delta m}{m}(15-15\eta+3\bar{\gamma})\Ss_++\left(-15+45\eta+\frac{3}{2}\eta^2-3\bar{\gamma}+6\eta\bar{\gamma}\right)\Ss_- \right.\nonumber \\
&\left.\qquad \quad -\frac{18}{\bar{\gamma}}\frac{\delta m}{m}(1-\eta)(\Ss_+\bar{\beta}_++\Ss_-\bar{\beta}_-)+\frac{18}{\bar{\gamma}}(1-3\eta)(\Ss_-\bar{\beta}_++\Ss_+\bar{\beta}_-)\right] \nonumber \\
&\qquad +\frac{G\alpha m}{r}\left[\frac{\delta m}{m}\left(-\frac{25}{2}-\frac{3}{2}\eta-6\bar{\gamma}-7\bar{\beta}_++7\frac{\delta m}{m}\bar{\beta}_-\right)\Ss_+ \right. \nonumber \\
&\qquad \quad +\left(\frac{25}{2}-\frac{47}{2}\eta+11\eta^2+6(1-2\eta)\bar{\gamma}+7(1-2\eta)\bar{\beta}_+-7(1-2\eta)\frac{\delta m}{m}\bar{\beta}_-\right)\Ss_-\nonumber \\
&\left.\left.\qquad \quad +\frac{8}{\bar{\gamma}}\frac{\delta m}{m}(1-\eta)(\Ss_+\bar{\beta}_++\Ss_-\bar{\beta}_-)-\frac{8}{\bar{\gamma}}(1-3\eta)(\Ss_-\bar{\beta}_++\Ss_+\bar{\beta}_-)\right]\right\} \nonumber \\
&\quad +(\mathbf{\hat{N}}\cdot\mathbf{v})^2(\mathbf{\hat{N}}\cdot\mathbf{\hat{n}})\frac{G\alpha m}{r}\dot{r}\left[\frac{\delta m}{m}\left(-18+\frac{21}{2}\eta-6\bar{\gamma}\right)\Ss_++\left(18-\frac{93}{2}\eta+6(1-2\eta)\bar{\gamma}\right)\Ss_- \right.\nonumber \\
&\left.\qquad +\frac{12}{\bar{\gamma}}\frac{\delta m}{m}(1-\eta)(\Ss_+\bar{\beta}_++\Ss_-\bar{\beta}_-)-\frac{12}{\bar{\gamma}}(1-3\eta)(\Ss_-\bar{\beta}_++\Ss_+\bar{\beta}_-)\right]\nonumber \\
&\quad +(\mathbf{\hat{N}}\cdot\mathbf{\hat{v}})^3\left\{v^2\left[\frac{1}{2}\frac{\delta m}{m}(1+\eta)\Ss_++\left(-\frac{1}{2}+\frac{1}{2}\eta+5\eta^2\right)\Ss_-\right]\right. \nonumber \\
&\qquad +\frac{G\alpha m}{r}\left[\frac{\delta m}{m}\left(1-\frac{5}{2}\eta\right)\Ss_++\left(-1+\frac{9}{2}\eta-6\eta^2\right)\Ss_- \right. \nonumber \\
&\left.\left.\qquad \quad -\frac{4}{\bar{\gamma}}\frac{\delta m}{m}(1-\eta)(\Ss_+\bar{\beta}_++\Ss_-\bar{\beta}_-)+\frac{4}{\bar{\gamma}}(1-3\eta)(\Ss_-\bar{\beta}_++\Ss_+\bar{\beta}_-)\right]\right\} \nonumber \\
&\quad +8(\mathbf{\hat{N}}\cdot\mathbf{\hat{n}})^2\eta\zeta \Ss_+\Ss_-^2\left(\frac{G\alpha m}{r}\right)^2\dot{r} \nonumber \\
&\quad +(\mathbf{\hat{N}}\cdot\mathbf{v})(\mathbf{\hat{N}}\cdot\mathbf{\hat{n}})\eta\zeta \Ss_-^2\left(\frac{G\alpha m}{r}\right)^2\left(-\frac{20}{3}\Ss_+-4\frac{\delta m}{m}\Ss_-\right) \nonumber \\
&\quad +(\mathbf{\hat{N}}\cdot\mathbf{\hat{n}})\frac{G\alpha m}{r}\dot{r}\left\{v^2\left[\frac{\delta m}{m}\left(\frac{47}{8}+\frac{7}{4}\eta+\frac{5}{2}\bar{\gamma}\right)\Ss_++\left(-\frac{17}{8}+\frac{55}{4}\eta-\frac{1}{2}\bar{\gamma}+4\eta\bar{\gamma}\right)\Ss_-\right.\right. \nonumber \\
&\left.\qquad \quad -\frac{2}{\bar{\gamma}}\frac{\delta m}{m}(1+3\eta)(\Ss_+\bar{\beta}_++\Ss_-\bar{\beta}_-)+\frac{2}{\bar{\gamma}}(1-5\eta)(\Ss_-\bar{\beta}_++\Ss_+\bar{\beta}_-)\right] \nonumber \\
&\qquad +\dot{r}^2\left[\frac{\delta m}{m}\left(-\frac{45}{8}-\frac{9}{4}\eta-3\bar{\gamma}\right)\Ss_++\left(\frac{3}{8}-\frac{15}{4}\eta\right)\Ss_-+\frac{6}{\bar{\gamma}}\frac{\delta m}{m}\eta(\Ss_+\bar{\beta}_++\Ss_-\bar{\beta}_-)+\frac{6}{\bar{\gamma}}\eta(\Ss_-\bar{\beta}_++\Ss_+\bar{\beta}_-)\right] \nonumber \\
&\qquad +\frac{G\alpha m}{r}\left[\frac{\delta m}{m}\left(-\frac{1}{4}-4\eta-2\bar{\gamma}+6\bar{\beta}_+-6\frac{\delta m}{m}\bar{\beta}_-\right)\Ss_++\left(-\frac{1}{4}+14\eta+12\eta\bar{\gamma}-2(1+4\eta)\bar{\beta}_++2\frac{\delta m}{m}\bar{\beta}_-\right)\Ss_- \right. \nonumber \\
&\qquad \quad -\frac{12}{\bar{\gamma}}\left(1-\frac{2}{3}\eta\right)\frac{\delta m}{m}(\Ss_+\bar{\beta}_++\Ss_-\bar{\beta}_-)+\frac{12}{\bar{\gamma}}(\Ss_-\bar{\beta}_++\Ss_+\bar{\beta}_-) \nonumber \\
&\left.\left.\qquad \quad -\frac{64}{\bar{\gamma}^2}\eta\Ss_-(\bar{\beta}_+^2-\bar{\beta}_-^2)+\frac{8}{\bar{\gamma}}\frac{\delta m}{m}(\Ss_+\bar{\chi}_++\Ss_-\bar{\chi}_-)-\frac{8}{\bar{\gamma}}(1-2\eta)(\Ss_-\bar{\chi}_++\Ss_+\bar{\chi}_-)\right]\right\} \nonumber \\
&\quad +(\mathbf{\hat{N}}\cdot\mathbf{v})\left\{v^4\left[\frac{1}{2}\frac{\delta m}{m}(1-5\eta)\Ss_++\frac{1}{4}(1-7\eta+11\eta^2)\Ss_-\right] \right. \nonumber \\
&\qquad +\frac{G\alpha m}{r}v^2\left[\frac{\delta m}{m}\left(\frac{1}{8}+\frac{7}{4}\eta+\frac{1}{2}\bar{\gamma}\right)\Ss_++\left(\frac{25}{8}-\frac{13}{4}\eta-3\eta^2+\frac{3}{2}\bar{\gamma}-2\eta\bar{\gamma}\right)\Ss_- \right.\nonumber \\
&\left.\qquad \quad +\frac{2}{\bar{\gamma}}\frac{\delta m}{m}(1+\eta)(\Ss_+\bar{\beta}_++\Ss_-\bar{\beta}_-)-\frac{2}{\bar{\gamma}}(1-3\eta)(\Ss_-\bar{\beta}_++\Ss_+\bar{\beta}_-)\right] \nonumber \\
&\qquad +\frac{G\alpha m}{r}\dot{r}^2\left[\frac{\delta m}{m}\left(\frac{29}{8}+\frac{7}{4}\eta+2\bar{\gamma}\right)\Ss_++\left(-\frac{19}{8}-\frac{23}{4}\eta-\eta^2-(1+4\eta)\bar{\gamma}\right)\Ss_-\right. \nonumber \\
&\left.\qquad \quad -\frac{2}{\bar{\gamma}}\frac{\delta m}{m}\eta(\Ss_+\bar{\beta}_++\Ss_-\bar{\beta}_-)-\frac{2}{\bar{\gamma}}\eta(\Ss_-\bar{\beta}_++\Ss_+\bar{\beta}_-)\right] \nonumber \\
&\qquad +\left(\frac{G\alpha m}{r}\right)^2\left[\frac{\delta m}{m}\left(3+\frac{1}{2}\eta+2\bar{\gamma}-\bar{\beta}_++\frac{\delta m}{m}\bar{\beta}_-\right)\Ss_+ \right. \nonumber \\
&\qquad \quad +\left(-4-\frac{25}{2}\eta+2\eta^2-2(1+4\eta)\bar{\gamma}+(1+2\eta)\bar{\beta}_+-(1-2\eta)\frac{\delta m}{m}\bar{\beta}_-\right)\Ss_-\nonumber \\
&\qquad \quad +\frac{4}{\bar{\gamma}}\frac{\delta m}{m}(1-\eta)(\Ss_+\bar{\beta}_++\Ss_-\bar{\beta}_-)-\frac{4}{\bar{\gamma}}(1+\eta)(\Ss_-\bar{\beta}_++\Ss_+\bar{\beta}_-) \nonumber \\
&\left.\left. \qquad \quad +\frac{32}{\bar{\gamma}^2}\eta\Ss_-(\bar{\beta}_+^2-\bar{\beta}_-^2)-\frac{4}{\bar{\gamma}}\frac{\delta m}{m}(\Ss_+\bar{\chi}_++\Ss_-\bar{\chi}_-)+\frac{4}{\bar{\gamma}}(1-2\eta)(\Ss_-\bar{\chi}_++\Ss_+\bar{\chi}_-)\right]\right\} \nonumber \\
&\quad +\left(\frac{G\alpha m}{r}\right)^2\dot{r}\left\{\frac{8}{3}\zeta\Ss_+^3+\frac{14}{3}\frac{\delta m}{m}\zeta\Ss_+^2\Ss_-+\frac{4}{3}(1+\eta)\zeta\Ss_+\Ss_-^2-\frac{2}{3}\frac{\delta m}{m}(1-2\eta)\zeta\Ss_-^3 \right. \nonumber \\
&\qquad +\frac{1}{\bar{\gamma}^2}\left[\frac{64}{3}\zeta\Ss_+^3+\frac{112}{3}\frac{\delta m}{m}\zeta \Ss_+^2\Ss_-+\frac{32}{3}\zeta\Ss_+\Ss_-^2-\frac{16}{3}\frac{\delta m}{m}\zeta\Ss_-^3+16\zeta^2\Ss_+^5+32\frac{\delta m}{m}\zeta^2\Ss_+^4\Ss_- \right. \nonumber \\
&\left.\qquad \quad -32\frac{\delta m}{m}\zeta^2\Ss_+^2\Ss_-^3-16\zeta^2\Ss_+\Ss_-^4\right]\bar{\beta}_+\nonumber \\
&\qquad +\frac{1}{\bar{\gamma}^2}\left[\frac{64}{3}\frac{\delta m}{m}\zeta\Ss_+^3+\frac{112}{3}\zeta \Ss_+^2\Ss_-+\frac{32}{3}\frac{\delta m}{m}\zeta\Ss_+\Ss_-^2-\frac{16}{3}\zeta\Ss_-^3+16\zeta^2\Ss_+^4\Ss_-+32\frac{\delta m}{m}\zeta^2\Ss_+^3\Ss_-^2 \right. \nonumber \\
&\left.\qquad \quad -32\frac{\delta m}{m}\zeta^2\Ss_+\Ss_-^4-16\zeta^2\Ss_-^5\right]\bar{\beta}_-\nonumber \\
&\qquad -\frac{64}{\bar{\gamma}^3}\zeta\Ss_+\left(\Ss_+^2+2\frac{\delta m}{m}\Ss_+\Ss_-+\Ss_-^2\right)\bar{\beta}_+^2-\frac{64}{\bar{\gamma}^3}\zeta\Ss_-\left(\frac{\delta m}{m}\Ss_+^2+2\Ss_+\Ss_-+\frac{\delta m}{m}\Ss_-^2\right)\bar{\beta}_-^2\nonumber \\
&\qquad -\frac{64}{\bar{\gamma}^3}\left(\frac{\delta m}{m}\zeta \Ss_+^3+3\zeta\Ss_+^2\Ss_-+3\frac{\delta m}{m}\zeta\Ss_+\Ss_-^2+\zeta\Ss_-^3\right)\bar{\beta}_+\bar{\beta}_- \nonumber \\
&\left.\qquad +(\lambda_1-\zeta)\frac{1}{\alpha^{1/2}}\left[\frac{16}{3}\Ss_+^2+4\frac{\delta m}{m}\Ss_+\Ss_--\frac{4}{3}\left(1-\frac{8}{3}\eta\right)\Ss_-^2-\frac{16}{\bar{\gamma}}\left(\Ss_++\frac{\delta m}{m}\Ss_-\right)(\Ss_+\bar{\beta}_++\Ss_-\bar{\beta}_-)\right]\right\} \, ,
\label{eq:Psi15N}
\end{align}

\begin{equation}
\begin{split}
\Psi_{1.5,\Cs-\Ns} &= \left(1+\frac{1}{2}\bar{\gamma}\right)\left\{\left(\Ss_+-2\frac{\delta m}{m}\Ss_-\right)\left(\frac{G\alpha m}{r}\right)^2[6\dot{r}(\mathbf{\hat{N}}\cdot\mathbf{\hat{n}})^2-8(\mathbf{\hat{N}}\cdot\mathbf{v})(\mathbf{\hat{N}}\cdot\mathbf{\hat{n}})] \right.\\
&\qquad +2\left(\Ss_+-\frac{\delta m}{m}\Ss_-\right)\int_0^\infty ds \left[\frac{(G\alpha m)^2}{r^3}\left(\left(3v^2-15\dot{r}^2+\frac{G\alpha m}{r}\right)(\mathbf{\hat{N}}\cdot\mathbf{\hat{n}})^2+18\dot{r}(\mathbf{\hat{N}}\cdot\mathbf{v})(\mathbf{\hat{N}}\cdot\mathbf{\hat{n}}) \right.\right. \\
&\left.\left.\left.\qquad \quad \vphantom{\frac{(G\alpha m)^2}{r^3}}-4(\mathbf{\hat{N}}\cdot\mathbf{v})^2\right)\right]_{\tau-s} \ln \frac{s}{2R+s} \right\} \\
&\quad +\left(1+\frac{1}{2}\bar{\gamma}\right)\left\{\frac{2}{3}\left(\Ss_+-2\frac{\delta m}{m}\Ss_-\right)\left(\frac{G\alpha m}{r}\right)^2\dot{r}\right. \\
&\left.\qquad +2\left[3\Ss_+-\frac{\delta m}{m}\Ss_--\frac{8}{\bar{\gamma}}(\Ss_+\bar{\beta}_++\Ss_-\bar{\beta}_-)\right]\int_0^\infty ds\left[\frac{(G\alpha m)^2}{r^3}\left(v^2-3\dot{r}^2-\frac{G\alpha m}{r}\right)\right]_{\tau-s}\ln \frac{s}{2R+s}\right\} \\
&\quad +(\lambda_1-\zeta)\frac{1}{\alpha^{1/2}}\left[-\frac{16}{3}\Ss_+^2-4\frac{\delta m}{m}\Ss_+\Ss_-+\frac{4}{3}\left(1-\frac{8}{3}\eta\right)\Ss_-^2 \right. \\
&\left.\qquad+\frac{16}{\bar{\gamma}}\left(\Ss_++\frac{\delta m}{m}\Ss_-\right)(\Ss_+\bar{\beta}_++\Ss_-\bar{\beta}_-)\right]\left(\frac{G\alpha m}{r}\right)^2\dot{r} \, .
\end{split}
\label{eq:Psi15R}
\end{equation}
\end{subequations}
These expressions agree with previous results to 0PN order \cite{wz89}.   

We saw in paper II that the tensor waves differ from their general relativistic version only by a relatively small number of parameters, $\zeta$, $\alpha$, $\bar{\gamma}$, $\bar{\beta}_\pm$, $\bar{\delta}_\pm$, $\bar{\chi}_\pm$, and $\Ss_\pm$.  We see here that the scalar waves depend only on this same set of parameters (obviously in addition to the masses, positions, and velocities of the compact objects).  Actually, expressions \eqref{eq:Psiminus05}--\eqref{eq:Psi15R} as given do not contain the parameters $\bar{\delta}_+$ and $\bar{\delta}_-$, though it may be possible to rewrite certain pieces in terms of them.  In Sec.\ \ref{sec:flux}, we will see the $\bar{\delta}_\pm$ parameters enter the energy flux through the equations of motion.

The relative two-body moments \eqref{eq:IsTB}--\eqref{eq:IsijkTB} also contain terms which depend explicitly on the parameter $\lambda_1$, whereas in paper II $\lambda_1$ appeared only within the $\bar{\beta}_\pm$ and $\bar{\chi}_\pm$ parameters.  It turns out that most of these terms cancel when the moments are contracted with the unit vectors $\hat{N}^i$ and summed together. Some terms depending on a ``bare'' $\lambda_1$ do survive until Eqs.\ (5.2).  Terms of this form in \eqref{eq:Psi15N} arise from the 2.5PN monopole moment, while terms like this in \eqref{eq:Psi15R} come from the $\tau_{s,{\text F1}}$ contribution to the radiation-zone integral.  Inspection of these terms reveals that they differ by a sign and thus cancel.  While there is no reason {\em a priori} to expect $\lambda_1$ to only appear within $\bar{\beta}_\pm$ and $\bar{\chi}_\pm$, these exact cancellations provide a nice sanity check for our results.  

In paper II, we discussed extensively the nature of the radiation-zone contributions to the tensor waves.  We found that they are made up of two types of terms, instantaneous and hereditary.  Instantaneous terms only depend on the state of the binary at the (retarded) time of interest, while hereditary terms depend on the entire integrated history of the binary.  In the case of tensor waves, there are two types of hereditary terms: ones with a logarithmic factor in the integral and ones without.  The terms with a logarithm arise from the scattering of radiation off the background spacetime; these are known as ``gravitational-wave tails.''  In paper II, we found that the same types of tail terms appear in scalar-tensor theory as in general relativity, with only a slight modification to some coefficients.  However, new nonlogarithmic hereditary integrals also appear, at lower order than their first appearance in general relativity.  The term of this kind at 1.5PN order has a zero-frequency (DC) component which creates a ``gravitational-wave memory'' effect.

The scalar waves, by contrast, have no nonlogarithmic hereditary terms.  All hereditary terms are tail terms.  The first occurs at 1PN order, one-half order lower than in the tensor waves, and arises from a coupling between the regular monopole and scalar dipole moments.  By contrast, the lowest order tensor tail comes from a monopole-quadrupole coupling.  At 1.5PN order, the scalar waves feature two tail pieces, one from a (regular) monopole-scalar quadrupole coupling and the other from a coupling between the 0PN regular monopole ($m$) and the 1PN scalar monopole ($m_{s1}$).

When the system consists of two black holes ($s_1 = s_2 = 1/2$ and all sensitivity derivatives zero), almost all of the scalar-tensor parameters vanish.  In particular, $\Ss_+$ and $\Ss_-$ vanish.  Since all pieces of \eqref{eq:Psiminus05}--\eqref{eq:Psi15R} depend on either $\Ss_+$ or $\Ss_-$, this implies that binary black holes do not emit scalar radiation.  This is consistent with results from papers I and II.  Those papers showed that, in the case of binary black holes, the equations of motion and tensor waveform are indistinguishable from those in general relativity, except for an unobservable mass rescaling.  The lack of scalar radiation is consistent with these findings.  Effectively, we have shown that Hawking's theorem that isolated black holes are identical in GR and ST theory \cite{h72} can be extended to binaries.  Our result, valid only to the post-Newtonian orders to which we have worked, has been supported by numerical results \cite{h11} and analytic results valid to all post-Newtonian orders but only to lowest order in the mass ratio \cite{ypc12}.  All of these results are, of course, dependent on the assumptions we made, including the lack of a potential and the choice of a time-independent scalar field at infinity.  More complicated scalar-tensor theories can support scalar ``hair'' on black holes.

Another interesting case is a mixed binary, in which one object (say, body 1) is a neutron star, while the other is a black hole (with $s_2 = 1/2$).  In papers I and II, we showed that for this case, the equations of motion and the tensor waves are identical to the general relativistic results (after a mass rescaling) through 1PN order.  At 1.5PN order and 2PN order (and 2.5PN order for the equations of motion), the deviations only depend on the parameter $Q \equiv \zeta(1-\zeta)^{-1}(1-2s_1)$, where $s_1$ is the sensitivity of the neutron star.  It is straightforward to show that the scalar waves in \eqref{eq:finalscalarwaves}--(5.2) are also significantly simplified in this case, at least up to 1PN order.  Each mass is multiplied by a factor $G(1-\zeta)$, which can be absorbed through a mass rescaling.  Then the 1PN waveform only depends on $Q$ (in the forms $Q^{1/2}$ and $Q^{3/2}$) and an overall factor of $[\zeta/(1-\zeta)]^{1/2}$.  This remarkable simplification is the result of the inability of the black hole to support its own scalar hair.  It could be helpful when using gravitational waves to test scalar-tensor gravity.  At 1.5PN order, the waveform is more complicated, with some terms depending on $\lambda_1$ and $s_1'$ in addition to $Q$.  These terms arise entirely from the 2.5PN scalar monopole moment.  In Sec.\ \ref{sec:flux}, we shall see that the 2.5PN monopole terms do not contribute to the 1PN flux.  Thus the 1PN flux depends only on $Q$ (plus the mass rescaling factor and an overall amplitude correction).

\section{Energy lost to tensor and scalar waves}
\label{sec:flux}

In this section, we compute the total energy flux as a result of the tensor and scalar radiation from the compact binary system.  The flux from the tensor waves is given by \cite{w93}

\begin{equation}
\frac{dE_T}{dt} = \frac{R^2}{32\pi}\phi_0\oint \dot{\tilde{h}}_\text{TT}^{ij}\dot{\tilde{h}}_\text{TT}^{ij} d^2\Omega \, ,
\label{eq:tensorfluxeqn}
\end{equation}
where the transverse-traceless projection of the tensor field is given by

\begin{equation}
\tilde{h}_\text{TT}^{ij} = \left(P^{ik}P^{jl}-\frac{1}{2}P^{ij}P^{kl}\right)\tilde{h}^{kl} \, .
\end{equation}
The projection tensor is

\begin{equation}
P^{ij} = \delta^{ij}-\hat{N}^i\hat{N}^j \, ,
\end{equation}
which obeys $P^{ii} = 2$, $P^{ij}P^{ij} = 2$, and $P^{ij}P^{ik} = P^{jk}$.  From these, one can show

\begin{equation}
\left(P^{ik}P^{jl}-\frac{1}{2}P^{ij}P^{kl}\right)\left(P^{im}P^{jn}-\frac{1}{2}P^{ij}P^{mn}\right) = P^{km}P^{ln}-\frac{1}{2}P^{kl}P^{mn} \, .
\end{equation}
This expression simplifies the calculation considerably.  Performing the time derivatives, plugging into \eqref{eq:tensorfluxeqn}, and evaluating the angular integrals (see Appendix B of paper II), we find the tensor flux

\begin{equation}
\begin{split}
\frac{dE_T}{dt} &= \frac{8}{15}\frac{G(1-\zeta)\mu^2}{r^2}\left(\frac{G\alpha m}{r}\right)^2\left\{\vphantom{\left(\frac{G\alpha m}{r}\right)^2}12v^2-11\dot{r}^2 \right.\\
&\quad +\frac{1}{28}\left[\vphantom{\left(\frac{G\alpha m}{r}\right)^2}(785-852\eta+336\bar{\gamma})v^4-2(1487-1392\eta+616\bar{\gamma})v^2\dot{r}^2 \right. \\
&\qquad +3(687-620\eta+280\bar{\gamma})\dot{r}^4-16\left(170-10\eta+63\bar{\gamma}+84\bar{\beta}_+-84\frac{\delta m}{m}\bar{\beta}_-\right)v^2\frac{G\alpha m}{r} \\
&\left.\left.\qquad +8\left(367-15\eta+140\bar{\gamma}+168\bar{\beta}_+-168\frac{\delta m}{m}\bar{\beta}_-\right)\dot{r}^2\frac{G\alpha m}{r}+16(1-4\eta)\left(\frac{G\alpha m}{r}\right)^2\right]\right\} \, .
\end{split}
\label{eq:tensorflux}
\end{equation}
In the limit $\zeta = \bar{\gamma} = \bar{\beta}_\pm = 0$, this reduces to the GR result in Eqs.\ (6.12)--(6.13) of \cite{ww96}.  Following our convention with the waveform, we call the lowest order piece of the tensor flux \eqref{eq:tensorflux} ``0PN.''   It results from multiplying the 0PN piece of $\dot{\tilde{h}}^{ij}_\text{TT}$ (i.e., the lowest order quadrupole contribution) by itself.  At 0.5PN order, there is no tensor flux because the product of the 0PN and 0.5PN $\dot{\tilde{h}}^{ij}_\text{TT}$ has an odd number of unit vectors $\hat{N}^i$.  [See paper II, Eqs.\ (7.2a)--(7.2b)].  The rest of the terms are at 1PN order, comprising 0PN--1PN and 0.5PN--0.5PN products.  The 1PN $\dot{\tilde{h}}^{ij}_\text{TT}$ includes contributions from both the two-index moment (which gives zero $\hat{N}^i$) and the four-index moment (which gives two $\hat{N}^i$).  Both of these can couple to the 0PN $\dot{\tilde{h}}^{ij}_\text{TT}$.  

The scalar flux is given by \cite{w93}

\begin{equation}
\frac{dE_S}{dt} = \frac{R^2}{32\pi}\phi_0(4\omega_0+6)\oint \dot{\Psi}^2 d^2\Omega \, .
\end{equation}
Carrying out this integral, we find that the total energy flux (including the tensor terms) can be written as

\begin{equation}
\frac{dE}{dt} = \dot{E}_{-1}+\dot{E}_0+\dot{E}_{0.5, \Cs}+\dot{E}_{0.5, \Cs-\Ns}+\dot{E}_1 \, ,
\label{eq:totalflux}
\end{equation}
where

\begin{subequations}

\begin{equation}
\dot{E}_{-1} = \frac{4}{3}\frac{\mu\eta}{r}\left(\frac{G\alpha m}{r}\right)^3\zeta\Ss_-^2 \, ,
\label{eq:minus1PNflux}
\end{equation}

\begin{equation}
\begin{split}
\dot{E}_0 &= \frac{8}{15}\frac{\mu\eta}{r}\left(\frac{G\alpha m}{r}\right)^3\left\{\frac{G\alpha m}{r}\left[-2\frac{\delta m}{m}\zeta \Ss_+\Ss_-+\left(-23+\eta-10\bar{\gamma}-10\bar{\beta}_++10\frac{\delta m}{m}\bar{\beta}_-\right)\zeta\Ss_-^2\right] \right. \\
&\qquad +v^2\left[12+6\bar{\gamma}+2\zeta\Ss_+^2+2\frac{\delta m}{m}\zeta\Ss_+\Ss_-+(6-\eta+5\bar{\gamma})\zeta\Ss_-^2-\frac{10}{\bar{\gamma}}\frac{\delta m}{m}\zeta\Ss_-(\Ss_+\bar{\beta}_++\Ss_-\bar{\beta}_-)\right. \\
&\left.\qquad \quad +\frac{10}{\bar{\gamma}}\zeta\Ss_-(\Ss_-\bar{\beta}_++\Ss_+\bar{\beta}_-)\right]\\
&\qquad +\dot{r}^2\left[-11-\frac{11}{2}\bar{\gamma}+\frac{23}{2}\zeta\Ss_+^2-8\frac{\delta m}{m}\zeta\Ss_+\Ss_-+\left(-\frac{37}{2}+9\eta-10\bar{\gamma}\right)\zeta\Ss_-^2 \right.\\
&\qquad \quad -\frac{80}{\bar{\gamma}}\zeta\Ss_+(\Ss_+\bar{\beta}_++\Ss_-\bar{\beta}_-)+\frac{30}{\bar{\gamma}}\frac{\delta m}{m}\zeta\Ss_-(\Ss_+\bar{\beta}_++\Ss_-\bar{\beta}_-) \\
&\left.\left.\qquad \quad-\frac{10}{\bar{\gamma}}\zeta\Ss_-(\Ss_-\bar{\beta}_++\Ss_+\bar{\beta}_-)+\frac{120}{\bar{\gamma}^2}\zeta(\Ss_+\bar{\beta}_++\Ss_-\bar{\beta}_-)^2\right]\right\} \, ,
\label{eq:0PNflux}
\end{split}
\end{equation}

\begin{equation}
\dot{E}_{0.5,\Cs} = -\frac{16}{9}\frac{\mu\eta}{r}\left(\frac{G\alpha m}{r}\right)^3(\zeta\Ss_-)^2\left(\Ss_+^2+2\frac{\delta m}{m}\Ss_+\Ss_-+\Ss_-^2\right)\frac{G\alpha m}{r}\dot{r} \, ,
\label{eq:05NZflux}
\end{equation}

\begin{equation}
\begin{split}
\dot{E}_{0.5,\Cs-\Ns} &= -\frac{16}{3}\frac{\mu \eta}{r}\left(1+\frac{1}{2}\bar{\gamma}\right)\zeta \Ss_-^2\left\{2\left(\frac{G\alpha m}{r}\right)^3\frac{G\alpha m}{r}\dot{r} \right. \\
&\left.\qquad +\frac{(G\alpha m)^3}{r} \hat{n}^k\int_0^\infty ds\ \left[\frac{G\alpha m}{r^4}\left(\left(3v^2-15\dot{r}^2-2\frac{G\alpha m}{r}\right)\hat{n}^k+6\dot{r}v^k\right)\right]_{\tau-s}\ln \frac{s}{2R+s}\right\} \, ,
\label{eq:RZflux}
\end{split}
\end{equation}

\begin{align}
\dot{E}_1 &= \frac{1}{105}\frac{\mu\eta}{r}\left(\frac{G\alpha m}{r}\right)^3\left\{v^4\left[\vphantom{\frac{\delta m}{m}}1570-1704\eta+1457\bar{\gamma}-852\eta\bar{\gamma}+336\bar{\gamma}^2+(219-444\eta+112\bar{\gamma})\zeta\Ss_+^2 \right. \right. \nonumber \\
&\qquad \quad -(78+320\eta+28\bar{\gamma}-560\bar{\beta}_+)\frac{\delta m}{m}\zeta\Ss_+\Ss_-+560\bar{\beta}_-\zeta\Ss_+\Ss_- \nonumber \\
&\qquad \quad +\left(139-372\eta-12\eta^2+336(1-\eta)\bar{\gamma}+140\bar{\gamma}^2+560\bar{\beta}_+-560\frac{\delta m}{m}\bar{\beta}_-\right)\zeta\Ss_-^2  \nonumber \\
&\qquad \quad +\frac{1344}{\bar{\gamma}}\eta \zeta\Ss_+(\Ss_+\bar{\beta}_++\Ss_-\bar{\beta}_-)+\frac{1}{\bar{\gamma}}(-1008+1568\eta)\frac{\delta m}{m}\zeta\Ss_-(\Ss_+\bar{\beta}_++\Ss_-\bar{\beta}_-)  \nonumber \\
&\qquad \quad +\frac{1008}{\bar{\gamma}}(1-\eta)\zeta\Ss_-(\Ss_-\bar{\beta}_++\Ss_+\bar{\beta}_-)+\frac{560}{\bar{\gamma}^2}(1-4\eta)\zeta(\Ss_+\bar{\beta}_++\Ss_-\bar{\beta}_-)^2  \nonumber \\
&\left.\qquad \quad -\frac{1120}{\bar{\gamma}^2}\frac{\delta m}{m}\zeta(\Ss_+\bar{\beta}_++\Ss_-\bar{\beta}_-)(\Ss_-\bar{\beta}_++\Ss_+\bar{\beta}_-)+\frac{560}{\bar{\gamma}^2}\zeta(\Ss_-\bar{\beta}_++\Ss_+\bar{\beta}_-)^2\right]  \nonumber \\
&\quad +v^2\dot{r}^2\left[\vphantom{\frac{\delta m}{m}}-5948+5568\eta-5438\bar{\gamma}+2784\eta\bar{\gamma}-1232\bar{\gamma}^2 \right.\nonumber \\
&\qquad \quad +(-2826-1824\eta-1904\bar{\gamma}+4480\bar{\beta}_+)\zeta\Ss_+^2+(-1404+2864\eta-616\bar{\gamma}+6160\bar{\beta}_+)\frac{\delta m}{m}\zeta\Ss_+\Ss_-+560\bar{\beta}_-\zeta\Ss_+\Ss_-  \nonumber \\
&\qquad \quad +\left(2830+2452\eta-720\eta^2+1400\bar{\gamma}+1008\eta\bar{\gamma}-3920\bar{\beta}_++6160\frac{\delta m}{m}\bar{\beta}_-\right)\zeta\Ss_-^2 \nonumber \\
&\qquad \quad +\frac{1}{\bar{\gamma}}(2800+13440\eta)\zeta\Ss_+(\Ss_+\bar{\beta}_++\Ss_-\bar{\beta}_-)  \nonumber \\
&\qquad \quad +\frac{1}{\bar{\gamma}}(12656-11312\eta)\frac{\delta m}{m}\zeta\Ss_-(\Ss_+\bar{\beta}_++\Ss_-\bar{\beta}_-)+\frac{336}{\bar{\gamma}}\frac{\delta m}{m}\zeta\Ss_+(\Ss_-\bar{\beta}_++\Ss_+\bar{\beta}_-)  \nonumber \\
&\qquad \quad +\frac{1}{\bar{\gamma}}(-9072+2016\eta)\zeta\Ss_-(\Ss_-\bar{\beta}_++\Ss_+\bar{\beta}_-)+\frac{1120}{\bar{\gamma}^2}(7-16\eta)\zeta(\Ss_+\bar{\beta}_++\Ss_-\bar{\beta}_-)^2  \nonumber \\
&\left.\qquad \quad +\frac{4480}{\bar{\gamma}^2}\frac{\delta m}{m}\zeta(\Ss_+\bar{\beta}_++\Ss_-\bar{\beta}_-)(\Ss_-\bar{\beta}_++\Ss_+\bar{\beta}_-)+\frac{1120}{\bar{\gamma}^2}\zeta(\Ss_-\bar{\beta}_++\Ss_+\bar{\beta}_-)^2\right]  \nonumber \\
&\quad +\dot{r}^4\left[\vphantom{\frac{\delta m}{m}}4122-3720\eta+3741\bar{\gamma}-1860\eta\bar{\gamma}+840\bar{\gamma}^2 \right.\nonumber \\
&\qquad \quad +(3387+3924\eta+2520\bar{\gamma}-6720\bar{\beta}_+)\zeta\Ss_+^2+(1962-3168\eta+756\bar{\gamma}-6720\bar{\beta}_+)\frac{\delta m}{m}\zeta\Ss_+\Ss_--3360\bar{\beta}_-\zeta\Ss_+\Ss_-  \nonumber \\
&\qquad \quad +\left(-2409-2712\eta+276\eta^2-1176\bar{\gamma}-1008\eta\bar{\gamma}+3360\bar{\beta}_+-6720\frac{\delta m}{m}\bar{\beta}_-\right)\zeta\Ss_-^2 \nonumber \\
&\qquad \quad-\frac{1}{\bar{\gamma}}(3024+24192\eta)\zeta\Ss_+(\Ss_+\bar{\beta}_++\Ss_-\bar{\beta}_-)  \nonumber \\
&\qquad \quad +\frac{1}{\bar{\gamma}}(-14784+11760\eta)\frac{\delta m}{m}\zeta\Ss_-(\Ss_+\bar{\beta}_++\Ss_-\bar{\beta}_-)-\frac{336}{\bar{\gamma}}\frac{\delta m}{m}\zeta\Ss_+(\Ss_-\bar{\beta}_++\Ss_+\bar{\beta}_-)  \nonumber \\
&\qquad \quad +\frac{1}{\bar{\gamma}}(8064-1008\eta)\zeta\Ss_-(\Ss_-\bar{\beta}_++\Ss_+\bar{\beta}_-)+\frac{1}{\bar{\gamma}^2}(-15120+33600\eta)\zeta(\Ss_+\bar{\beta}_++\Ss_-\bar{\beta}_-)^2  \nonumber \\
&\left.\qquad \quad -\frac{3360}{\bar{\gamma}^2}\frac{\delta m}{m}\zeta(\Ss_+\bar{\beta}_++\Ss_-\bar{\beta}_-)(\Ss_-\bar{\beta}_++\Ss_+\bar{\beta}_-)-\frac{1680}{\bar{\gamma}^2}\zeta(\Ss_-\bar{\beta}_++\Ss_+\bar{\beta}_-)^2\right]  \nonumber \\
&\quad + v^2\frac{G\alpha m}{r}\left[-5440+320\eta-4736\bar{\gamma}+160\eta\bar{\gamma}-1008\bar{\gamma}^2-2688\bar{\beta}_+-1344\bar{\gamma}\bar{\beta}_++2688\frac{\delta m}{m}\bar{\beta}_-+1344\bar{\gamma}\frac{\delta m}{m}\bar{\beta}_- \right. \nonumber \\
&\qquad \quad +\left(-960+320\eta-336\bar{\gamma}-448\bar{\beta}_++448\frac{\delta m}{m}\bar{\beta}_-\right)\zeta\Ss_+^2+\left(200+384\eta+280\bar{\gamma}+448\bar{\beta}_+\right)\frac{\delta m}{m}\zeta\Ss_+\Ss_- \nonumber \\
&\qquad \quad -(448+2688\eta)\bar{\beta}_-\zeta\Ss_+\Ss_- +\left(-1830+46\eta+90\eta^2-2184\bar{\gamma}-280\eta\bar{\gamma}-560\bar{\gamma}^2-1400\bar{\beta}_++616\eta\bar{\beta}_+ \vphantom{\frac{\delta m}{m}}\right. \nonumber \\
&\left. \qquad \qquad -560\bar{\gamma}\bar{\beta}_++1400\frac{\delta m}{m}\bar{\beta}_--56\eta\frac{\delta m}{m}\bar{\beta}_- +560\bar{\gamma}\frac{\delta m}{m}\bar{\beta}_-+\frac{8960}{\bar{\gamma}^2}\eta(\bar{\beta}_+^2-\bar{\beta}_-^2)\right)\zeta\Ss_-^2 \nonumber \\
&\qquad \quad +\frac{448}{\bar{\gamma}}(1-3\eta)\zeta\Ss_+(\Ss_+\bar{\beta}_++\Ss_-\bar{\beta}_-)+\frac{1}{\bar{\gamma}}\left(4144-1456\eta+1120\bar{\beta}_+-1120\frac{\delta m}{m}\bar{\beta}_-\right)\frac{\delta m}{m}\zeta\Ss_-(\Ss_+\bar{\beta}_++\Ss_-\bar{\beta}_-)  \nonumber \\
&\qquad \quad -\frac{448}{\bar{\gamma}}\frac{\delta m}{m}\zeta\Ss_+(\Ss_-\bar{\beta}_++\Ss_+\bar{\beta}_-)+\frac{1}{\bar{\gamma}}\left(-4144-1008\eta-1120\bar{\beta}_++1120\frac{\delta m}{m}\bar{\beta}_-\right)\zeta\Ss_-(\Ss_-\bar{\beta}_++\Ss_+\bar{\beta}_-)  \nonumber \\
&\left.\qquad \quad -\frac{1120}{\bar{\gamma}}\frac{\delta m}{m}\zeta\Ss_-(\Ss_+\bar{\chi}_++\Ss_-\bar{\chi}_-)+\frac{1120}{\bar{\gamma}}(1-2\eta)\zeta\Ss_-(\Ss_-\bar{\chi}_++\Ss_+\bar{\chi}_-)\right]  \nonumber \\
&\quad +\dot{r}^2\frac{G\alpha m}{r}\left[5872-240\eta+5176\bar{\gamma}-120\eta\bar{\gamma}+1120\bar{\gamma}^2+2688\bar{\beta}_++1344\bar{\gamma}\bar{\beta}_+-2688\frac{\delta m}{m}\bar{\beta}_--1344\bar{\gamma}\frac{\delta m}{m}\bar{\beta}_- \right. \nonumber \\
&\qquad \quad +\left(3128+2392\eta+3360\bar{\gamma}-12992\bar{\beta}_++4032\frac{\delta m}{m}\bar{\beta}_--\frac{17920}{\bar{\gamma}^2}(\bar{\beta}_+^2-\bar{\beta}_-^2)\right)\zeta\Ss_+^2 \nonumber \\
&\qquad \quad +\left(1640-3000\eta+168\bar{\gamma}-3360\bar{\beta}_+-\frac{13440}{\bar{\gamma}^2}(\bar{\beta}_+^2-\bar{\beta}_-^2)\right)\frac{\delta m}{m}\zeta\Ss_+\Ss_-+(-10080+13440\eta)\bar{\beta}_-\zeta\Ss_+\Ss_-  \nonumber \\
&\qquad \quad +\left(4682-6402\eta+1550\eta^2+4592\bar{\gamma}-1624\eta\bar{\gamma}+1120\bar{\gamma}^2+3192\bar{\beta}_+-3920\eta\bar{\beta}_++1120\bar{\gamma}\bar{\beta}_+ \vphantom{\frac{\delta m}{m}}\right. \nonumber \\
&\left.\qquad \qquad -7672\frac{\delta m}{m}\bar{\beta}_-+1680\eta\frac{\delta m}{m}\bar{\beta}_--1120\bar{\gamma}\frac{\delta m}{m}\bar{\beta}_-+\frac{4480}{\bar{\gamma}^2}(1-8\eta)(\bar{\beta}_+^2-\bar{\beta}_-^2)\right)\zeta\Ss_-^2  \nonumber \\
&\qquad \quad +\frac{1}{\bar{\gamma}}\left(9632-17024\eta+13440\bar{\beta}_+-13440\frac{\delta m}{m}\bar{\beta}_-\right)\zeta\Ss_+(\Ss_+\bar{\beta}_++\Ss_-\bar{\beta}_-)  \nonumber \\
&\qquad \quad+\frac{53760}{\bar{\gamma}^3}(\bar{\beta}_+^2-\bar{\beta}_-^2)\zeta\Ss_+(\Ss_+\bar{\beta}_++\Ss_-\bar{\beta}_-)  \nonumber \\
&\qquad \quad +\frac{1}{\bar{\gamma}}\left(-17360+8624\eta-1120\bar{\beta}_+\right)\frac{\delta m}{m}\zeta\Ss_-(\Ss_+\bar{\beta}_++\Ss_-\bar{\beta}_-)+\frac{1}{\bar{\gamma}}(1120-17920\eta)\bar{\beta}_-\zeta\Ss_-(\Ss_+\bar{\beta}_++\Ss_-\bar{\beta}_-) \nonumber \\
&\qquad \quad +\frac{53760}{\bar{\gamma}^3}(\bar{\beta}_+^2-\bar{\beta}_-^2)\frac{\delta m}{m}\zeta\Ss_-(\Ss_+\bar{\beta}_++\Ss_-\bar{\beta}_-)  \nonumber \\
&\qquad \quad +\frac{2016}{\bar{\gamma}}\frac{\delta m}{m}\zeta\Ss_+(\Ss_-\bar{\beta}_++\Ss_+\bar{\beta}_-)+\frac{1}{\bar{\gamma}}\left(5712+2352\eta+1120\bar{\beta}_+-1120\frac{\delta m}{m}\bar{\beta}_-\right)\zeta\Ss_-(\Ss_-\bar{\beta}_++\Ss_+\bar{\beta}_-)  \nonumber \\
&\qquad \quad +\frac{1}{\bar{\gamma}^2}(-49280+31360\eta)\zeta(\Ss_+\bar{\beta}_++\Ss_-\bar{\beta}_-)^2-\frac{4480}{\bar{\gamma}^2}\frac{\delta m}{m}\zeta(\Ss_+\bar{\beta}_++\Ss_-\bar{\beta}_-)(\Ss_-\bar{\beta}_++\Ss_+\bar{\beta}_-)  \nonumber \\
&\qquad \quad -\frac{4480}{\bar{\gamma}}\zeta\Ss_+(\Ss_+\bar{\chi}_++\Ss_-\bar{\chi}_-)+\frac{3360}{\bar{\gamma}}\frac{\delta m}{m}\zeta\Ss_-(\Ss_+\bar{\chi}_++\Ss_-\bar{\chi}_-)  \nonumber \\
&\qquad \quad +\frac{4480}{\bar{\gamma}}\frac{\delta m}{m}\zeta\Ss_+(\Ss_-\bar{\chi}_++\Ss_+\bar{\chi}_-)+\frac{1}{\bar{\gamma}}(-3360+8960\eta)\zeta\Ss_-(\Ss_-\bar{\chi}_++\Ss_+\bar{\chi}_-)  \nonumber \\
&\left. \qquad \quad +\frac{13440}{\bar{\gamma}^2}\zeta(\Ss_+\bar{\beta}_++\Ss_-\bar{\beta}_-)(\Ss_+\bar{\chi}_++\Ss_-\bar{\chi}_-)-\frac{13440}{\bar{\gamma}^2}\frac{\delta m}{m}\zeta(\Ss_+\bar{\beta}_++\Ss_-\bar{\beta}_-)(\Ss_-\bar{\chi}_++\Ss_+\bar{\chi}_-)\right]  \nonumber \\
&\quad +\left(\frac{G\alpha m}{r}\right)^2\left[32(1-4\eta)+32(1-4\eta)\zeta\Ss_+^2+\left(984-176\eta+280\bar{\gamma}+896\bar{\beta}_+-896\frac{\delta m}{m}\bar{\beta}_-\right)\frac{\delta m}{m}\zeta\Ss_+\Ss_- \right.  \nonumber \\
&\qquad \quad +\left(5914+3112\eta-22\eta^2+5320\bar{\gamma}+1120\eta\bar{\gamma}+1190\bar{\gamma}^2-140\eta\bar{\gamma}^2+5264\bar{\beta}_++1512\eta\bar{\beta}_++2240\bar{\gamma}\bar{\beta}_+ \vphantom{\frac{4480}{\bar{\gamma}^2}}\right.  \nonumber \\
&\qquad \qquad -5264\frac{\delta m}{m}\bar{\beta}_-+168\eta\frac{\delta m}{m}\bar{\beta}_--2240\bar{\gamma}\frac{\delta m}{m}\bar{\beta}_-+560\bar{\beta}_+^2+560(1-4\eta)\bar{\beta}_-^2-1120\frac{\delta m}{m}\bar{\beta}_+\bar{\beta}_-+280(1-2\eta)\bar{\delta}_+  \nonumber \\
&\left.\qquad \qquad +280\frac{\delta m}{m}\bar{\delta}_--\frac{6720}{\bar{\gamma}^2}\eta(\bar{\beta}_+^2-\bar{\beta}_-^2)-\frac{4480}{\bar{\gamma}^2}\eta(\bar{\beta}_+^2-\bar{\beta}_-^2)-560(1-2\eta)\bar{\chi}_++560\frac{\delta m}{m}\bar{\chi}_-\right)\zeta\Ss_-^2  \nonumber \\
&\qquad \quad +\frac{1}{\bar{\gamma}}(-1008+448\eta)\frac{\delta m}{m}\zeta\Ss_-(\Ss_+\bar{\beta}_++\Ss_-\bar{\beta}_-)+\frac{1}{\bar{\gamma}}(1008-224\eta)\zeta\Ss_-(\Ss_-\bar{\beta}_++\Ss_+\bar{\beta}_-) \nonumber \\
&\left.\left.\qquad \quad +\frac{560}{\bar{\gamma}}\frac{\delta m}{m}\zeta\Ss_-(\Ss_+\bar{\chi}_++\Ss_-\bar{\chi}_-)-\frac{560}{\bar{\gamma}}(1-2\eta)\zeta\Ss_-(\Ss_-\bar{\chi}_++\Ss_+\bar{\chi}_-)\right]\right\} \, .
\label{eq:1PNflux}
\end{align}
\end{subequations}
The tensor flux terms are easily spotted as the ones without factors of $\Ss_+$ or $\Ss_-$.  To incorporate \eqref{eq:tensorflux} into this expression, we used the relation $1-\zeta = \alpha(1+\bar{\gamma}/2)$.  

Since we have defined the tensor flux to begin at 0PN order, the lowest order piece of the total flux is actually at $-1$PN order, resulting from multiplying the $-0.5$PN piece of $\dot{\Psi}$ by itself.  Thus scalar-tensor theory predicts an energy flux contribution at lower post-Newtonian order than general relativity.  This term is well known and has been used to set strong limits on the theory.  It was also calculated in paper I, Eq.\ (6.16), using the conserved energy $E$ and the equations of motion.  Our result in \eqref{eq:minus1PNflux} agrees except for a choice of sign; paper I considers the energy {\em lost} by the system, while we consider the energy flux at infinity.

There is no contribution to the flux at $-0.5$PN order because the product of the $-0.5$PN and 0PN pieces of $\dot{\Psi}$ has an odd number of $\hat{N}^i$.  At 0PN order, we have ($-0.5$PN)--(+0.5PN) and 0PN--0PN contributions from the scalar waveform, in addition to the contributions from the tensor waveform.  Generally speaking, because the scalar waves begin at $-0.5$PN order, the calculation of the $N$th order flux requires at least some pieces of the $(N+1/2)$th order scalar waveform.  For the 0PN flux, we thus need $\Psi_{0.5}$.  We also need the 1PN correction to the time derivative of $\Psi_{-0.5}$.  In general, calculating the $N$th order flux requires knowing the $(N+1)$th order equations of motion for purposes of evaluating $\dot{\Psi}$.

The 0PN total flux was previously calculated by Will and Zaglauer \cite{wz89}, but they neglected to include any of the ($-0.5$PN)--(+0.5PN) contributions.  Thus their final expression [Eqs.\ (2.23)--(2.24) of \cite{wz89}] is in error.  When those contributions are added to the Will-Zaglauer flux, the result is in agreement with \eqref{eq:0PNflux} \cite{will}.  On the other hand, the 0PN expression for the flux as determined by Damour and Esposito-Far\`{e}se \cite{de92} does include all necessary contributions and, after considerable effort to translate notations, can be shown to agree with Eq.\ \eqref{eq:0PNflux}.

The 0PN flux was also calculated in paper I, Eqs.\ (6.19)--(6.20) using the conserved energy $E$.  At this order, comparing the two methods is a bit tricky.  Our result \eqref{eq:0PNflux} depends on $G\alpha m/r^5$, whereas the paper I result does not.  In fact, we can extract from \eqref{eq:0PNflux} a term proportional to

\begin{equation}
\frac{v^2}{r^4}-4\frac{\dot{r}^2}{r^4}-\frac{G\alpha m}{r^5} \, ,
\end{equation}
which is, to lowest order, the time derivative of $\dot{r}/r^3$.  Therefore, the time integral of the term can be absorbed into the definition of $E$ at 2.5PN order.  Applying this redefinition and a few other minor simplifications, we find for the 0PN flux

\begin{equation}
\begin{split}
\dot{E}_0 &= \frac{8}{15}\frac{\mu\eta}{r}\left(\frac{G\alpha m}{r}\right)^3\left\{v^2\left[12+5\bar{\gamma}-5(3+\bar{\gamma}+2\bar{\beta}_+)\zeta\Ss_-^2+\frac{10}{\bar{\gamma}}\zeta\Ss_-(\Ss_-\bar{\beta}_++\Ss_+\bar{\beta}_-)\right. \right. \\
&\left.\qquad \quad +10\frac{\delta m}{m}\zeta\Ss_-^2\bar{\beta}_--\frac{10}{\bar{\gamma}}\frac{\delta m}{m}\zeta\Ss_-(\Ss_+\bar{\beta}_++\Ss_-\bar{\beta}_-)\right] \\
&\qquad +\dot{r}^2\left[-11-\frac{45}{4}\bar{\gamma}+40\bar{\beta}_++5(17+\eta+6\bar{\gamma}+8\bar{\beta}_+)\zeta\Ss_-^2-\frac{90}{\bar{\gamma}}\zeta\Ss_-(\Ss_-\bar{\beta}_++\Ss_+\bar{\beta}_-) \right. \\
&\left.\left. \qquad \quad -40\frac{\delta m}{m}\zeta\Ss_-^2\bar{\beta}_-+\frac{30}{\bar{\gamma}}\frac{\delta m}{m}\zeta\Ss_-(\Ss_+\bar{\beta}_++\Ss_-\bar{\beta}_-)+\frac{120}{\bar{\gamma}^2}\zeta(\Ss_+\bar{\beta}_++\Ss_-\bar{\beta}_-)^2\right]\right\} \, .
\label{eq:revised0PNflux}
\end{split}
\end{equation}
This agrees with paper I.  Note that if we use \eqref{eq:revised0PNflux} for the 0PN flux instead of \eqref{eq:0PNflux}, we must be careful to modify \eqref{eq:1PNflux} by higher order corrections to $d/dt(\dot{r}/r^3)$.

There are two pieces of the flux at 0.5PN order.  Both pieces arise from multiplying together the $-0.5$PN and 1PN pieces of $\dot{\Psi}$.  However, in one case, the 1PN piece comes from the near-zone integral, while in the other, it comes from the radiation-zone integral.  These products integrate to a nonzero quantity because the 1PN $\dot{\Psi}$ contains at least a few terms with just one factor of $\hat{N}^i$.  (In fact, the entire radiation-zone piece depends on a single $\hat{N}^i$.)  By contrast, no 0PN--0.5PN products survive the integration process.  All 0PN pieces of $\dot{\Psi}$ include an even number of $\hat{N}^i$, while all 0.5PN pieces have an odd number of these unit vectors.

The 0.5PN flux can also be simplified from the forms in \eqref{eq:05NZflux}--\eqref{eq:RZflux} by absorbing total time derivatives into the definition of $E$.  In fact, all of \eqref{eq:05NZflux} is a time derivative, so we can write $\dot{E}_{0.5,\Ns} = 0$.  In \eqref{eq:RZflux}, the first piece (which was generated by the instantaneous term of $\Psi_{1,\Cs-\Ns}$) is also a total time derivative.  The second piece, involving the tail term, can be integrated by parts, generating one more total time derivative and one surviving term,

\begin{equation}
\dot{E}_{0.5,\Cs-\Ns} = -\frac{4}{3}\frac{1-\zeta}{\zeta}m\dddot{\Is}_s^k\int_0^\infty ds\ \dddot{\Is}_s^k(\tau-s)\ln \frac{s}{2R+s} \, .
\label{eq:revisedRZflux}
\end{equation}
Written in this way, the radiation-zone contribution to the flux matches the form used in \cite{ww96} (for the GR flux).

The 1PN piece of the flux comes from ($-0.5$PN)--1.5PN, 0PN--1PN, and 0.5PN--0.5PN products of $\dot{\Psi}$, plus the tensor contribution.  Careful examination of these products reveals that neither $\Psi_{1,\Cs-\Ns}$ nor $\Psi_{1.5,\Cs-\Ns}$  contributes to the flux at this order, hence the lack of tail integrals in \eqref{eq:1PNflux}.  Similarly, the monopole and quadrupole pieces of $\Psi_{1.5,\Ns}$ do not contribute.  As always, the relevant issue is the parity of unit vectors $\hat{N}^i$.  One could attempt to manipulate the form of \eqref{eq:1PNflux} by absorbing total time derivatives into the definition of $E$, but it is not clear that any miraculous simplification would result.

It is useful to think about what we need to calculate further pieces of the energy flux.  With the 2PN tensor waveform, calculated in paper II, we could easily compute the 2PN tensor flux.  However, the 2PN scalar flux is not so easy.  Computing it would require certain pieces of the 2.5PN scalar waveform, specifically those which have an odd number of $\hat{N}^i$.  This set includes the contribution from the scalar dipole moment, meaning that we would need to know $\Is_s^i$ to relative 3PN order.  As we discussed earlier, even the 3PN monopole moment requires a great deal of effort to calculate.  The 3PN dipole moment is even more difficult, due to the presence of $x^i$ in the integrals.  We also need the 3PN equations of motion in order to evaluate the time derivative of $\Psi_{-0.5}$ to the appropriate order.  Paper I only computed the equations of motion to 2.5PN order.

With the energy flux and the conserved energy [paper I, Eq.\ (6.4)], we can determine the evolution of the orbital separation, frequency, and phase.  These will allow the generation of ``ready-to-go'' templates for gravitational-wave studies.  From these templates, we can use parameter estimation techniques to study how well the advanced detectors will be able to measure deviations from general relativity, as well as how the new complexity of the waveforms will affect measurements of astrophysical parameters.  This is the subject of future work.

\acknowledgments

We are extremely grateful to Clifford Will for his many useful insights into the calculations and careful reading of the manuscript.  This work was primarily supported by the National Science Foundation (NSF), Grants No.\ PHY-1260995 and No.\ PHY-1306069.  The final preparation of the manuscript was supported by NSF Grant No.\ PHY-1300903 and NASA Grant No.\ NNX13AH44G, as well as a Fortner Fellowship at the University of Illinois at Urbana-Champaign.  The software {\em Mathematica} was used to check or perform many of the calculations.

\bibliographystyle{apsrev}
\bibliography{scalarwaves}

\end{document}